\documentclass[iop,revtex4]{emulateapj}
\slugcomment{{\sc Accepted to ApJ:} February 28, 2014}

\usepackage{graphics,graphicx,subfigure,apjfonts,color,aas_macros,fancybox,wrapfig,appendix,lscape}

\def\cxo{{\em Chandra}}
\def\xmm{{\em XMM}}
\def\asca{{\em ASCA}}
\def\chic{ChIcAGO}

\def\ns{$\log N - \log S$}
\def\erg{erg~cm$^{-2}$~s$^{-1}$}

\def\etal{et al.}

\begin{document}

\title{Chasing the Identification of \textit{ASCA} Galactic Objects (ChIcAGO) - An X-ray Survey of Unidentified Sources in the Galactic Plane. I: Source Sample and Initial Results}

\shorttitle{The \chic\ Survey}

\author{Gemma E. Anderson\altaffilmark{1, $\dagger$ }, 
B. M. Gaensler\altaffilmark{1}, 
David L. Kaplan\altaffilmark{2}, 
Patrick O. Slane\altaffilmark{3},
Michael P. Muno\altaffilmark{4, $\ddagger$ }, 
Bettina Posselt\altaffilmark{5}, 
Jaesub Hong\altaffilmark{3}, 
Stephen S. Murray\altaffilmark{6},
Danny T. H. Steeghs\altaffilmark{7}
Crystal L. Brogan\altaffilmark{8},
Jeremy J. Drake\altaffilmark{3}, 
Sean A. Farrell\altaffilmark{1},
Robert A. Benjamin\altaffilmark{9}, 
Deepto Chakrabarty\altaffilmark{10},
Janet E. Drew\altaffilmark{11},
John P. Finley\altaffilmark{12},
Jonathan E. Grindlay\altaffilmark{3}, 
T. Joseph W. Lazio\altaffilmark{13}, 
Julia C. Lee\altaffilmark{3}, 
Jon C. Mauerhan\altaffilmark{14},
Marten H. van Kerkwijk\altaffilmark{15}
}

\altaffiltext{1}{Sydney Institute for Astronomy, School of Physics, The University of Sydney, NSW 2006, Australia: gemma.anderson@astro.ox.ac.uk}
\altaffiltext{2}{Department of Physics, University of Wisconsin, Milwaukee, WI 53201, USA}
\altaffiltext{3}{Harvard-Smithsonian Center for Astrophysics, Cambridge, MA 02138, USA}
\altaffiltext{4}{Space Radiation Laboratory, California Institute of Technology, Pasadena, CA 91125, USA}
\altaffiltext{5}{Department of Astronomy and Astrophysics, Pennsylvania State University, PA 16802, USA}
\altaffiltext{6}{Department of Physics and Astronomy, John Hopkins University, Baltimore, MD 21218, USA}
\altaffiltext{7}{Department of Physics, University of Warwick, Coventry CV4 7AL, UK}
\altaffiltext{8}{National Radio Astronomy Observatory, Charlottesville, VA 22903, USA}
\altaffiltext{9}{Department of Physics, University of Wisconsin, Whitewater, WI 53190, USA}
\altaffiltext{10}{MIT Kavli Institute for Astrophysics and Space Research and Department of Physics, Massachusetts Institute of Technology, Cambridge, MA 02139, USA}
\altaffiltext{11}{Centre for Astrophysics Research, STRI, University of Hertfordshire, Hatfield AL10 9AB, UK}
\altaffiltext{12}{Department of Physics, Purdue University, West Lafayette, IN 47907, USA}
\altaffiltext{13}{Jet Propulsion Laboratory, California Institute of Technology, Pasadena, CA 91109, USA}
\altaffiltext{14}{Spitzer Science Center, California Institute of Technology, Pasadena, CA 91125, USA}
\altaffiltext{15}{Department of Astronomy and Astrophysics, University of Toronto, Toronto, ON M5S 3H4, Canada}
\altaffiltext{$\dagger$}{Current address: Department of Physics, Astrophysics, University of Oxford, Denys Wilkinson Building, Oxford, OX1 3RH, UK}
\altaffiltext{$\ddagger$}{Current address: Lincoln Laboratory, Massachusetts Institute of Technology, Lexington, MA 02420-9108, USA}

\begin{abstract}

We present the {\em \underline{Ch}asing the \underline{I}dentifi\underline{c}ation of \underline{A}SCA \underline{G}alactic \underline{O}bjects} (\chic) survey, which is designed to identify the unknown X-ray sources discovered during the \asca\ Galactic Plane Survey (AGPS). Little is known about most of the AGPS sources, especially those that emit primarily in hard X-rays ($2-10$ keV) within the $F_{x} \sim 10^{-13} \mathrm{~to~} 10^{-11}$ \erg\ X-ray flux range. In ChIcAGO, the subarcsecond localization capabilities of \cxo\ have been combined with a detailed multi-wavelength follow-up program, with the ultimate goal of classifying the $>100$ unidentified sources in the AGPS. Overall to date, 93 unidentified AGPS sources have been observed with \cxo\ as part of the \chic\ survey. A total of 253 X-ray point sources have been detected in these \cxo\ observations within $3'$ of the original \asca\ positions. We have identified infrared and optical counterparts to the majority of these sources, using both new observations and catalogs from existing Galactic plane surveys. X-ray and infrared population statistics for the X-ray point sources detected in the \cxo\ observations reveal that the primary populations of Galactic plane X-ray sources that emit in the $F_{x} \sim 10^{-13} \mathrm{~to~} 10^{-11}$ \erg\ flux range are active stellar coronae, massive stars with strong stellar winds that are possibly in colliding-wind binaries, X-ray binaries, and magnetars. There is also a fifth population that is still unidentified but, based on its X-ray and infrared properties, likely comprise partly of Galactic sources and partly of active galactic nuclei. 

\end{abstract}

\section{Introduction}

From 1996 to 1999, the \textit{Advanced Satellite for Cosmology and Astrophysics} (\asca) performed the \asca\ Galactic plane survey (AGPS), which was designed to study 40 deg$^{2}$ of the X-ray sky, over the Galactic coordinates $|l| \lesssim 45^{\circ}$ and $|b| \lesssim 0^{\circ}.4$, in the $0.7-10$ keV energy range \citep{sugizaki01}. This survey resulted in a catalog of 163 discrete X-ray sources with X-ray fluxes between $F_{x} \sim 10^{-13} \mathrm{~and~} 10^{-11}$ \erg, many of which are much harder and more absorbed than any other X-ray source previously detected in the Galactic plane. While the AGPS yielded the first ever \ns\ distribution of hard ($2-10$ keV) Galactic plane X-ray sources, \asca's limited spatial resolution ($3'$) and large positional uncertainty ($1'$) left $>100$ of the AGPS sources unidentified. Even in the era of the \textit{Chandra X-ray Observatory} and the \textit{XMM-Newton} telescope, a substantial fraction of the AGPS source catalog, and therefore a large fraction of the Galactic plane X-ray population, still remain unidentified.

For the last few years, new and archival multi-wavelength data have been used to improve the general understanding of the Galactic X-ray sources detected in the AGPS. Recent work has demonstrated that unidentified \asca\ sources represent a whole range of unusual objects. For example \citet{gelfand07} used new and archival \cxo\ and \xmm\ observations to identify the AGPS source \object[AX J1550.8-5418]{AX J155052--5418} (also known as \object[PSR J1550-5418]{1E 1547.0--5408}) as a magnetar sitting at the center of a faint and small, previously unidentified, radio supernova remnant (SNR) called G327.24--0.13. Investigations of archival \xmm\ data allowed \citet{kaplan07} to identify the AGPS source \object[AX J1835.4-0737]{AX J183528--0737} as a likely symbiotic X-ray binary (SyXB) comprising of a late-type giant or supergiant and a neutron star (NS) with a 112s pulse period. \citet{gaensler08} identified the AGPS source \object[AX J1721.0-3726]{AX J172105--3726} as the X-ray emission associated with the radio \object[SNR G350.1-00.3]{SNR G350.1--0.3}. The \xmm\ X-ray spectrum, combined with the presence of non-thermal, polarized, radio emission, showed G350.1--0.3 to be a very young and luminous SNR. A central compact object was also resolved in these X-ray observations and identified as a NS. \xmm\ \citep{funk07} and \cxo\ \citep{lemiere09} observations have demonstrated that the AGPS source \object[AX J1640.7-4632]{AX J164042--4632} is an X-ray pulsar wind nebula (PWN) located at the center of the radio \object[SNR G338.3-00.0]{SNR G338.3--0.0}. \cxo\ results, discussed in \citet{anderson11}, have revealed that two AGPS sources, \object[AX J1632.8-4746]{AX J163252--4746} and \object[AX J1847.6-0156]{AX J184738--0156}, are massive stars in colliding wind binaries (CWBs). New \cxo, \xmm, and ATCA observations have also been used to identify the AGPS source \object[PSR J1622-4950]{AX J162246--4946} as the radio and X-ray emitting magnetar, \object[PSR J1622-4950]{PSR J1622--4950}, and have exposed the likely X-ray transient nature of this source \citep{anderson12}. These identifications over the last 8 years have therefore demonstrated that many of the unidentified AGPS sources are unusual and rare Galactic plane X-ray objects.

The most comprehensive X-ray survey to-date, in terms of area coverage, was performed by the \textit{ROSAT}  X-ray Satellite \citep[for example see][]{voges99}, which mapped the soft X-ray source population ($0.1-2.4$ keV) down to a flux sensitivity of a few $10^{-13}$ \erg. Projects that focused on the \textit{ROSAT} data covering the Galactic plane \citep[e.g. the ROSAT Galactic Plane Survey;][]{motch97,motch98} demonstrated that stars and AGN dominate the soft X-ray sky. However, performing a similar Galactic plane survey to include those sources with energies up to $10$ keV, sensitive to the $F_{x} \sim 10^{-13} \mathrm{~to~} \sim10^{-11}$ \erg\ flux range, would be impractical to achieve with the current X-ray telescopes \cxo\ and \xmm\ due to their limited fields of view.  Astronomers have therefore had to rely upon characterizing the distribution of the harder X-ray source populations within much smaller regions of the Galactic plane \citep[e.g.][]{hands04,ebisawa05,grindlay05}. For example, \citet{motch10} used the XGPS \citep{hands04} to determine the contributions of active stellar coronae and accreting X-ray source populations in the Galactic plane for $F_{x} \lesssim 10^{-12}$ \erg. The Chandra Multiwavelength Plane survey \citep[ChaMPlane;][]{grindlay05} has now surveyed 7 deg$^{2}$ of the Galactic plane and bulge with \cxo\ \citep{vandenberg12}, identifying the contributions of magnetic cataclysmic variables (CVs) to the Galactic ridge X-ray emission \citep{hong12b}.

The key to obtaining a complete understanding of the Galactic plane X-ray source populations, from $0.3-10$ keV, that make up the $F_{x} \sim 10^{-13} \mathrm{~to~} 10^{-11}$ \erg\ X-ray flux range is to identify the unidentified AGPS sources, as \asca\ covered a much larger area of the Galactic plane ($\sim40$ deg$^{2}$) than other X-ray surveys \citep[for example the XGPS and ChaMPlane;][]{hands04,motch10,grindlay05,vandenberg12}. In order to identify the AGPS sources, the {\em \underline{Ch}asing the \underline{I}dentifi\underline{c}ation of \underline{A}SCA \underline{G}alactic \underline{O}bjects} (\chic) survey was conceived. In this survey the subarcsecond capabilities of \cxo\ are used to localize the unidentified AGPS sources listed by \citet{sugizaki01}. Once the positions of these sources have been determined, an extensive multi-wavelength program is activated, which is aimed at determining the identities of the sources and the nature of their X-ray emission. 

In this paper, we present the results of \cxo\ observations of 93 unidentified AGPS sources, along with the multi-wavelength follow-up that has allowed the identification of optical, infrared and radio counterparts. Section 2 explains the \cxo\ observing strategy employed to localize the unidentified AGPS sources. To begin the identification process, we automated the \cxo\ data analysis and preliminary multi-wavelength follow-up, which involves comparisons with existing optical, near-infrared (NIR) and infrared (IR) surveys. X-ray spectral modeling using ``quantile analysis'' \citep{hong04} and \citet{cash79} statistics, and further multi-wavelength observations in the optical, infrared and radio bands required to ultimately classify each source, are also described. Section 3 details the results of each AGPS position observed with \cxo. These results include details on the individual X-ray sources detected, the parameters of their likely X-ray spectral shapes, and the names and magnitudes of their infrared, optical and radio counterparts. The possibility of short term variability or periodicity is also explored. In Section 4, we discuss the AGPS sources that have been identified through a visual inspection of radio Galactic plane surveys. The X-ray fluxes and NIR and IR magnitudes of the remaining unidentified sources, reported in Section 3, are then used to conduct X-ray and infrared population statistics. Resulting flux and color-color diagrams allow the identification of likely Galactic plane X-ray populations with infrared counterparts. This analysis is followed by a discussion of particularly interesting individual sources that have been identified as a result of this work. The final part of this section includes a tabulated summary of all the 163 AGPS sources along with their confirmed identifications (obtained from the literature and the present paper) or their tentative identifications that are based on our \chic\ survey statistical results. In Section 5 we summarize the results from this paper, with a particular focus on our statistical findings.

\section{Method}

\subsection{\cxo\ Observations}

The main goal of the \chic\ survey is to localize the positions of the unidentified AGPS sources, so that multi-wavelength follow-up can be used to identify them. It is therefore necessary to design an experiment that will allow each source to be localized precisely enough to identify counterparts in the crowded Galactic plane. \cxo\ can provide subarcsecond localization as it has an intrinsic astrometric precision accuracy of $0\farcs6$ at 90\% confidence within $2'$ of the aim-point \citep{weisskopf03}. 

For all but the brightest targets, \cxo's Advanced CCD Imaging Spectrometer \citep[ACIS;][]{garmire03} was used as it provides simultaneous positional, temporal and spectroscopic information. The ACIS-S configuration was chosen as it fully encompasses the AGPS positional uncertainties of up to $3'$ \citep{sugizaki01}. (The aimpoint of the ACIS-I configuration is near a chip gap.)  The High Resolution Camera \citep[HRC;][]{murray00} in the I focal plane array was used to observe those sources with a predicted ACIS count-rate $>0.2$ counts s$^{-1}$, to avoid positional, spectral, and temporal degradation associated with pile-up \citep{davis01}.

At \cxo's high angular resolution, only a small number of X-ray counts are required to localize each source sufficiently to overcome confusion from IR field stars in the Galactic plane. We first considered the number density of such stars in the $K_{s}$-band at low Galactic latitudes. Figure 24 of \citet{kaplan04} shows that $0.2$ stars arcsec$^{-2}$ are expected with a magnitude $K_{s} \lesssim 19$. For there to be a $<25\%$ chance of random alignment of the \cxo\ source with an infrared field star of this magnitude, a total astrometric error $<0\farcs7$ is required. Using 2MASS as a guide, given its extremely high positional precision ($0\farcs1$ $1\sigma$ error) and \cxo's 90\% absolute astrometry error of $0\farcs6$, a centroiding error of $<0\farcs4$ with 95\% accuracy is required for \cxo. The \cxo\ Interactive Analysis of Observations (\texttt{CIAO})\footnote{http://cxc.harvard.edu/ciao/index.html} software tool \texttt{wavdetect} \citep{freeman02} was chosen to detect the point sources in our fields. Equation (5) of \citet{hong05} provides the 95\% confidence position error circle of a point source detected with \texttt{wavdetect} for a given number of source counts at a given off axis angle. At the maximum off-axis angle expected for an \asca\ source localization ($<3'$), $\sim100$ X-ray counts are required to ensure that a source's centroiding error is below $0\farcs4$.

Using the count rates and power-law spectral fits calculated by \citet{sugizaki01} for each AGPS source and the \cxo\ Proposal Planning Toolkit (PIMMS)\footnote{http://cxc.harvard.edu/toolkit/pimms.jsp}, the exposure time required to detect $\sim100$ counts with \cxo\ for each source was estimated. For those AGPS sources that were too faint for \citet{sugizaki01} to calculate spectral fit parameters, an absorbed power-law model with a photon index $\Gamma=2$ and an absorption  $N_{H}=10^{22}$ cm$^{-2}$ was used, which are representative values of a non-thermal X-ray source and typical Galactic plane absorption.

In order to select the AGPS source candidates to be observed with \cxo, each source was investigated individually. First those AGPS sources that have already been conclusively identified, either by \citet{sugizaki01} or by other groups in the literature, were removed from the \cxo\ target list. Based on this criteria, a total of 43 AGPS sources were identified and therefore rejected for \cxo\ follow-up (these sources are described in Appendix A). The \asca\ images of each of the remaining unidentified AGPS sources were then studied to determine if any sources appeared to be too extended for \cxo\ to successfully localize in a short amount of time. These sources were also rejected for \cxo\ follow-up. The remaining, unidentified AGPS sources, were then prioritized for \cxo\ follow-up based on their absorbed X-ray flux or count-rate that was listed by \citet{sugizaki01}. 

A total of 93 AGPS sources have been observed with \cxo\ as part of the \chic\ survey, of which 84 were imaged with ACIS-S and 9 were imaged with HRC-I. The \chic\ \cxo\ observations took place over a three and a half year period, from 2007 January to 2010 July. The \cxo\ exposure times ranged from $\sim1-10$ ks. All the details of these \cxo\ observations are listed in Table~\ref{Tab1}. The initial automated analysis of these \cxo\ observations was conducted using the \chic\ Multi-wavelength Analysis Pipeline, described in Section 2.2. We then performed a more detailed X-ray analysis and counterpart study for those 74 sources with $>20$ X-ray counts, as such sources are approximately within the original AGPS sources X-ray flux range (see Sections 3.2 and 3.3).

\subsection{ChIcAGO Multi-wavelength Analysis Pipeline (MAP)}

It is crucial to the efficiency of the project to automate the analysis of the \cxo\ observations, such as the detection and extraction of sources, as well as the search for multi-wavelength counterparts. We therefore created the \chic\ Multi-wavelength Analysis Pipeline (MAP) for this task. \chic\ MAP takes the ACIS-S or HRC-I \cxo\ observation of an AGPS source field and detects and analyzes all point sources within $3'$, equivalent to the largest likely position error, for the original AGPS source positions supplied by \citet{sugizaki01}. From hereon we refer to all point sources detected in the \cxo\ observations of the AGPS fields as ``\chic\ sources''. The X-ray analysis component of this pipeline uses the \texttt{CIAO} software, version 4.3, with CALDB version 4.5.5, and follows standard reduction recipes given in the online \texttt{CIAO} 4.3 Science Threads.\footnote{http://cxc.harvard.edu/ciao4.3/threads/index.html}

\chic\ MAP carries out the following steps, all of which are explained in more detail below. These steps apply to both ACIS-S and HRC-I datasets unless otherwise stated.

\begin{figure}[htp]
\begin{center}
\subfigure{\includegraphics[width=0.4\textwidth, angle=270]{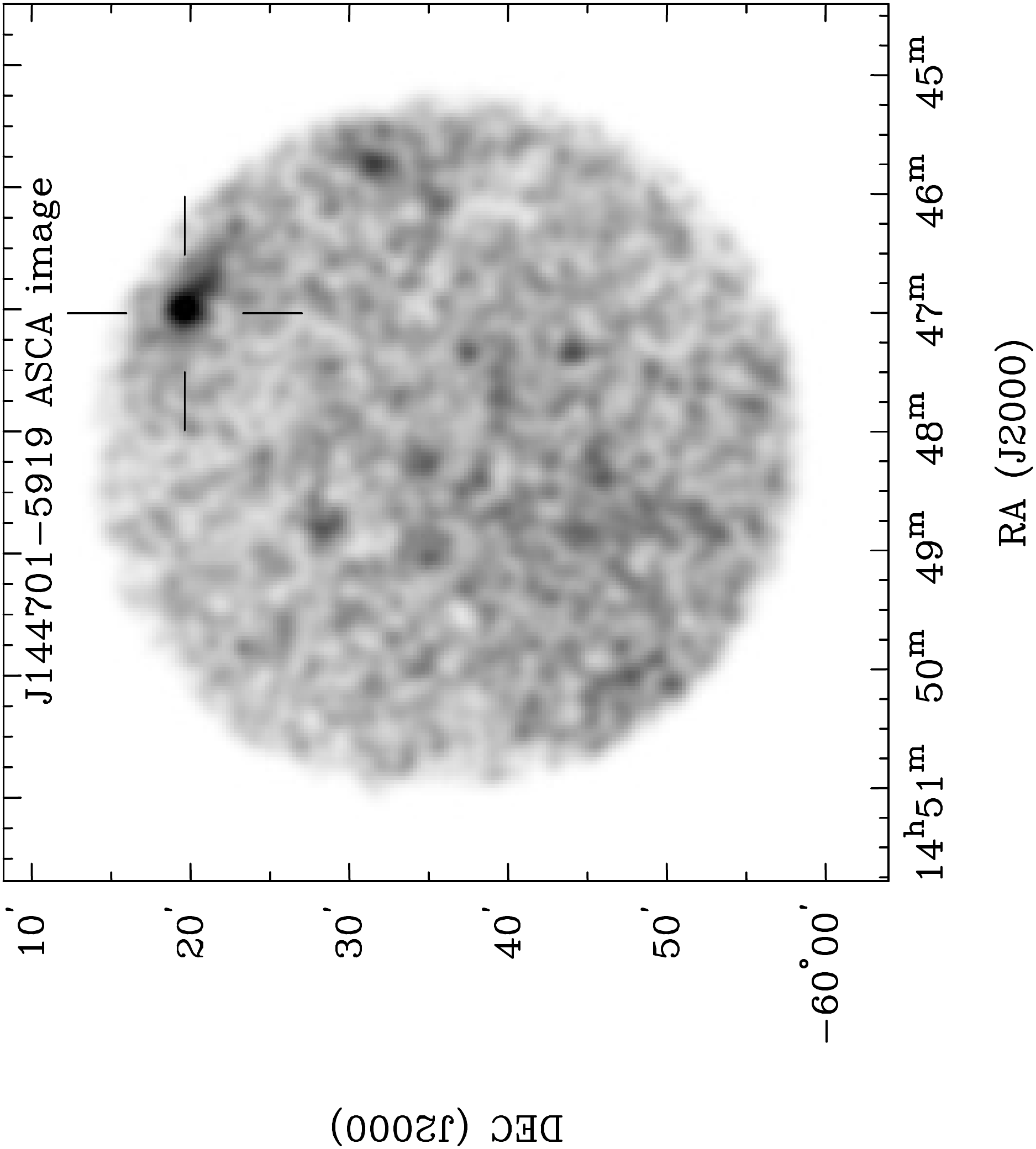}}
\subfigure{\includegraphics[width=0.4\textwidth, angle=270]{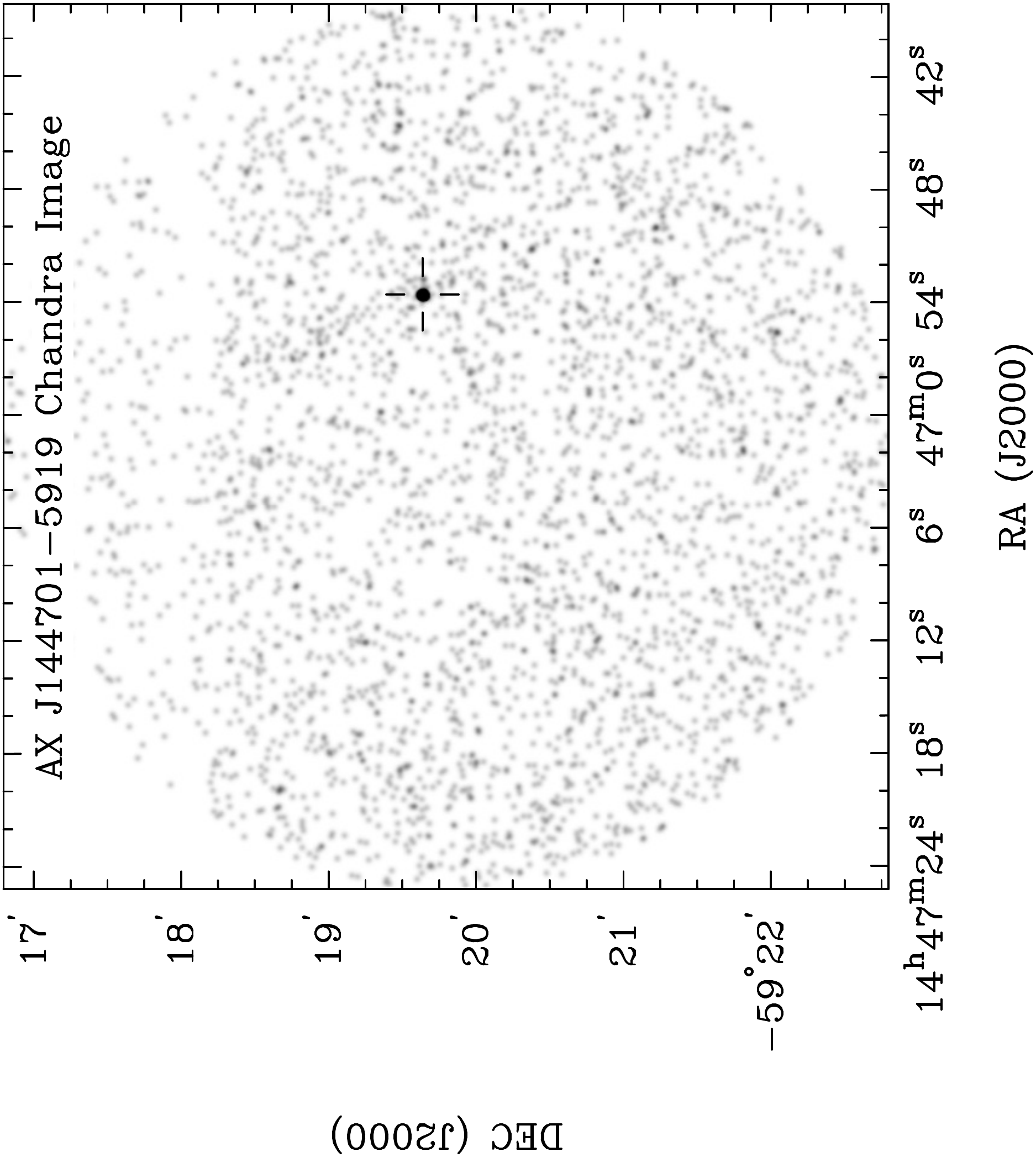}}
\caption{X-ray images of AX J144701--5919 as output by \chic\ MAP. Top: The \asca\ GIS detection of the AGPS source AX J144701--5919. The circular field-of-view has a $50'$ diameter and the position of this AGPS source, as listed by \citet{sugizaki01}, is indicated by the black cross-hair. Bottom: The $3'$ radius field-of-view of the \cxo\ observation of AX J144701--5919, which is centered on the position of this AGPS source listed by \citet{sugizaki01}. The one \chic\ source detected in this field is indicated by a black cross-hair, and is likely the \cxo\ counterpart to AX J144701--5919.}
\label{Fig1}
\end{center}
\end{figure}

\begin{figure}[htp]
\begin{center}
\includegraphics[width=0.4\textwidth]{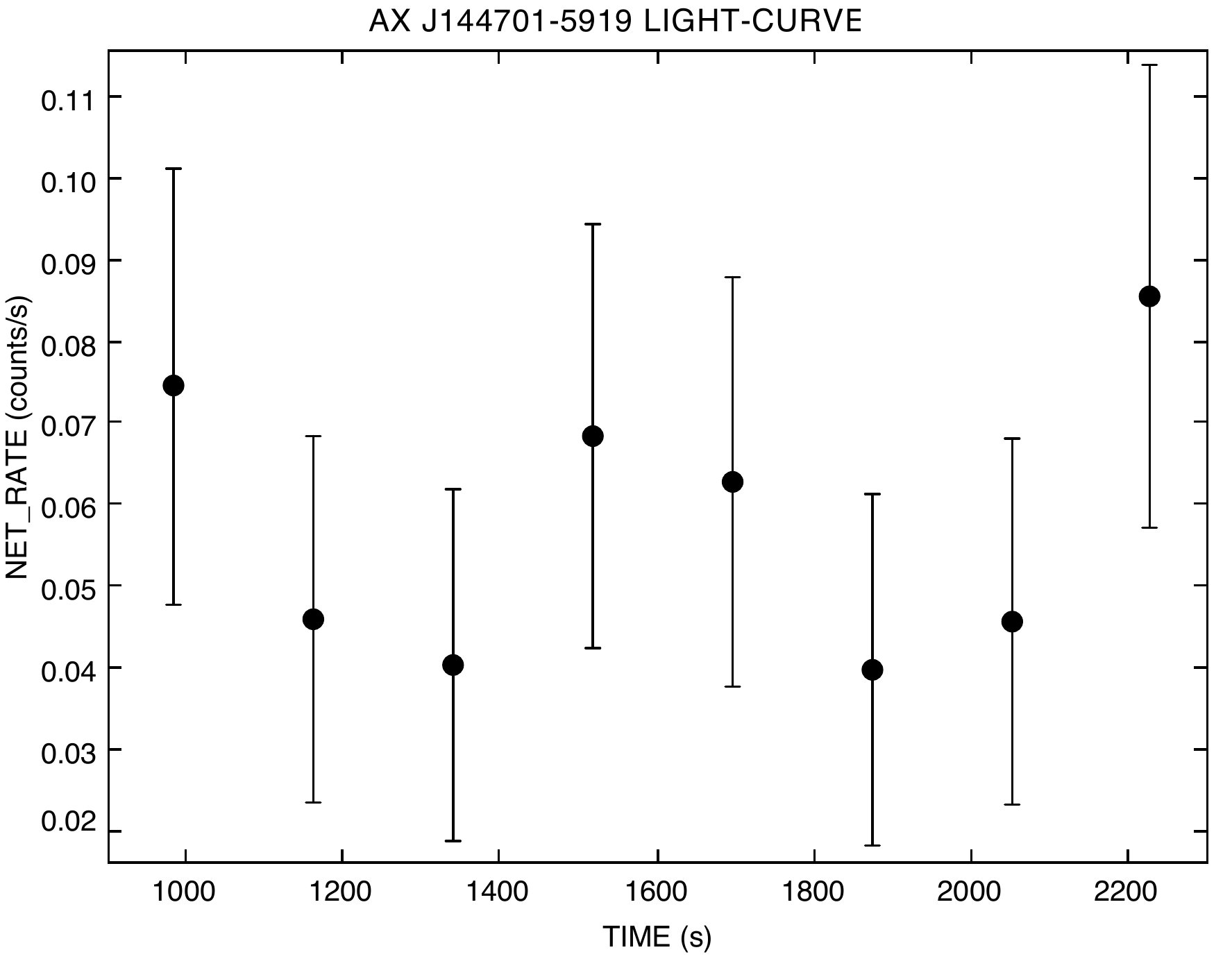}
\caption{The \cxo\ ACIS-S light-curve of the \chic\ source AX J144701--5919 as output by \chic\ MAP. This \chic\ source does not exhibit evidence for short-term variability as its resulting $\chi^{2} = 3.4$ for 7 degrees of freedom is below the confidence threshold of 99.9\%.}
\label{Fig2}
\end{center}
\end{figure}

\begin{figure}[htp]
\begin{center}
\subfigure{\includegraphics[width=0.3\textwidth, angle=270]{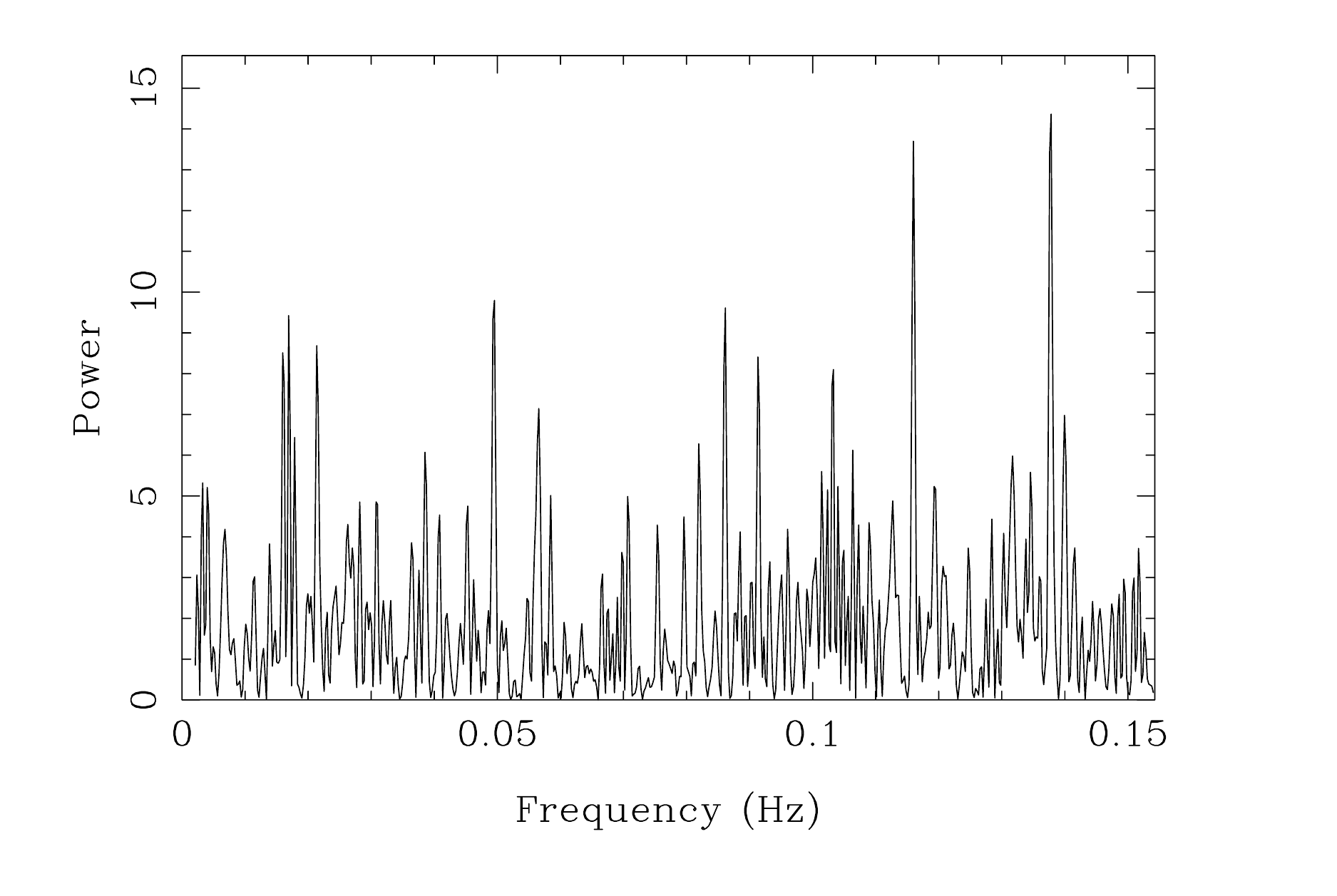}}
\subfigure{\includegraphics[width=0.3\textwidth, angle=270]{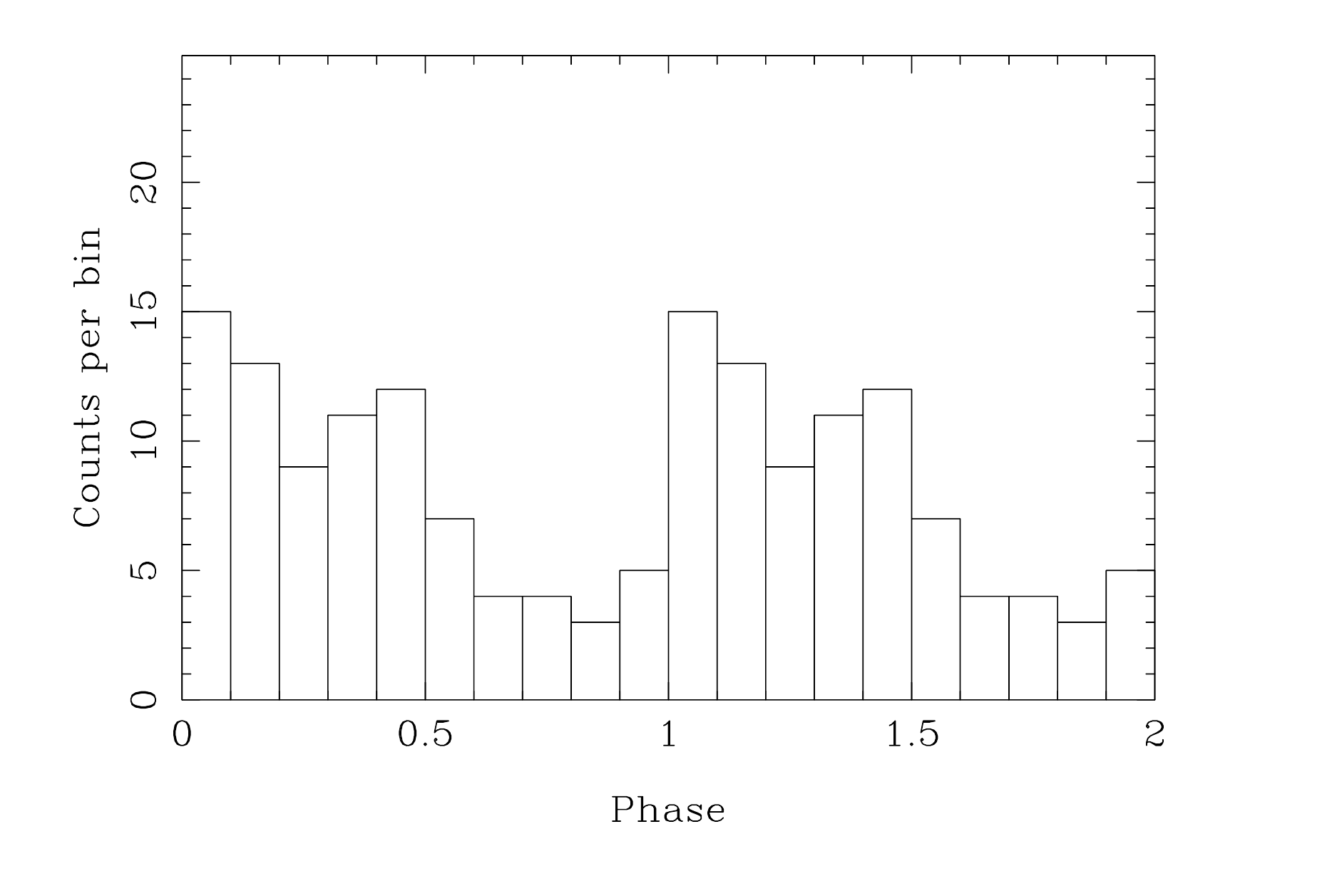}}
\caption{The power spectrum and folded light curve resulting from a $Z_{1}^{2}$ statistical search for periodicity from AX J144701--5919 as output by \chic\ MAP. Top: The power spectrum (power vs. frequency (Hz)) of AX J144701--5919. The power spectrum peaks at a value of $Z_{1}^{2}=14.4$ at a frequency of 0.1378 Hz. The probability of this power being random noise is 0.16. It is therefore unlikely to be a real signal. Bottom: The folded light curve (counts per bin vs phase) that has been folded at 0.1378 Hz.}
\label{Fig3}
\end{center}
\end{figure}

\begin{figure*}[htp]
\begin{center}
\subfigure{\includegraphics[width=0.3\textwidth, angle=270]{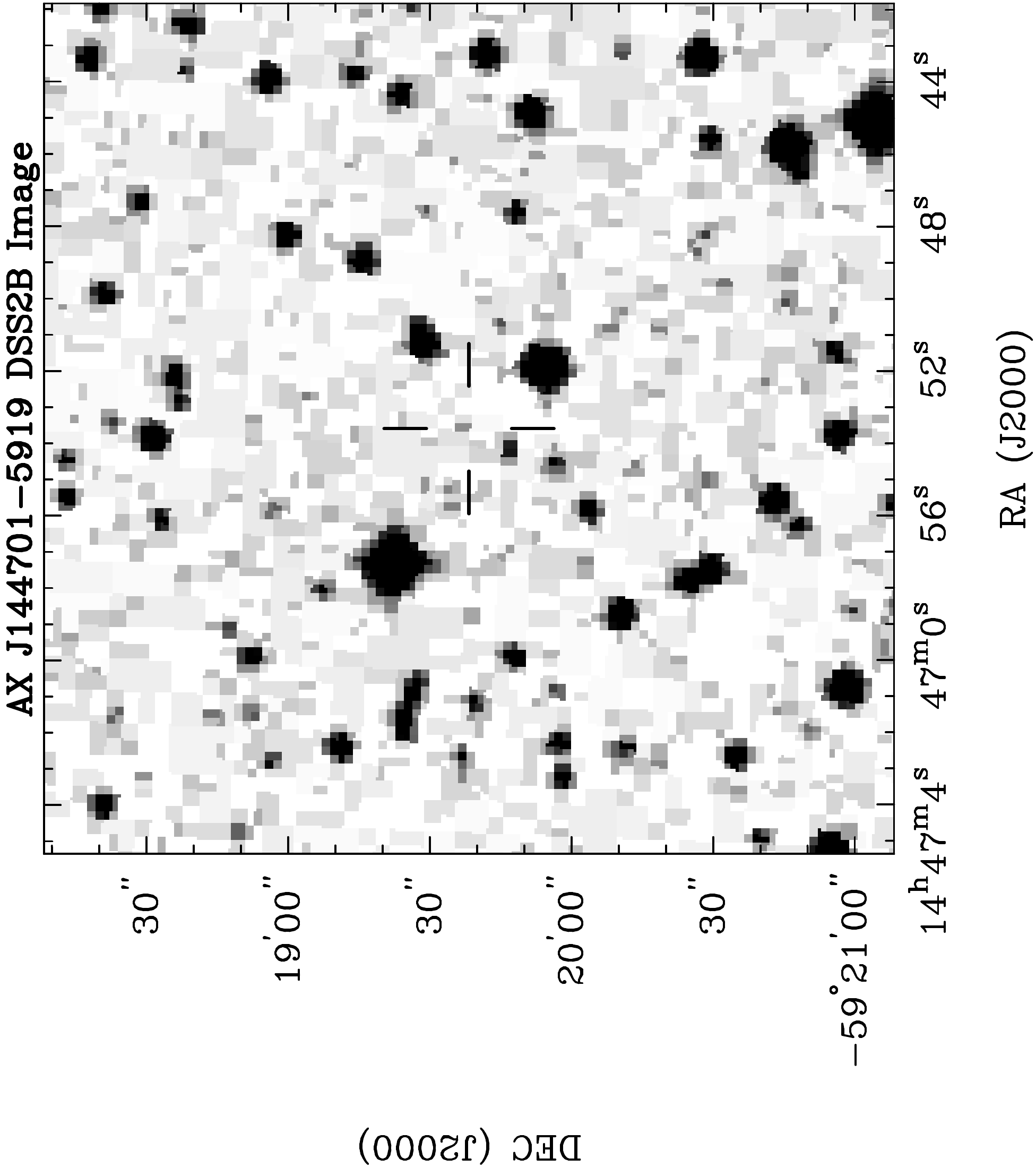}}
\subfigure{\includegraphics[width=0.3\textwidth, angle=270]{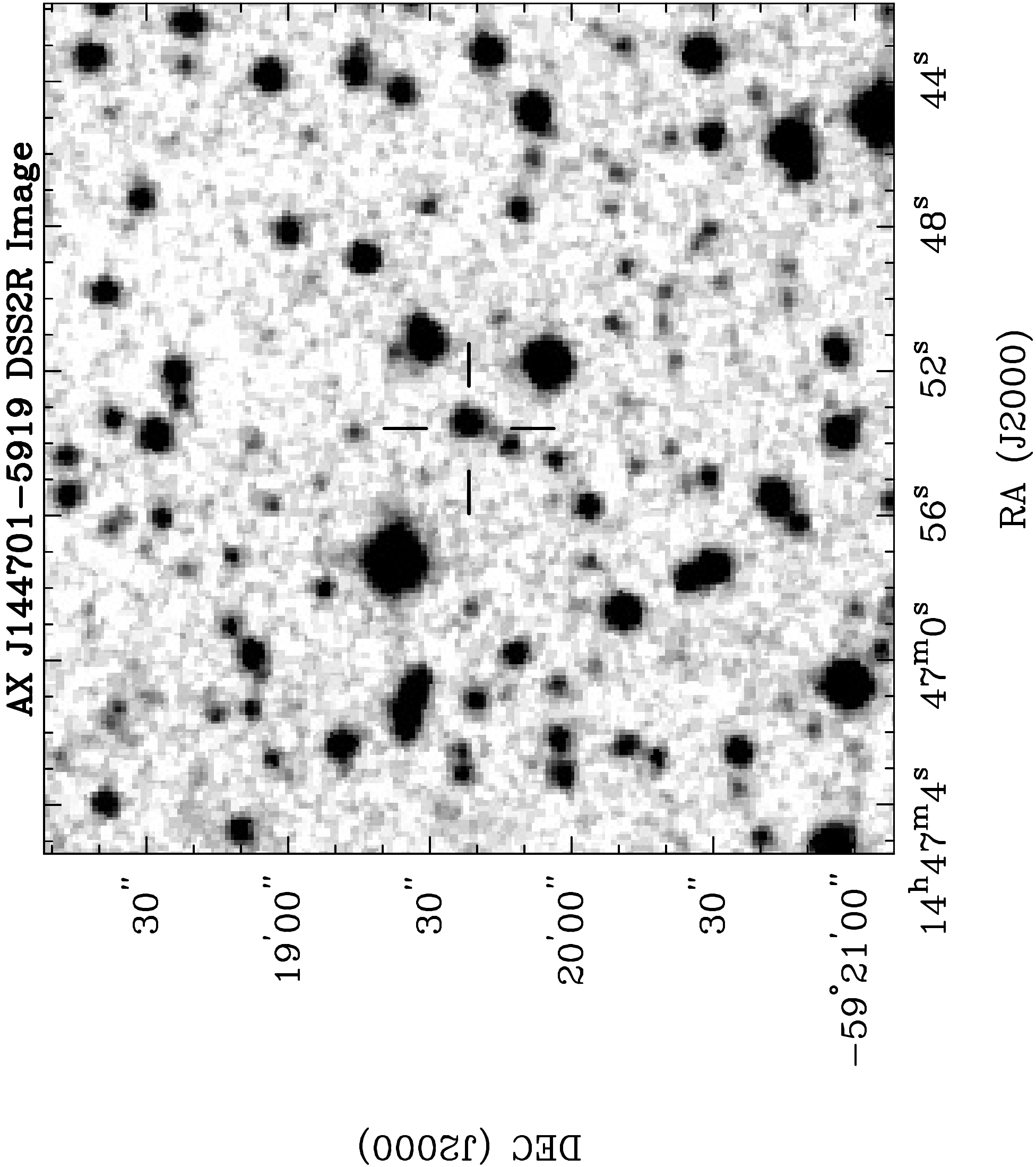}}
\subfigure{\includegraphics[width=0.3\textwidth, angle=270]{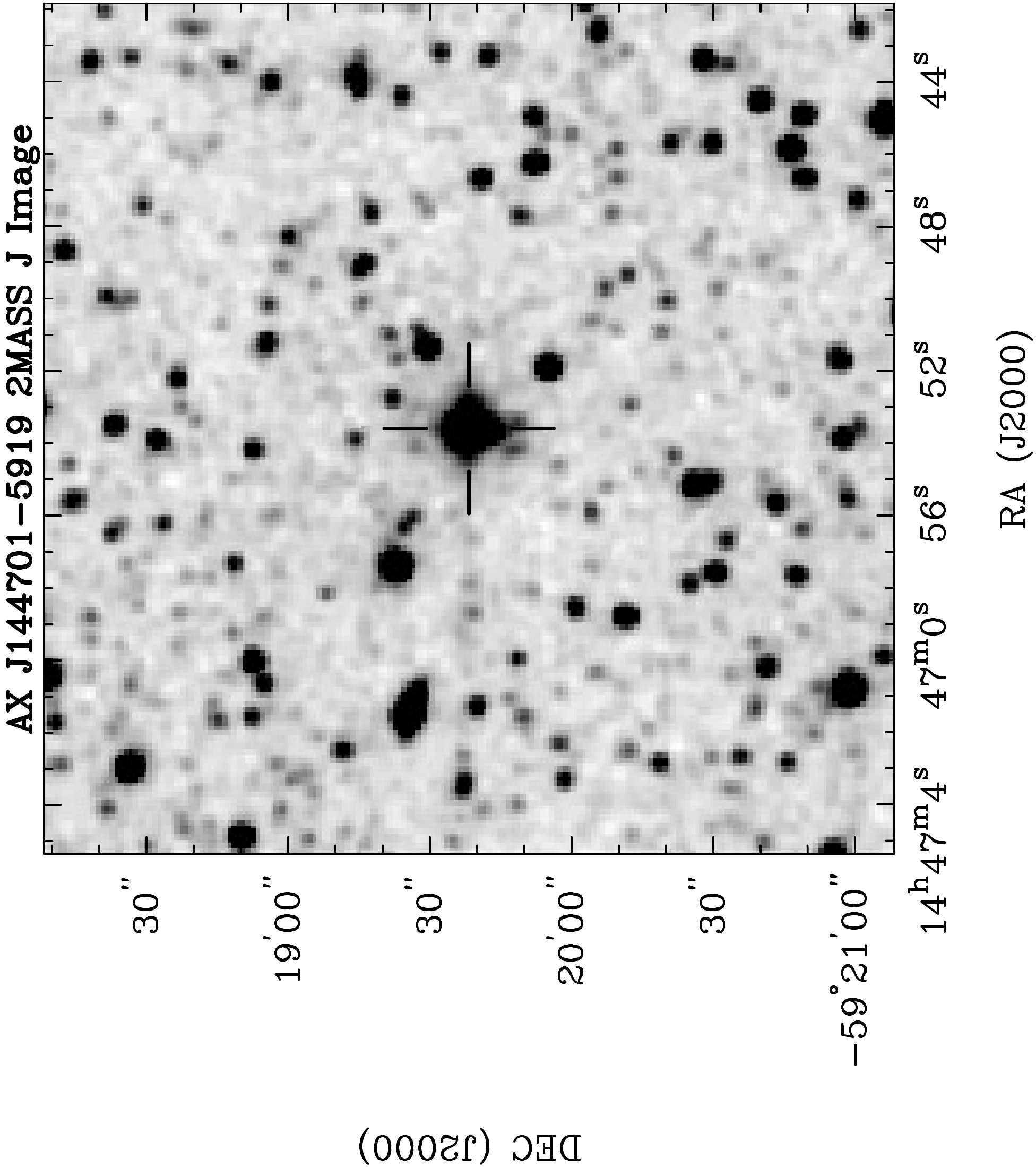}}
\subfigure{\includegraphics[width=0.3\textwidth, angle=270]{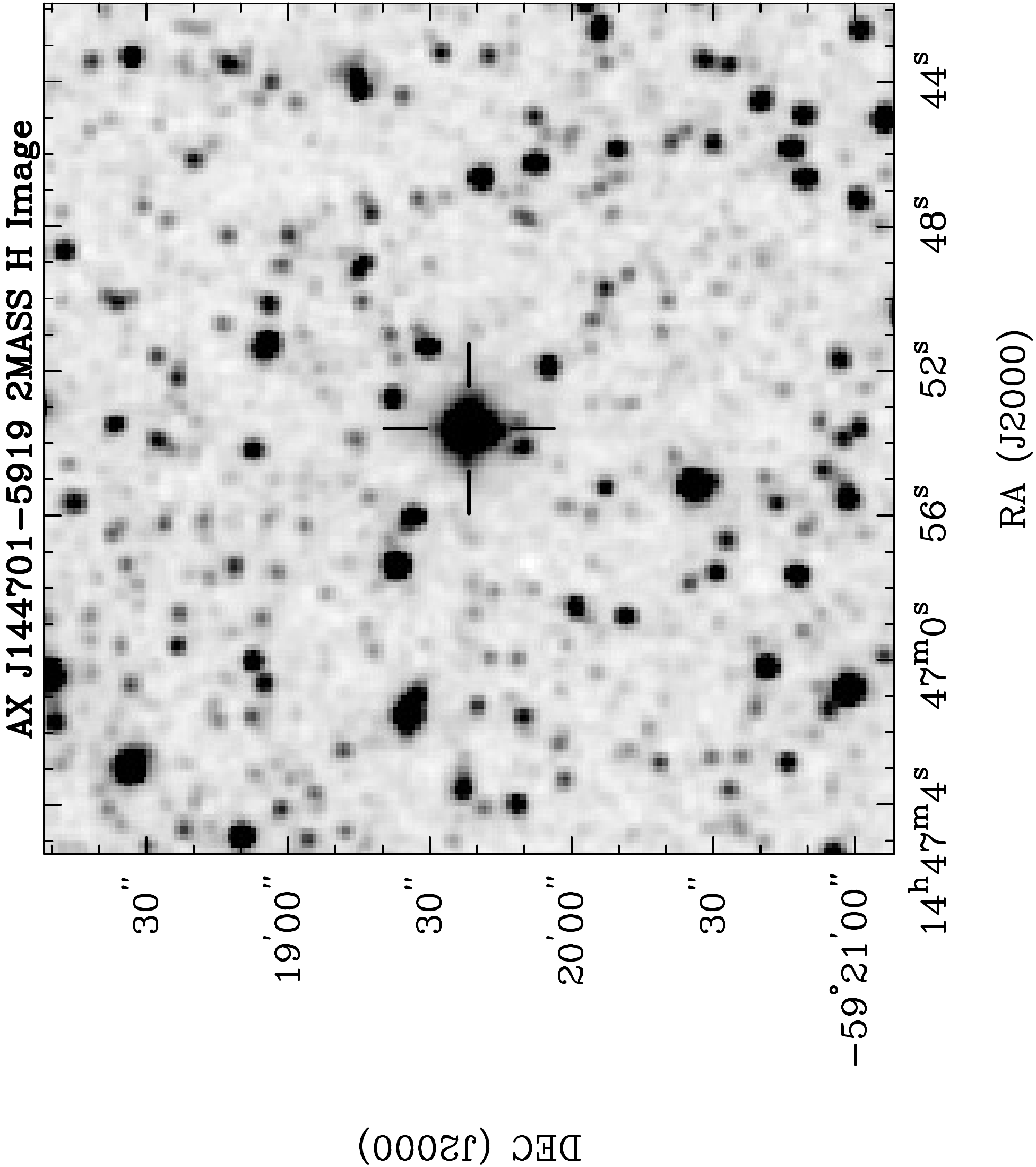}}
\subfigure{\includegraphics[width=0.3\textwidth, angle=270]{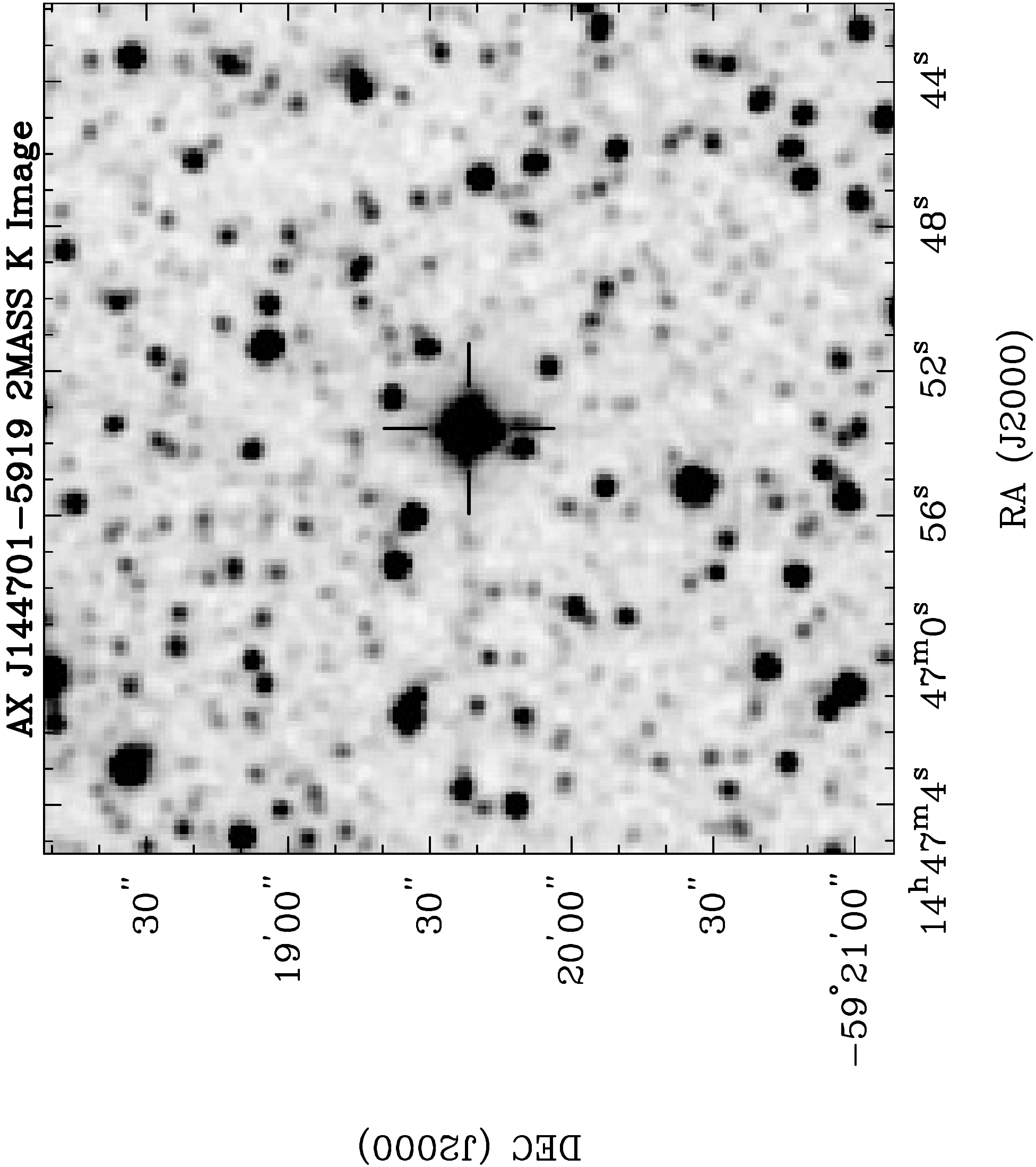}}
\subfigure{\includegraphics[width=0.3\textwidth, angle=270]{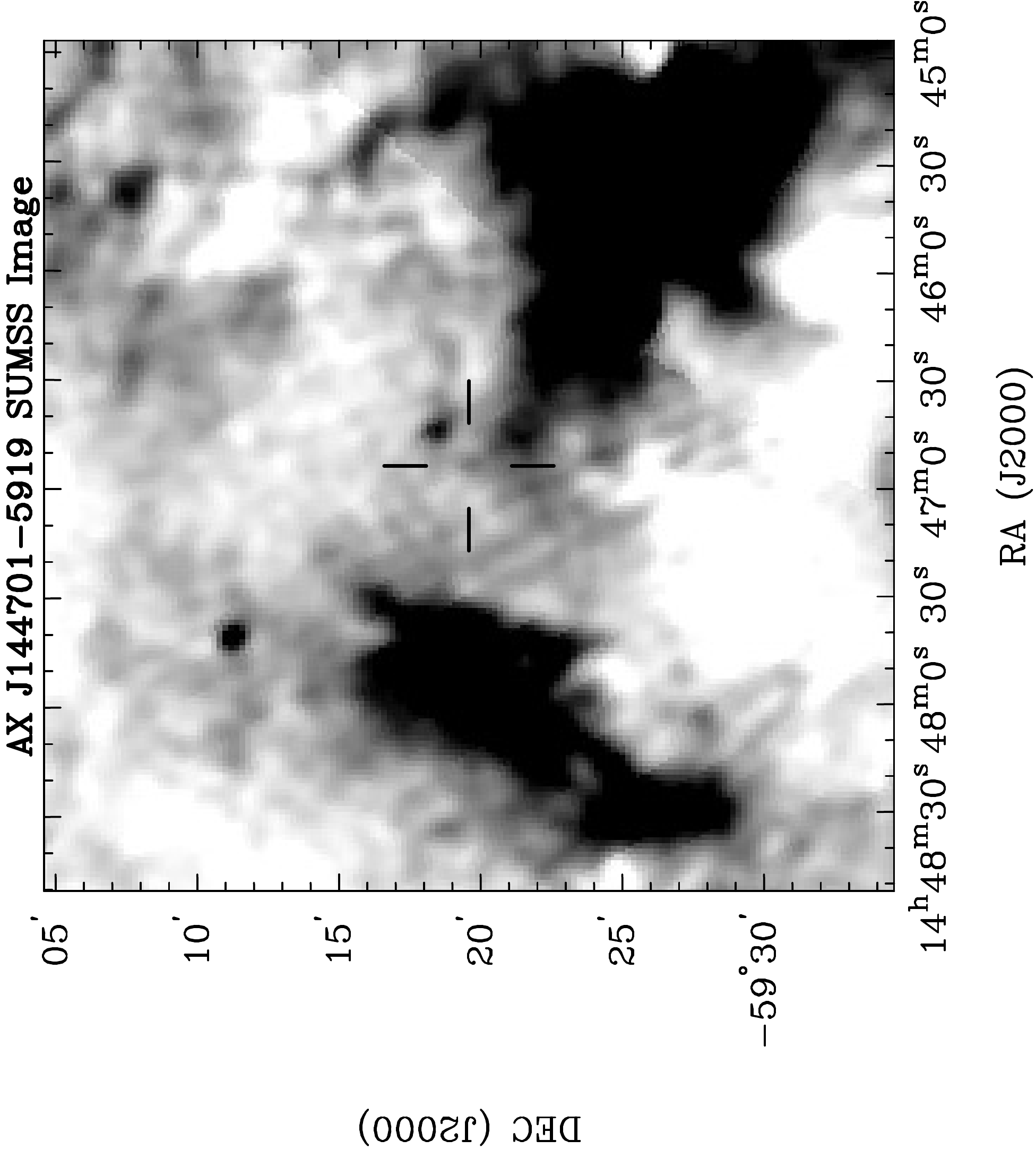}}
\caption{The DSS2B, DSS2R, 2MASS and SUMSS image cutouts of the region surrounding the AGPS source AX J144701--5919, which has been localized by \cxo, as output by \chic\ MAP. In all images the \cxo\ position of AX J144701--5919 is indicated by a black cross-hair. The survey name and filter band of each image cutout is listed at the top of each Figure.}
\label{Fig4}
\end{center}
\end{figure*}

\begin{figure*}[htp]
\begin{center}
\subfigure{\includegraphics[width=0.3\textwidth, angle=270]{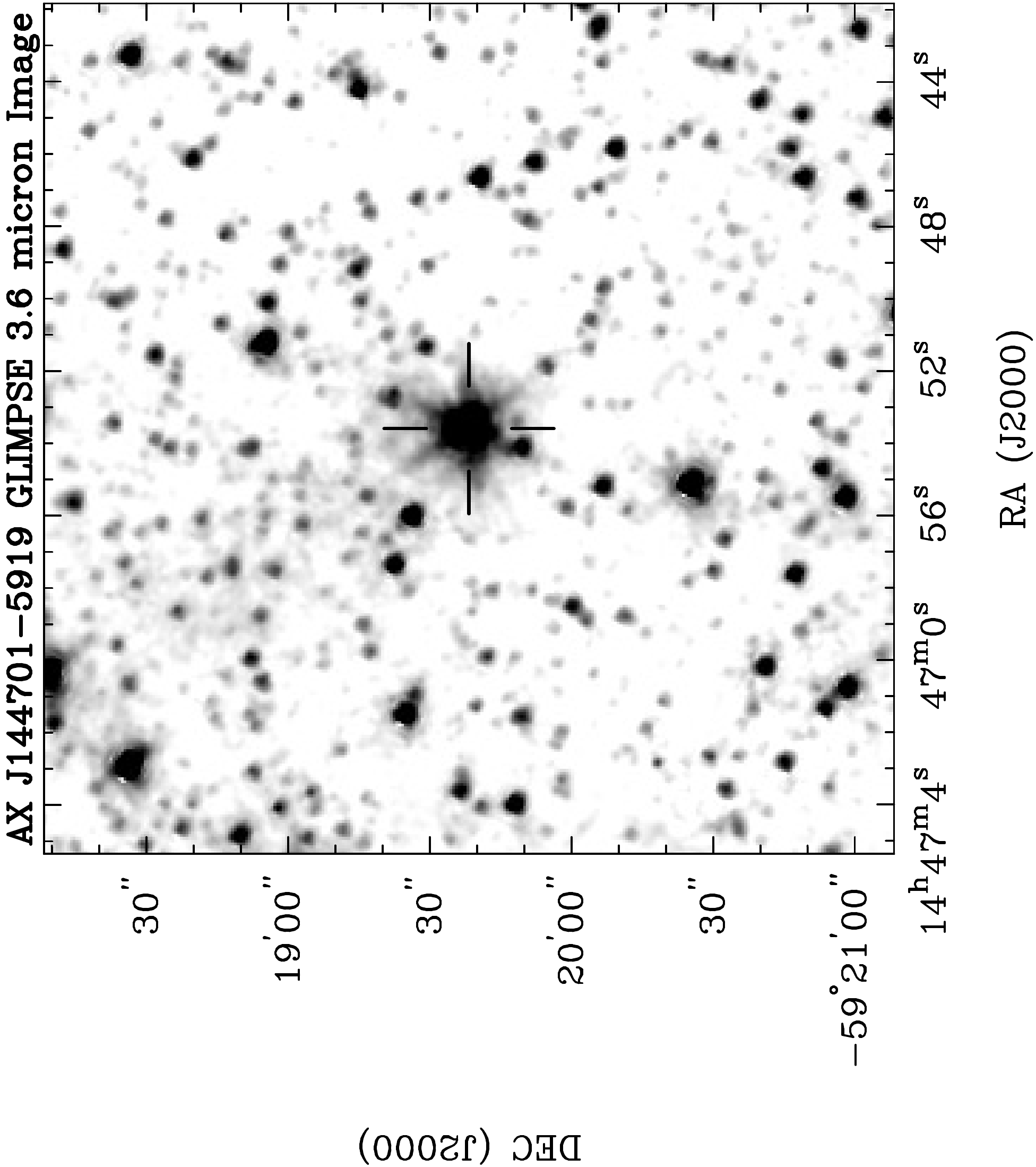}}
\subfigure{\includegraphics[width=0.3\textwidth, angle=270]{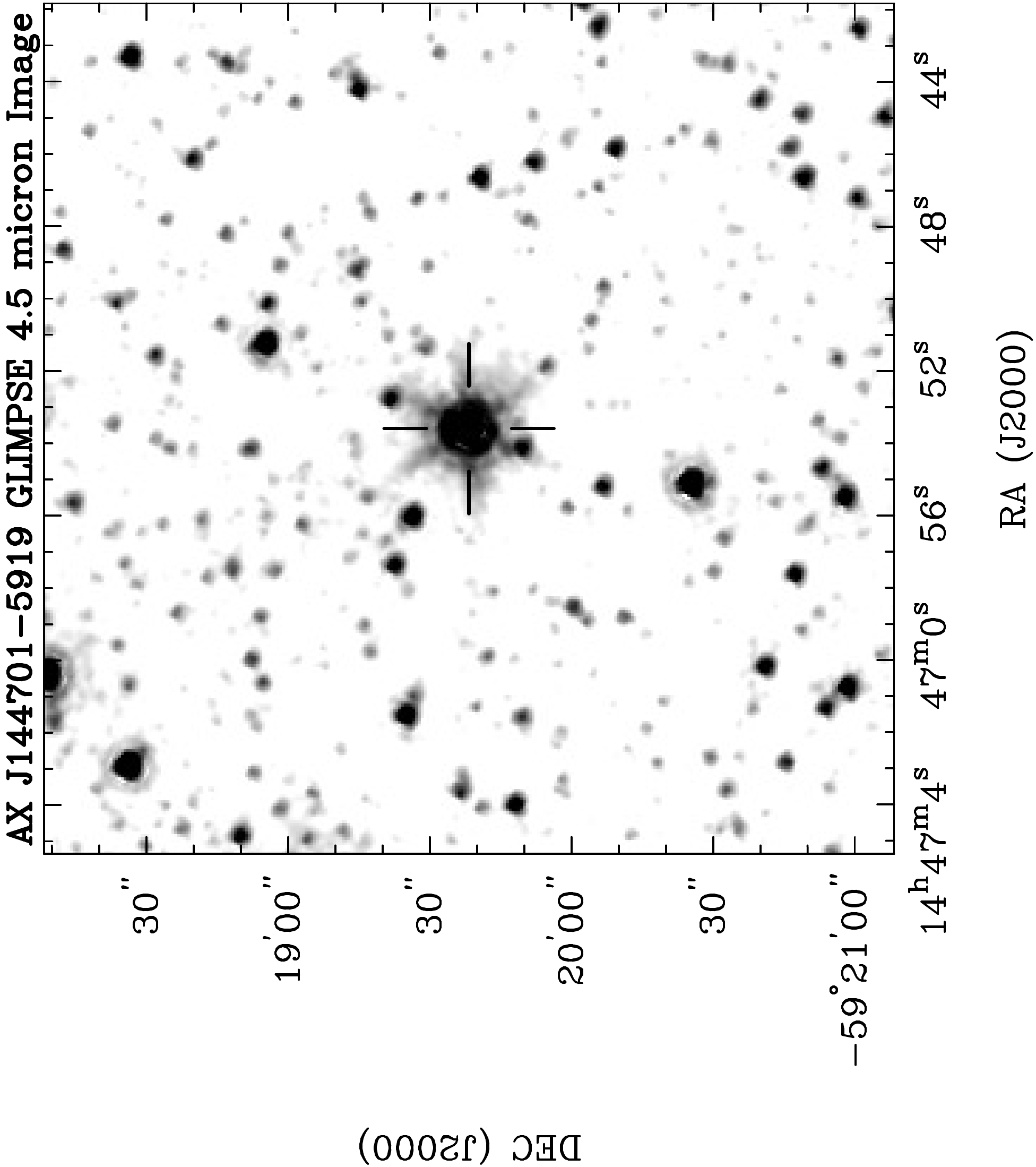}}
\subfigure{\includegraphics[width=0.3\textwidth, angle=270]{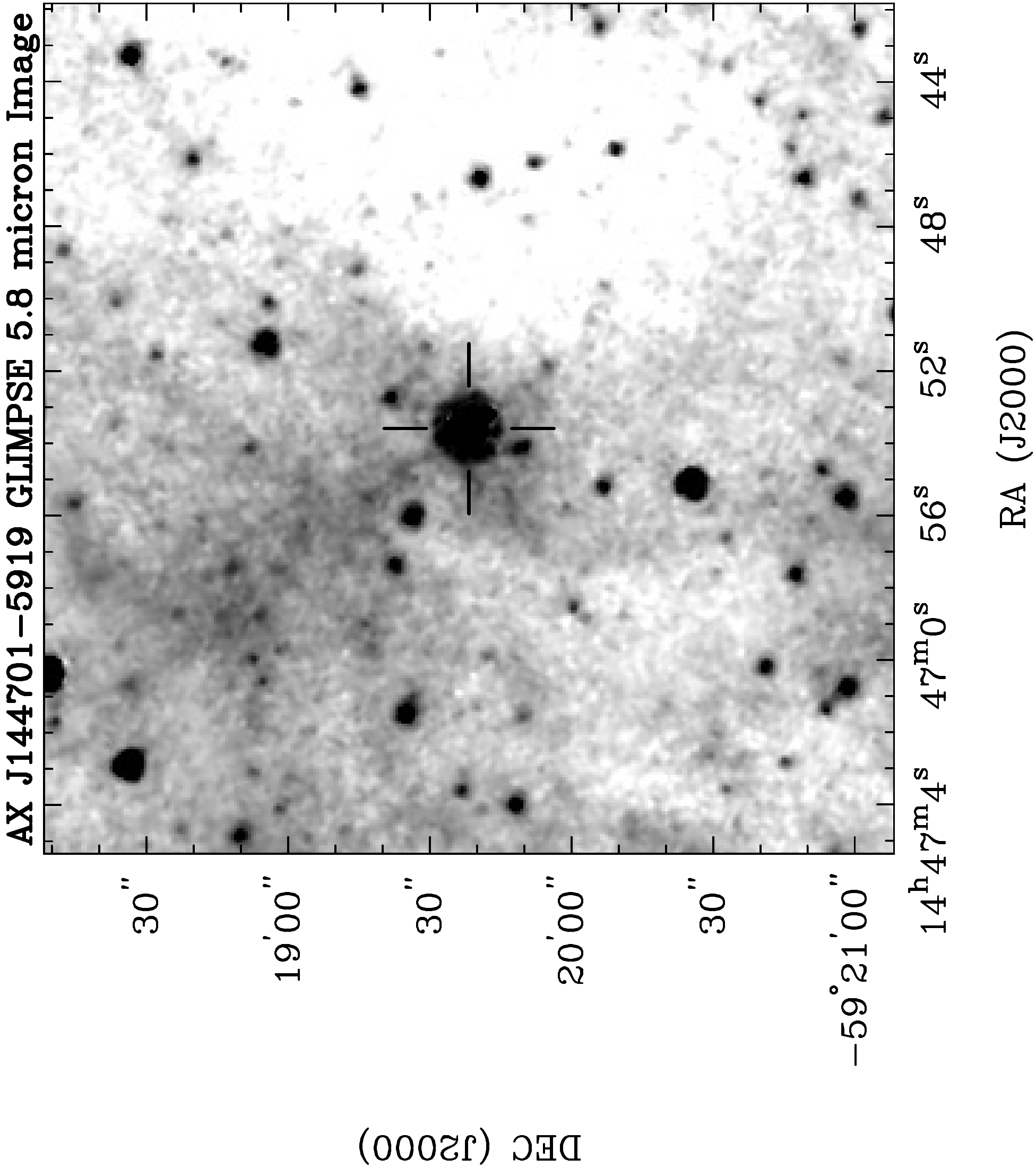}}
\subfigure{\includegraphics[width=0.3\textwidth, angle=270]{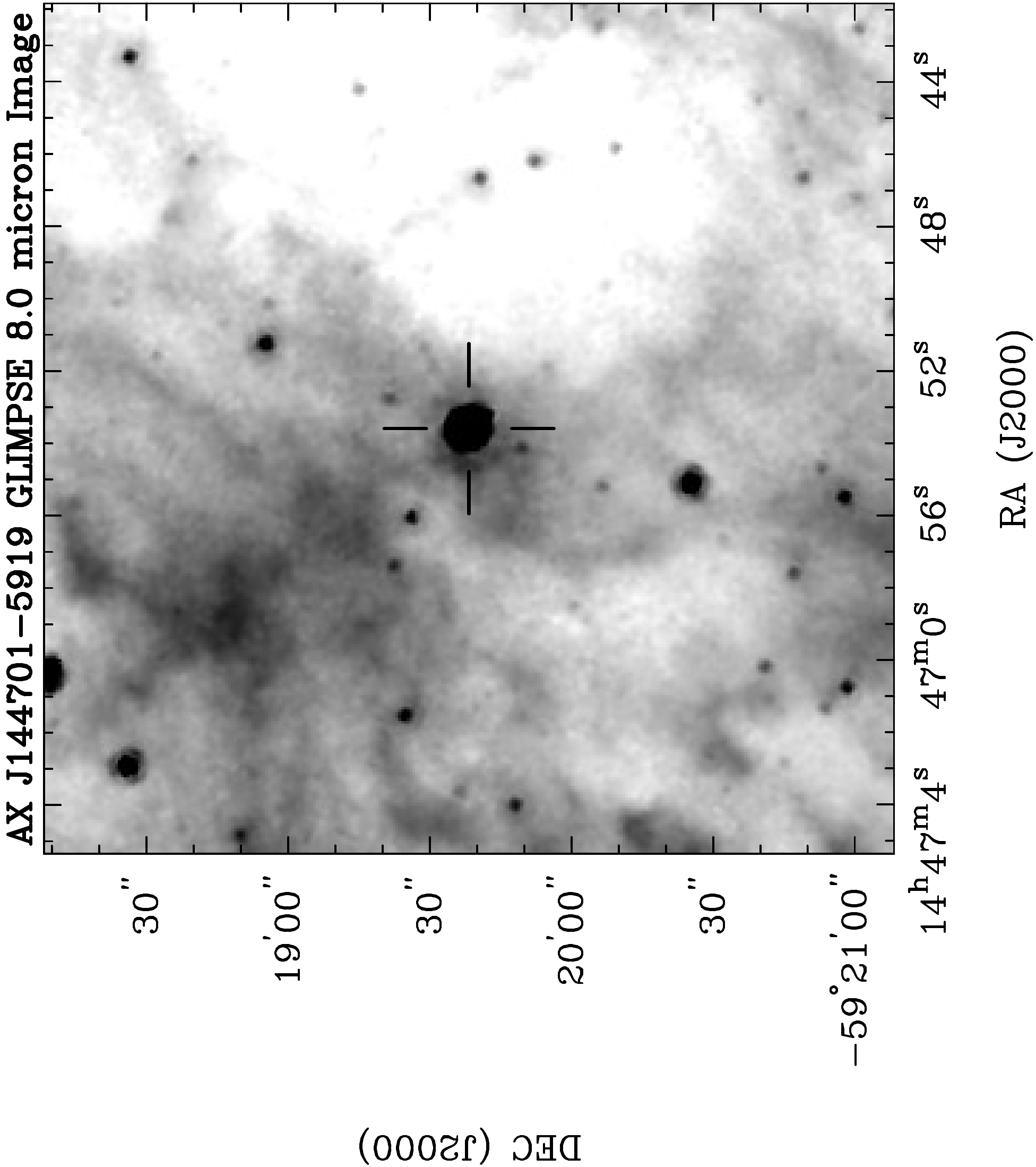}}
\caption{As for Figure~\ref{Fig4}, but showing the GLIMPSE 3.6, 4.5, 5.8, and 8.0 $\mu m$ bands.}
\label{Fig5}
\end{center}
\end{figure*}

\begin{itemize}
\item An image of the original \asca\ detection of the AGPS source is created (for example see the top image of Figure~\ref{Fig1}).
\item The \texttt{CIAO} tool \texttt{chandra\_repro} is run to reprocess the \cxo\ data.
\item The new event file is filtered to only include photons with energies in the range $0.3-8.0$ keV.
\item The \texttt{CIAO} tool \texttt{wavdetect} is used to detect all X-ray point sources (\chic\ sources) within $3'$ of the AGPS position. A \cxo\ image is then created for each \chic\ source (for example see the bottom image of Figure~\ref{Fig1}). If no sources are detected, \chic\ MAP ends.
\item The position, source counts, and associated errors are calculated for each \chic\ source detected. If the dataset is an ACIS-S observation then the total counts are obtained in the $0.3-8.0$, $0.5-2.0$, and $2.0-8.0$ keV energy ranges, and the energy quartiles ($E_{25}$, $E_{50}$, and $E_{75}$), which are used in quantile analysis (see Section 2.3.1), are calculated.\footnote{The HRC instrument has very poor spectral resolution so this, and the following step, are not conducted on these datasets.}
\item The \texttt{CIAO} tool \texttt{specextract} is run on ACIS-S datasets to obtain the source and background spectrum files and their corresponding redistribution matrix file (RMF) and the ancillary response file (ARF) for each \chic\ source. These files are used in quantile analysis and spectral modelling (see Section 2.3).
\item A timing analysis is conducted on each \chic\ source (detected with either the ACIS or HRC instruments) to search for short term variability and periodicity. 
\begin{itemize}
\item A light-curve with 8 bins is constructed using the \texttt{CIAO} tool \texttt{dmextract}. The $\chi^{2}$ is calculated for this light-curve in order to test for short-term variability (e.g. Figure~\ref{Fig2}). 
\item The $Z^{2}_{1}$ statistic is calculated to search for sinusoidal periodicity \citep{buccheri83}. This process creates a power spectrum and folded light curve (e.g. Figure~\ref{Fig3}), which predicts the most likely pulsed frequency, its corresponding power, and the probability that the power is random noise. 
\end{itemize}
\item Multi-wavelength follow-up and catalog searches are conducted to identify likely optical, infrared, and radio counterparts to each \chic\ source.
\begin{itemize}
\item The USNO B1 \citep{monet03}, 2MASS PSC \citep{skrutskie06}, and GLIMPSE I and II \citep{benjamin03} catalogs are accessed to obtain a list of all optical and infrared sources within $4''$ of the \chic\ source's \texttt{wavdetect} position.
\item Small sized ($6'$ by $6'$) image cutouts, centered on the \chic\ source's \texttt{wavdetect} position, are obtained from the 2nd Digitized Sky Survey \citep[Red: DSS2R and Blue: DSS2B;][]{mclean00}, 2MASS, and the GLIMPSE I and II surveys. A $30'$ by $30'$ radio image cutout is obtained from the Sydney University Molonglo Sky Survey \citep[SUMSS;][]{bock99}. Examples of all the image cutouts can be found in Figures~\ref{Fig4} and \ref{Fig5}.
\end{itemize}
\end{itemize}

\chic\ MAP first generates an image of the AGPS source as originally detected by the Gas Imaging Spectrometer \citep[GIS;][]{ohashi96} onboard \asca. The GIS has a circular field-of-view with a $50'$ diameter. The top image of Figure~\ref{Fig1} shows the \asca\ GIS detection of the AGPS source \object[AX J1447.0-5919]{AX J144701--5919}. We have chosen AX J144701--5919 as the example source for illustrating the output of \chic\ MAP because it was the first AGPS source observed with \cxo\ as part of the \chic\ survey, and it is also an interesting source with a bright counterpart \citep[as demonstrated by its identification as an X-ray emitting WR star in][]{anderson11}.

\chic\ MAP reprocesses all of the \cxo\ observations, both ACIS and HRC, using the \texttt{CIAO} \texttt{chandra\_repro} script, which creates a new level=2 event file and bad pixel file. The \cxo\ ACIS data are filtered to only include events with energies in the range $0.3-8.0$ keV in order to avoid high-energy and cosmic ray particle backgrounds. 

We have chosen to optimize \chic\ MAP for the detection of point sources as our short \cxo\ observations ($<10$ ks) are not very sensitive to extended sources. \chic\ MAP uses the \texttt{CIAO} wavelet detection algorithm, \texttt{wavdetect}, which we have set to search for all sources with wavelet scales of radii 1, 2, 4, 8 and 16 pixels. This does, however, introduce a source selection bias as it is possible that AGPS sources that were unresolved with the \asca\ PSF could be extended, and therefore resolvable with \cxo. Re-running \chic\ MAP using \texttt{wavdetect} with larger scales more appropriate for extended sources, or using the \texttt{CIAO} Voronoi Tessellation and Percolation source detection algorithm \texttt{vtdetect} \citep{ebeling93}, could be conducted in the future to detect extended sources in the \chic\ \cxo\ images. 

ChIcAGO MAP utilizes \texttt{wavdetect} to detect all sources within $3'$ of the original \asca\ position (for example see the \cxo\ detection of AX J144701--5919 in the bottom image of Figure~\ref{Fig1}). This search radius is based on the position accuracy and spatial resolution of \asca\ and therefore designed to ensure that the majority of contributing X-ray sources are encompassed. (However, it is also possible that there are associated X-ray sources beyond $3'$ from the AGPS position. This could be due to the inaccuracy of the \asca\ positions of those sources that are blended or near the edge of the field-of-view. This is further explored in Section 3.1 and Appendix B.) In many cases more than one X-ray source could have been contributing to the total X-ray flux from an AGPS source originally detected with \asca. (Note this is not the case for the example of AX J144701--5919.) The positional accuracy of \texttt{wavdetect} has been well investigated and tested in previous \cxo\ surveys \citep[i.e. ChaMPlane;][]{hong05}.

The position of each \chic\ source as output by \texttt{wavdetect} is obtained and Equation (5) of \citet{hong05} is used to calculate the 95\% confidence position error circle.\footnote{This equation was constructed by running \texttt{wavdetect} on ACIS-I and ACIS-S data. It is therefore unknown whether this position error equation is applicable to the positions calculated from running \texttt{wavdetect} on HRC data sets. However, since the \texttt{wavdetect} algorithm is not instrument specific, it is likely that this equation estimates reasonable errors for sources detected in HRC observations.} This \texttt{wavdetect} error is then added in quadrature to the absolute astrometry error of \cxo\ to obtain the total position error of each \chic\ source.

The source regions used to calculate the total number of source counts, which are centered on the source position as output by \texttt{wavdetect}, have a radius equivalent to 95\% of the point-spread function (PSF) at 1.5 keV. The background subtraction is performed using an annulus whose size is between two and five times the above 95\% PSF radius, centered on the source position. Given the low number of counts detected, usually $<100$, the 1$\sigma$ lower and upper confidence limits of the total number of counts are calculated using \citet{gehrels86} statistics. The following energy based analysis is then performed on the ACIS-S observations. The total number of counts are calculated for the $0.3-8.0$, $0.5-2.0$, and $2.0-8.0$ keV energy ranges. ChIcAGO MAP then uses the extracted ACIS-S counts and corresponding energies to calculate the 25\%, 50\% and 75\% photon fractions ($E_{25}$, $E_{50}$ and $E_{75}$), the energies below which 25\%, 50\% and 75\% of the photon energies are found, respectively. These median ($E_{50}$) and quartile ($E_{25}$ and $E_{75}$) values can immediately characterize the hardness of a source without using conventional hardness ratios, which are not versatile enough to account for diverse X-ray spectral types in the Galactic plane. These quartile fractions can then be used to employ quantile analysis \citep{hong04}, which uses a quantile based color-color diagram to classify spectral features and shapes of low count sources (see Section 2.3.1). 

The \texttt{CIAO} tool, \texttt{specextract}, is also run on the ACIS-S detected \chic\ sources to generate source and background spectrum files and their corresponding RMF and the ARF. These files are used to perform spectral interpolation with quantile analysis and to conduct spectral modelling and fitting with the \texttt{CIAO} package \texttt{Sherpa}. Further details on the spectral investigations of the \chic\ sources can be found in Section 2.3

\chic\ MAP then performs a timing analysis on all the \chic\ sources, detected with either the ACIS or HRC instruments, to search for evidence for short term variability and periodicity. In each case a light-curve is extracted using \texttt{dmextract} where the counts are divided into 8 bins (for example see Figure~\ref{Fig2}). The Gehrels approximation to confidence limits for a Poisson distribution is used to estimate the errors as there are $<20$ X-ray counts in each bin. The $\chi^{2}$ statistics are adopted to test for variability \citep[for example see][]{gaensler00}. (It should be noted that as \texttt{dmextract} uses the upper (larger) Gehrels confidence limit to estimate the count-rate errors, the resulting $\chi^{2}$ output by \chic\ MAP may be underestimated.) For 7 degrees of freedom, $\chi^{2} \gtrsim 24.3$ is required for a source to be considered variable at 99.9\% confidence. 

After correcting the photon arrival times to the Solar System barycenter, the target source is then investigated for evidence of periodicity using the $Z_{n}^{2}$ test \citep{buccheri83}, equivalent to the Rayleigh statistic when $n$, the chosen number of harmonics, is set to 1. \chic\ MAP uses $n=1$ for the sake of simplicity and because such a test is sensitive to sinusoidal distributions. $Z_{1}^{2}$ has a probability density function equivalent to the $\chi^{2}$ statistic with $2$ degrees of freedom. ChIcAGO MAP searched for periodicity down to 6.48s \citep[twice the frame-time resolution;][]{weisskopf03} for sources detected with ACIS, and down to 0.01s for sources detected with HRC.\footnote{The lower-limit of 0.01s for the HRC observations is chosen because a wiring error in the detector degrades the time resolution accuracy from $16\mu s$ to the mean time between events. See http://cxc.harvard.edu/proposer/POG/html/chap7.html\#sec:hrc\_anom} A power spectrum (power vs. frequency) and folded light curve (counts per bin vs. phase) of each source is generated, predicting the pulsed frequency of the highest $Z_{1}^{2}$ power, and the probability that this power was random noise for a given number of trials (for example see Figure~\ref{Fig3}). A 99.9\% confidence was required for a source to have a significant level of periodicity, which corresponds approximately to $Z_{1}^{2} \gtrsim 26$ and $Z_{1}^{2} \gtrsim 36$ for ACIS and HRC observations, respectively. 

It is also possible that many of the AGPS sources are transient or undergo long-term variability, so have changed significantly in flux since the original \asca\ observations. The detailed analysis of any periodic, variable, and transient sources is beyond the scope of this paper. We only flag those source that may fit into one of the above categories for the purpose of future investigations.

The next step in the \chic\ identification process is to search for multi-wavelength counterparts to the \chic\ sources. \chic\ MAP accesses the US Naval Observatory B Catalog, version 1.0 \citep[USNO B1, visual magnitude bands $B$, $R$, and $I$;][]{monet03}, the Two Micron All Sky Survey Point Source Catalog \citep[2MASS PSC, near-infrared magnitude bands $J$, $H$, and $K_{s}$;][]{skrutskie06}, and the Galactic Legacy Infrared Mid-Plane Survey Extraordinaire I and II Spring 07' Catalogs (highly reliable) and Archives (more complete, less reliable) \citep[GLIMPSE, infrared magnitude bands 3.6, 4.5, 5.8, and 8.0 $\mu$m;][]{benjamin03} to obtain a list of all the optical and infrared sources within $4''$ of the \texttt{wavdetect} position of each \chic\ source. The information extracted from these surveys includes the position of the source, the offset from the \cxo\ position and the magnitudes listed in the given survey or dataset.\footnote{The catalog information is downloaded via a generic URL from the VizieR Service (http://vizier.cfa.harvard.edu/viz-bin/VizieR) and the NASA/IPAC Infrared Science Archive (http://irsa.ipac.caltech.edu/).}

Small sized images cutouts ($6'$ by $6'$) from optical and infrared surveys, centered on the \chic\ source \texttt{wavdetect} position, are also downloaded to enable a visual inspection of likely counterparts and their surrounding environments. The $B$ and $R$ magnitude band images are obtained from the 2nd Digitized Sky Survey \citep[Blue: DSS2B and Red: DSS2R;][]{mclean00} and the $J$, $H$, and $K$ magnitude band images are obtained from 2MASS. Image cutouts of the infrared magnitude bands 3.6, 4.5, 5.8 and 8.0 $\mu$m are obtained from the GLIMPSE I and II, version 3.5, surveys.\footnote{http://irsa.ipac.caltech.edu/data/SPITZER/GLIMPSE/} A $30'$ by $30'$ image cutout of the 843 MHz radio sky is also generated from SUMSS, which is a survey conducted with the Molonglo Observatory Synthesis Telescope (MOST) and has a resolution of $43'' \times 43'' cosec |\mathrm{dec}|$ \citep{bock99}.\footnote{The images from the DSS2R, DSS2B, 2MASS and SUMSS surveys are downloaded using a generic URL from the virtual observatory SkyView, which is a service of the Astrophysics Science Division at NASA/GSFC and the High Energy Astrophysics Division of the Smithsonian Astrophysical Observatory (SAO), http://skyview.gsfc.nasa.gov/} An example of all the image cutouts generated by \chic\ MAP for AX J144701--5919 can be seen in Figures~\ref{Fig4} and \ref{Fig5}.

\subsection{\chic\ Sources Spectral Investigation}

Deducing the best spectral model fit to the \chic\ sources is difficult using standard X-ray spectral fitting techniques due to the small number of X-ray source counts detected in the \cxo\ observations (usually $<100$). We therefore implement two different techniques for predicting the best spectral parameters for the brighter ($>20$ X-ray counts) \chic\ sources. The first technique is ``quantile analysis" \citep{hong04}, which has been recently developed to address some of the problems associated with spectral modeling of sources with low number statistics. Quantile analysis allows for the interpolation of likely spectral shapes of X-ray sources with as few as $10$ counts. The second technique is utilizing the \texttt{CIAO} spectral fitting tool \texttt{Sherpa} to obtain best fit spectral parameters using \citet{cash79} statistics. These statistics are based on Poisson distributed data and are therefore ideal for modeling spectra with a limited number of source counts. Both methods are described in detail below. 

\subsubsection{Quantile Analysis}

Quantile analysis uses predetermined fractions of the total number of energy source counts, such as the median ($E_{50}$) and quartile energies ($E_{25} \mathrm{~and~} E_{75}$), to construct a quantile-based phase space that can be overlayed with grid patterns of common spectral models. This quantile phase space is more sensitive to the wide range of Galactic X-ray sources than those constructed from conventional hardness ratios. This is because there is no count-dependent sensitivity bias toward any spectral type, which is inherent to the choice of sub-energy bands in the conventional spectral hardness or X-ray color analysis. In the \chic\ survey, quantile analysis is adopted to calculate potential spectral shapes of the \chic\ sources.

The ACIS-S detected \chic\ sources selected for quantile analysis have $>20$ (net) X-ray counts in the $0.3-8.0$ keV energy band, as these are likely to be bright enough to be the original AGPS sources (see Section 2.1). The net counts were obtained by summing the total number of counts inside a source region that is $6$ times the error radius in size and subtracting the background normalized counts calculated from an annulus that has an outer radius of $\sim15''$ with the source region subtracted. The three quartiles ($E_{25}, E_{50}, \mathrm{~and~} E_{75}$) and their corresponding errors were calculated as outlined by \citet{hong04}. 

The quantile phase space, suggested by \citet[][see their Figure 4]{hong09}, used to calculate the likely spectral shapes, was constructed for each \chic\ source. This phase space consists of the normalized logarithmic median $(\mathrm{log}(E_{50}/E_{low})/\mathrm{log}(E_{high}/E_{low}))$ and the normalized quartile ratio ($3 \times (E_{25}-E_{low})/(E_{75}-E_{low}))$, where $E_{low} = 0.3$ keV and $E_{high}=8.0$ keV, equivalent to the energy range explored. The logarithmic median phase space takes advantage of the higher sensitivity at low energies in typical X-ray telescope CCDs, while keeping the spectral discernibility more or less uniform throughout the full range of the phase space. In order to compensate for the spatial change of the detector response in Chandra/ACIS, the RMF and the ARF, appropriate to the observed location of each \chic\ source in the CCD, are calculated using the \texttt{CIAO} tools.

The data points of the \chic\ sources in the quantile diagram were compared with simple power-law and thermal bremsstrahlung spectral models to extract the most plausible spectral parameter values for each source. It should be noted that quantile analysis cannot evaluate which model is more likely unless the estimated parameters of the model turn out to be unphysical (e.g. $\Gamma>4$). The estimate of the spectral parameters are limited to $-2 \leq \Gamma \leq 4$ for a power law model and $0.1 \mathrm{~keV} \leq kT \leq 10 \mathrm{~keV}$ for a bremsstrahlung model. The explored extinction ($N_{H}$) covers the range $0.01 - 100 \times 10^{22}$ cm$^{-2}$. If the data point for a source in quantile phase space sits outside of the model grid set by these parameter ranges, the model is considered incompatible with the observed spectrum of the source. The quantile errors allow the spectral parameter uncertainties to be calculated for each \chic\ source \citep{hong04}. 

\subsubsection{Spectral Modeling}

The spectral modeling of the \chic\ sources was conducted using the \texttt{CIAO} 4.5 spectral fitting package \texttt{Sherpa} with the statistics set to the \texttt{XSpec} \citep{dorman01} implementation of \citet{cash79} statistics. Cash statistics apply a maximum likelihood ratio test that can be performed on sources with a low number of source counts per bin. We chose to restrict all spectral modeling to those \chic\ sources with $>50$ X-ray counts as attempted modeling of those sources with less counts usually did not converge.

Cash statistics cannot be performed on a background subtracted spectrum. The source and background spectrum must instead be modeled simultaneously. However, in the case of the \cxo\ observations of \chic\ sources the background is extremely low and cannot be described by a generic spectral model. As a result attempting to model the background spectrum does not improve the overall spectral fit. We therefore only model the source spectrum of the \chic\ sources. 

Both an absorbed power law and an absorbed thermal bremsstrahlung model are applied to the \chic\ source spectra so that they can be directly compared to the quantile analysis spectral interpolation results. The parameter errors are calculated using the \texttt{Sherpa} ``projection" function, which estimates the $1\sigma$ confidence intervals. The absorbed and unabsorbed flux (plus errors) are calculated using the \texttt{Sherpa} function ``sample\_flux", which is new to \texttt{CIAO} 4.5. The overall goodness of fit measure is defined by the value of the Cash statistic divided by the number of degrees of freedom and should be of order 1. 

\subsection{Multi-wavelength Follow-up Observations}

While the X-ray morphology and spectrum can provide information on the nature of a \chic\ source, the key to identification is usually through an extensive multi-wavelength follow-up campaign. \chic\ MAP identifies the possible optical and infrared counterparts in existing multi-wavelength Galactic plane surveys. There are, however, many cases where the counterparts are too faint to be detected in these surveys due to the high absorption in the Galactic plane. It is also possible that the high object density in the Galactic plane may result in confusion with nearby sources. In these cases, further optical and infrared photometric observations  were conducted with large telescopes to obtain detections of faint counterpart candidates and to separate likely blends. If the first photometric observing attempt was unsuccessful at detecting or separating a counterpart candidate then deeper imaging using longer exposure times was conducted. Those X-ray sources that remain undetected at optical and infrared wavelengths will need to be further investigated in the X-ray band or at other wavelengths.

The radio wavelength band is also a useful diagnostic for identifying X-ray sources in the Galactic plane. Comparing the X-ray source positions with radio surveys can indicate if there is a likely radio counterpart or whether the X-ray source lies in a diffuse region of radio emission in the Galactic plane. Interferometric radio observations of a small subset of \chic\ sources were obtained in order to resolve confusing regions of radio emission and allow for the detection or confirmation of compact radio counterparts.

\subsubsection{Optical and Near Infrared Observations with Magellan}

The optical and NIR photometric observations presented in this paper were obtained using instruments on the twin 6.5m, Baade and Clay, Magellan telescopes, located at Las Campanas Observatory, Chile. The NIR photometry, $1-2.5 \mu$m, was obtained using the Persson's Auxiliary Nasmyth Infrared Camera \citep[PANIC;][]{martini04,osip08} on Baade. PANIC was used primarily to detect the counterparts of \chic\ sources with no cataloged counterparts or to utilize this instrument's high angular resolution to eliminate possible blendings. Observations were obtained in the $J$, $H$ and $K_{s}$ photometric bands, using short exposures ($10-30$s) that were dithered to account for the high sky background inherent to NIR observations. 

The PANIC NIR imaging data were reduced using the Image Reduction and Analysis facility \citep[IRAF;][]{tody86,tody93} and the PANIC Data Reduction Package for \texttt{IRAF} \citep{martini04}, taking into account the corresponding darks, sky flats and bad pixel maps obtained for each respective night. The absolute astrometry for the PANIC observations was derived using the 2MASS PSC and the Graphical Astronomy and Image Analysis Tool \citep[\texttt{GAIA};][]{draper09}. The positional accuracy of the 2MASS PSC is $0\farcs1$ \citep[$1\sigma$,][]{skrutskie06}, and since there are usually many 2MASS sources in the field, a similar order of astrometric accuracy at the target positions was reached. \texttt{SExtractor} \citep{bertin96} was used for source detection, and the calibration of the photometry was performed by applying 2MASS PSC photometry to known 2MASS sources in the target fields. No correction to the atmospheric extinction was applied as this effect is very small in the NIR ($\sim0.06$~magnitudes in the $H$-band). The errors were obtained by comparing the NIR-magnitudes with 2MASS PSC magnitudes, and reflected average deviations from the 2MASS catalog magnitudes.

Optical photometric observations of the counterparts were obtained utilizing several Magellan instruments depending on availability. These included the Inamori Magellan Areal Camera and Spectrograph \citep[IMACS;][]{dressler06,osip08} and the Raymond and Beverly Sackler Magellan Instant Camera \citep[MagIC;][]{osip08} on the Baade telescope. These instruments provide access to different photometric filters including Bessel $B$, $V$ and $R$, and CTIO $I$.

The IMACS $B$, $V$, $R$ and $I$ imaging data were reduced using \texttt{IRAF}, in which the data were trimmed, overscan-corrected and flat-fielded. A dark current subtraction was not applied since the dark images showed it to be negligible. Standard stars were observed throughout in several bands, although the weather during some of the observing nights varied. The absolute astrometry was again computed using \texttt{GAIA} with comparisons to the USNO B1 Catalog or the 2MASS PSC. The photometry of each counterpart was calculated using \texttt{SExtractor} and was calibrated using the USNO B1 Catalog.

MagIC observations were obtained in the $V$, $R$, and $I$ bands. Short 30\,s exposures were obtained in each filter, which were later combined to make sure that the brighter stars (used for astrometric referencing) were not saturated. These data were then reduced following standard procedures in \texttt{IRAF}: overscan subtraction for each amplifier, flatfielded using dome flats, and the separate exposures combined. The astrometry was applied by referencing the observations to the 2MASS PSC, resulting in an rms residual of $0\farcs1$ in each coordinate. Photometric calibration was derived using observations of the \citet{stetson00} standard fields. The measured photometry for the airmass terms appropriate to Las Campanas Observatory were corrected and the zero-points with scatters of $0.02$ magnitudes in each filter were obtained. Similarly to the PANIC and IMACS observations, the photometry of each counterpart was then measured using \texttt{SExtractor}. 

\subsubsection{Radio follow-up and ATCA Observations}

Radio survey data already exist for all of the AGPS source regions, via the first and second epoch Molonglo Galactic Plane Surveys at 843 MHz \citep[MGPS1 and MGPS2 respectively, $43'' \times 43'' cosec |\mathrm{dec}|$ resolution;][]{green99,murphy07}, the 90cm Multi-configuration Very Large Array Survey of the Galactic Plane \citep[$42''$ resolution;][]{brogan06}, The Multi-Array Galactic Plane Imaging Survey at 1.4 GHz \citep[MAGPIS, $6''$ resolution;][]{helfand06}, and the Very Large Array (VLA) Galactic Plane Survey at 1.4 GHz \citep[VGPS, $1'$ resolution;][]{stil06}. To search for possible radio counterparts we first visually inspected the above surveys at the position of each \chic\ source. Any coincident radio emission was then categorized (see Section 3.4). 

Follow-up Australia Telescope Compact Array (ATCA) observations were conducted to identify the nature of any radio counterparts to the \chic\ sources found through this visual inspection. The ATCA was also used to resolve any diffuse radio emission surrounding a \chic\ source, allowing the detection of any underlying, compact radio counterpart. The high resolution ATCA observations were instrumental in confirming a positional coincidence between the radio counterpart and the \cxo\ position of the \chic\ source, to measure accurate radio fluxes and spectral indices, to broadly characterize variability with respect to earlier epochs, and to constrain the object's spatial extent. The \cxo\ positions of ten \chic\ sources were observed with the ATCA on 2008 January 21 and on 2008 April 11. Each source was observed for $\sim1$ hour at each of 1.4, 2.4, 4.8 and 8.6 GHz over a 12 hour period with a 6km baseline configuration.

\section{Results}

\subsection{\cxo\ Results}

\begin{figure}[htp]
\begin{center}
\includegraphics[width=0.4\textwidth]{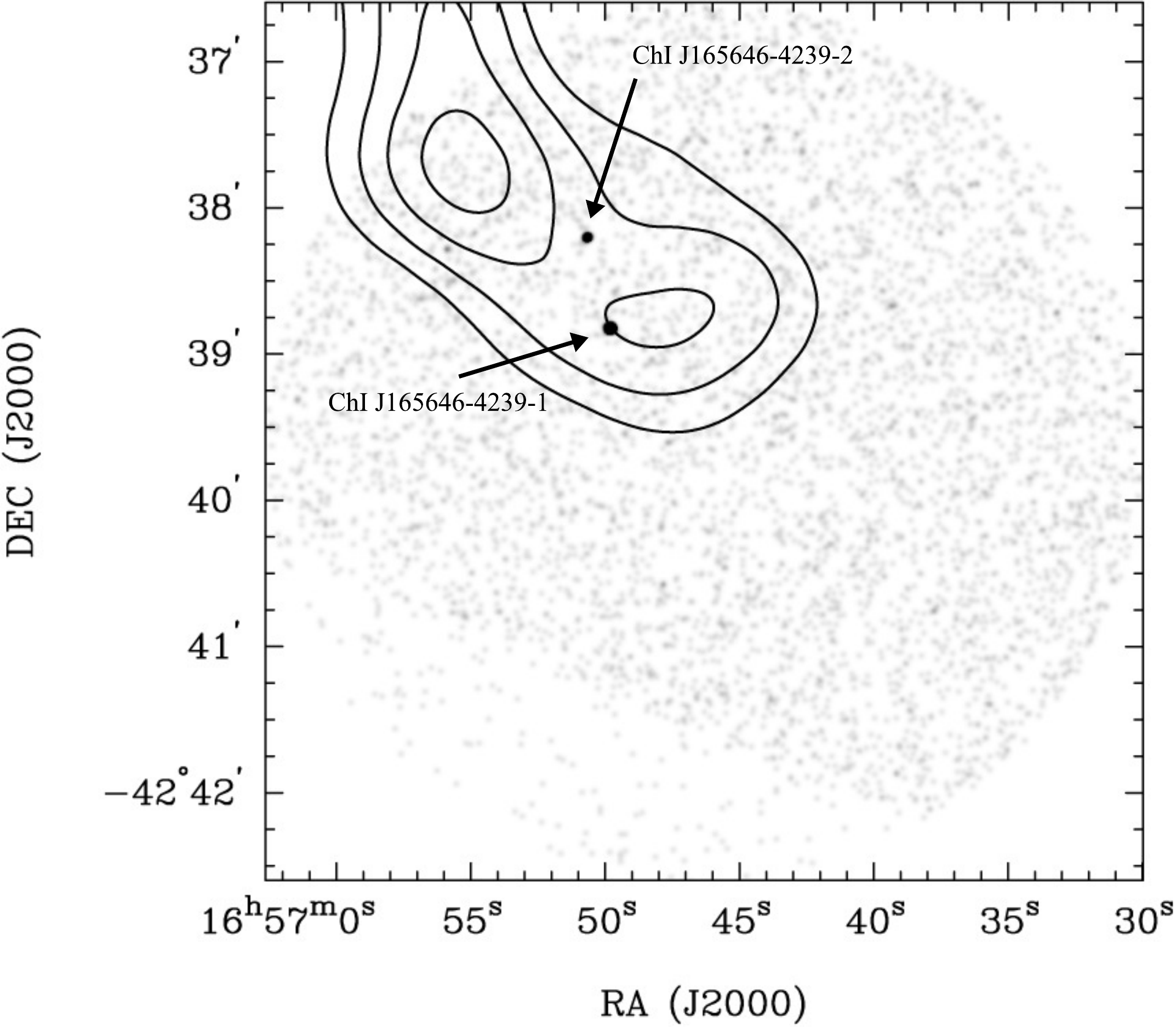}
\caption{The $3'$ radius field-of-view of the \cxo\ observation of the AGPS source AX J165646--4239, which is centered on the position of AX J165646--4239 as listed by \citet{sugizaki01}. Two \chic\ sources, ChI J165646--4239\_1 and ChI J165646--4239\_2, which are black dots indicated by arrows, were detected with \cxo\ and are labeled on this Figure. The black contours represent the smoothed \asca\ GIS detection of AX J165646--4239 at 65\%, 75\%, 85\%, and 95\% of the peak count-rate.}
\label{Fig6}
\end{center}
\end{figure}

We present results on 93 APGS sources that have been observed with \cxo\ as part of the \chic\ survey. In many cases more than one source was detected within $3'$ of the original AGPS source positions so, as mentioned in Section 2.2, we will refer to these all as \chic\ sources. We therefore detected a total of 253 \chic\ sources in these 93 \cxo\ observations. The naming convention we have adopted is to call each source by the AGPS coordinate name, using ChI as the prefix, with a suffix between 1 to $i$, where $i$ is equal to the number of sources detected in the $3'$ field. (The source name order is based on the order that \texttt{wavdetect} detected and output the sources in \chic\ MAP.) For example, two \chic\ sources, ChI J165646--4239\_1 and ChI J165646--4239\_2, were detected with \cxo\ in the vicinity of the AGPS source \object[AX J1656.7-4239]{AX J165646--4239}. These two \chic\ sources are the black dots in Figure~\ref{Fig6}. The \asca\ GIS detection of AX J165646--4239 is overlayed on Figure~\ref{Fig6} in the form of contours, demonstrating that both ChI J165646--4239\_1 and ChI J165646--4239\_2 may have contributed to the X-ray emission originally detected for this source in the AGPS.

The \cxo\ observations of the 93 AGPS sources are summarized in Table~\ref{Tab1}. This information includes the observation identification, the date of the observation, instrument (ACIS-S or HRC-I), and the exposure time. Table~\ref{Tab1} also lists the \chic\ sources detected by \chic\ MAP within $3'$ of the \asca\ position and their corresponding \texttt{wavdetect} position, the offset of this position from the original AGPS position, the 95\% \texttt{wavdetect} and total \chic\ source position error, the background corrected net counts in the $0.3-8.0$, $0.5-2.0$ and $2.0-8.0$ keV energy ranges,\footnote{The net counts were calculated separately for each energy range so there are several cases where the total number of counts in the $0.3-8.0$ keV energy range is slightly different to the sum of counts detected in the $0.5-2.0$ and $2.0-8.0$ keV energy ranges. These differences are due to rounding errors, as well as the exclusion of very soft counts that were detected between $0.3-0.5$ keV. (The $0.3-0.5$ keV energy range was excluded as the $0.5-2$ keV energy band is more commonly used than the $0.3-2$ keV energy band.)} and the median energy quartile ($E_{50}$) and quartile ratio ($3\times(E_{25}-0.3)/(E_{75}-0.3)$) used in quantile analysis (see Section 2.3.1). 

The variability analysis performed by \chic\ MAP detected only one \chic\ source, ChI J170444--4109\_1, with a $\chi^{2}$ exceeding the 99.9\% short term variability threshold ($\chi^{2}=80.6$). However, this variability is not real as the ACIS-S detection of ChI J170444--4109\_1 fell in the chip gap between the CCDs. This resulted in its light-curve displaying the \cxo\ spacecraft's built-in dither as the source moved on and off the chip at regular intervals.\footnote{see http://cxc.harvard.edu/ciao/why/dither.html} No other \chic\ source had a $\chi^{2}$ exceeding the 99.9\% confidence threshold for short term variability or periodicity. This does not, however, rule out the possibility that some of these sources are periodic as the $Z^{2}_{1}$ period search technique is extremely limited for $<200$ X-ray counts.

Several faint \chic\ sources were detected with \texttt{wavdetect} in the \cxo\ observations that are not suspected of being significant contributors to the \asca\ counts detected in the AGPS. (All these faint sources are included in Table~\ref{Tab1}.) In order to determine the detection significance of these sources, we independently calculated the probability of a false detection (Pfd) for each source based on the Poisson statistics. We expect there to be $\sim100000$ trials in a $3'$ radius ACIS CCD field-of-view if a detection cell size of a $1''$ radius circle is assumed with 2D Nyquist sampling \citep{weisskopf07}. (A generic detection cell size of $1''$ for this detection significance calculation was suggested by \citet{weisskopf07} since the \texttt{CIAO} source detection algorithms usually use a detection cell size of between $40-80\%$ of the PSF at a given energy.\footnote{See the \texttt{CIAO} detect manual http://cxc.harvard.edu/ciao/download/doc/detect\_manual/}) Pfd $=10^{-5}$ implies that there is one false source in the $3'$ radius ACIS \cxo\ field. The probability of a false detection was calculated for each source \citep[see footnote 13 in][]{kashyap10}. Only 11 sources (out of 253) have a Pfd $>10^{-5}$ but all have Pfd $<7 \times 10^{-4}$. Several of these sources also have a likely optical and/or infrared counterpart, increasing the significance of these detections. As this technique for calculating source significance is highly theoretical (e.g. the effective detection cell size of ACIS is likely to be slightly smaller than the $1''$ radius circles), without any consideration for possible counterparts, we will include these 11 marginally significant sources in Table~\ref{Tab1} (denoted by a * symbol in the first column).

Several of the sources detected in these observations have also been listed in the \cxo\ Source Catalog \citep[CSC;][]{evans10}. We assume that a CSC source and a \chic\ source are the same if the separation between their two positions is less than the quadratic sum of their 95\% error radii plus a constant term of $0\farcs7$, which accounts for the 95\% absolute astrometry error of \cxo\ assuming the errors follow a Gaussian distribution\footnote{The \cxo\ astrometric error is included in this calculation as the source positions listed in the CSC can be obtained from more than one dataset.} \citep[see Equation (6) of][]{hong05}. We have conducted the same comparison with the fifth public release of the Second \textit{XMM-Newton} Serendipitous Source Catalogue \citep[2XMMi-DR3;][]{watson09}\footnote{http://xmmssc-www.star.le.ac.uk/Catalogue/2XMMi-DR3/}. The CSC and 2XMMi-DR3 names that correspond to any \chic\ sources are listed in Table~\ref{Tab1}. Further analysis of the \xmm\ observations will be conducted in future work.

The purpose of the \chic\ survey is to study those X-ray sources detected in the \chic\ \cxo\ observations that fall within the flux range of the AGPS sources ($F_{x} \sim 10^{-13}\mathrm{~to~}10^{-12}$ \erg). In 62 out of the 93 AGPS sources observed with \cxo, \chic\ MAP found 74 sources with $>20$ X-ray counts that fall within the investigated $3'$ field-of-view that is centered on the AGPS positions. For the purposes of this study we will focus on these 74 \chic\ sources, listed in Table~\ref{Tab2}, as they are approximately within the original AGPS flux range. The detailed analysis of those 179 \chic\ sources with $<20$ counts will be deferred to future work. There were 6 AGPS sources where no \chic\ source was detected by \cxo: \object[AX J1659.8-4209]{AX J165951--4209}, \object[AX J1753.5-2538]{AX J175331--2538}, \object[AX J1808.2-2021]{AX J180816--2021}, \object[AX J1836.1-0756]{AX J183607--0756}, \object[AX J1859.0+0333]{AX J185905+0333}, and \object[AX J1910.7+0917]{AX J191046+0917}. At least three of these AGPS sources, AX J180816--2021, AX J185905+0333, and AX J191046+0917, may be transient given how bright they were in the original \asca\ detections \citep{sugizaki01}. The remaining 25 AGPS fields observed with \cxo\ only have faint sources with $<20$ X-ray counts. There are also 6 AGPS sources, \object[AX J1445.3-5949]{AX J144519--5949}, \object[AX J1510.0-5824]{AX J151005--5824}, \object[AX J1549.0-5420]{AX J154905--5420}, \object[AX J1943.1+2318]{AX J194310+2318}, \object[AX J1943.5+2323]{AX J194332+2323}, and \object[AX J1950.1+2628]{AX J195006+2628}, where multiple \chic\ sources detected in each region (with $\lesssim30$ X-ray counts) sum to $\gtrsim60$ counts, which are close to the number of X-ray counts that were expected to be detected with \cxo\ from each AGPS source (see Section 2.1). Many of the \chic\ sources in these field also have optical and/or infrared counterparts and may therefore be members of star clusters (see Section 4.1 for further discussions). 

The contribution of X-ray emission beyond the $3'$ search radius was also investigated using \texttt{wavdetect} to identify all the point sources with $>20$ X-ray counts in the \cxo\ observations that lie between $3'-5'$ from the original AGPS position. Only 14 X-ray point sources were found in the $3'-5'$ annulus surrounding 11 AGPS sources, which demonstrates that the $3'$ search radius used by \chic\ MAP is reasonable and has likely allowed us to identify the majority of \chic\ sources. These 14 X-ray point sources are further discussed in Appendix B where we explore the likelihood of whether they could be associated with their nearby AGPS source.

\subsection{Quantile Analysis and Spectral Modeling Results}

Quantile analysis and spectral modeling using Cash statistics were both techniques used to infer the most likely absorbed power-law and absorbed thermal bremsstrahlung spectrum of the bright \chic\ source detected by ACIS-S.\footnote{\chic\ sources with $>20$ X-ray counts for quantile analysis and $>50$ X-ray counts for spectral modeling.} The spectral parameters and the absorbed and unabsorbed X-ray fluxes for an absorbed power law and absorbed thermal bremsstrahlung model are listed in Tables~\ref{Tab3} and \ref{Tab4}, respectively. There are also six HRC-I detected \chic\ sources with $>20$ counts (ChI J144042--6001\_1, ChI J153818--5541\_1, ChI J163252--4746\_2, ChI J163751--4656\_1, ChI J165420--4337\_1, and ChI J172642--3540\_1) that are not included in Tables~\ref{Tab3} and \ref{Tab4} as their X-ray observations do not contain any spectral information. The possible counterparts to these six sources are, however, explored in detail in Section 3.3.

There is no measure of goodness for the quantile analysis derived absorbed power-law and absorbed bremsstrahlung spectral parameters listed for the \chic\ sources in Tables~\ref{Tab3} and \ref{Tab4}. Instead a spectral interpolation is classified as unreasonable if the resulting parameters are outside $-2<\Gamma<4$ for a power-law model and $0.1< kT <10$ keV for a bremsstrahlung model. These parameter limits can allow for the grouping of the \chic\ sources into existing categories that are based on the physical understanding of X-ray sources that emit in the \cxo\ energy range ($0.3-8$ keV). While these selection criteria explicitly excludes the identification of new types of X-ray sources with unusual spectra, such an investigation is beyond the scope of this work. In all cases if one of the quantile analysis spectral interpolations was disregarded for a given \chic\ source, the other is reasonable when considering the above criteria.

As a way to further explore the goodness of the spectral interpolations, the \texttt{CIAO} spectral fitting tool \texttt{Sherpa} was used to fit the X-ray spectrum of each \chic\ source with spectral parameters derived from quantile analysis in Tables~\ref{Tab3} and \ref{Tab4}. The  $\chi^{2}_{red}$ was calculated for each \chic\ source where $>40$ X-ray counts was detected with ACIS-S. In many cases we found $\chi^{2}_{red} < 1$, which is not unexpected given the low number of X-ray counts detected, indicating that any reasonable model is a decent fit. However, there are also a few cases where $\chi^{2}_{red} > 2$ indicating that the spectral parameters do not adequately describe the spectrum. The values of $\chi^{2}_{red}$ corresponding to the quantile interpolated spectral parameters are included in Tables~\ref{Tab3} and \ref{Tab4}.

The parameters and fluxes resulting from the absorbed power law and absorbed thermal bremsstrahlung spectral modeling of the \chic\ sources have also been included in Tables~\ref{Tab3} and \ref{Tab4} so that they can be directly compared to the quantile analysis results. Best fit absorbed power law parameters were obtained for all the ACIS-S detected \chic\ sources with $>50$ X-ray counts (see Table~\ref{Tab3}). However, there were several cases where the absorbed thermal bremsstrahlung fitting was unsuccessful as the \texttt{Sherpa} modeling algorithm hit the hard maximum limit on the $kT$ parameter ($10$ keV). These unsuccessful fits were not included in Table~\ref{Tab4}. 

All of the parameters and absorbed fluxes derived from the Cash statistics spectral fitting agree within $3\sigma$ of those derived from the quantile analysis spectral interpolation, the majority of which agree within the $1\sigma$ errors. The only exception is the $kT$ parameter value for ChI J171910--3652\_2. However, the unabsorbed fluxes were far less agreeable between the two techniques as there were several \chic\ sources for which this value differed by $>3\sigma$. These include ChI J183356--0822\_2 and ChI J184738--0156\_1 from the power law spectral analysis and ChI J144547--5931\_1,  ChI J144701--5919\_1, ChI J165646--4239\_1, ChI J171910--3652\_2, ChI J172050--3710\_1, and ChI J185608+0218\_1 from the bremsstrahlung spectral analysis. Both these techniques therefore appear to be successful in constraining the spectral parameters and absorbed fluxes for each of the investigated \chic\ sources but less successful in constraining the unabsorbed fluxes. It should also be noted that the majority of the reduced Cash statistics from the spectral modeling were systematically higher than the corresponding $\chi^{2}_{red} $ derived from the spectral parameters interpolated through quantile analysis. 

\subsection{Infrared and Optical Counterparts}

Infrared and optical follow-up were primarily performed on those \chic\ sources with $>20$ X-ray counts (see Table~\ref{Tab2}). In order to determine which optical and infrared sources are counterparts to \chic\ sources, we used a similar technique to that described by \citet{zhao05}, using their Equation (11). If the separation between a \chic\ source's \texttt{wavdetect} position and its possible counterpart is less than the quadratic sum of their 3$\sigma$ position errors and the 3$\sigma$ \cxo\ pointing error\footnote{http://cxc.harvard.edu/cal/ASPECT/celmon/}, then the X-ray and optical (or infrared) sources are likely to be associated. The 1$\sigma$ position errors for all sources in 2MASS PSC and the GLIMPSE\footnote{See GLIMPSE documents http://www.astro.wisc.edu/sirtf/docs.html} catalogs are $0\farcs1$ \citep{skrutskie06} and $0\farcs3$, respectively. USNO B has an astrometric accuracy of $<0\farcs25$ \citep{monet03}. We have assumed that the error distributions of the \cxo\ observations, \cxo\ pointing, and USNO B Catalog are all Gaussian for the purposes of identifying possible counterparts to the \chic\ sources. While this assumption is not necessarily correct in every case, the full examination required to obtain the Gaussian errors would involve a very complicated approach. However, other \cxo\ Galactic plane X-ray surveys that search for multi-wavelength counterparts assuming Gaussian errors for cross-correlation purposes have had successful results \citep[e.g. ChaMPlane;][]{zhao05}. Based on these results we feel that Gaussian errors are an acceptable assumption for the purpose of identifying optical and infrared counterparts to the \chic\ sources.

The infrared properties, such as the names and magnitudes of any likely 2MASS or GLIMPSE counterparts, together with the NIR magnitudes ($J,~H,~\mathrm{and}~K$) obtained from Magellan PANIC observations, are listed in Table~\ref{Tab2} along with the date of each observation. We assume that those \chic\ sources with no listed 2MASS (or PANIC) counterpart have 2MASS PSC limiting magnitudes $J>15.8$, $H>15.1$, and $K>14.3$ \citep{skrutskie06}. In Table~\ref{Tab5}, optical magnitudes have also been provided for those 44 \chic\ sources with $>20$ X-ray counts that have optical counterparts in the USNO B1 Catalog or in one of the IMACS or MagIC Magellan observations. Two other \chic\ sources (ChI J170017--4220\_1 and ChI J181213--1842\_7) with magnitude limits obtained with either IMACS or MagIC are also included in Table~\ref{Tab5}. 

Of the 74 \chic\ sources with $>20$ X-ray counts listed in Table~\ref{Tab2}, 59 have a NIR counterpart, 44 of which are 2MASS sources and 15 of which were detected in PANIC observations. Looking into the mid-infrared wavelength bands, we find that 41 of these 2MASS sources and 3 of the PANIC sources also have GLIMPSE counterparts. NIR magnitude limits were obtained for 4 other PANIC-observed \chic\ sources, since any possible counterparts were too faint to be detected or, in the case of ChI J181116--1828\_2 and ChI J185643+0220\_2, the counterpart appeared to be blended. (ChI J181116--1828\_2 does, however, have a unique GLIMPSE counterpart.) If we include ChI J181116--1828\_2 and ChI J185643+0220\_2, for which we have detected a counterpart but the magnitudes are only an upper limit due to blending, then 89\% of our PANIC observations of \chic\ sources have yielded a detection in one of more of the $J$, $H$, or $K$ filter bands. There are also a few \chic\ sources where the 2MASS counterpart magnitudes are listed as 95\% confidence upper limits due to a non-detection or inconsistent deblending. We were therefore able to use PANIC to obtain more accurate magnitudes for 4 \chic\ sources for which have limited 2MASS magnitude information in one or more bands. (These 4 \chic\ sources can be identified as those with three letters listed in the ``Data'' column of Table~\ref{Tab2}.)

All of the 46 \chic\ sources listed in Table~\ref{Tab5} (44 \chic\ sources with optical counterparts and 2 with limiting magnitudes obtained with Magellan instruments) have NIR counterparts detected with either 2MASS or PANIC. Of the 44 with optical counterparts, 41 have USNO B1 counterparts, for 4 of which extra magnitude measurements were obtained with one of the two Magellan optical imagers. A further 3 \chic\ sources, uncataloged in USNO B1, were also detected in the optical with these Magellan instruments. Of the 74 \chic\ sources with $>20$ X-ray counts, 14 do not have a detected optical, NIR or IR counterpart.

We conducted an experiment similar to that outlined in \citet{kaplan04} to quantify the probability that the optical and infrared survey counterparts quoted in Tables~\ref{Tab2} and \ref{Tab5}, are a random chance association with the $>20$ X-ray count \chic\ source of which they are coincident. We searched for all 2MASS, GLIMPSE and USNO B1 sources brighter than the possible counterpart listed in Tables~\ref{Tab2} and \ref{Tab5} within $10'$ of each \chic\ source position. (We only searched for survey sources that have a brighter $K_{s}$, $3.6 \mu$m, and second epoch $R$ band magnitude for the 2MASS, GLIMPSE and USNO B1 surveys, respectively.) We then used the resulting statistics to determine the number of survey sources brighter than the listed counterpart that are likely to be detected within a region the same size as the \chic\ source's 95\% position error circle. We did this for each \chic\ source individually as the density of sources can vary dramatically across the Galactic plane. In most cases the resulting chance of a random association is very low ($<0.01$). Those \chic\ sources that have a random chance of association $>0.01$ in either 2MASS or GLIMPSE are listed in Table~\ref{Tab6}. For each \chic\ source the chance of a random association with a USNO B1 is $<0.01$.  

We refer to the \chic\ sources that have $<20$ counts as ``secondary'' sources. In Table~\ref{Tab7} we list the names of any USNO B1, 2MASS and GLIMPSE sources that appear to be coincident with a secondary \chic\ source based on our position agreement criteria outlined in Section 2.4.1. This Table also includes the offset in arcseconds between the secondary \chic\ source's \texttt{wavdetect} position and the position of the coincident survey source. (Only those secondary \chic\ sources that have a coincident survey source have been included in Table~\ref{Tab7}.) A summary of the fraction of \chic\ sources with a coincident source in the 2MASS, GLIMPSE and USNO B1 catalogs, can be found in Table~\ref{Tab8}. This Table includes the fraction of the total number of \chic\ sources, as well the fraction of just the $>20$ X-ray count \chic\ sources, with a coincident survey source. (Note that we assume the coincident survey sources to the $>20$ X-ray count \chic\ sources are counterparts based on the very low chance of random associations as demonstrated by Table~\ref{Tab6}.)

\subsection{Radio Counterparts}

As mentioned in Section 2.4.2, the position of each \chic\ source (above and below $20$ X-ray counts) was visually inspected for any possibly associated radio emission in the MGPS, MAGPIS, VGPS and the 90cm Multi-configuration Very Large Array Survey of the Galactic Plane. The results of this inspection are listed in Table~\ref{Tab9}. \chic\ source position comparisons were also made with the \citet{green09} catalog of Galactic supernova remnants. Any possibly counterparts that are known objects, such as supernova remnants (SNRs), H {\sc ii} regions, infrared dark clouds \citep[IRDCs;][]{peretto09}, colliding-wind binaries (CWBs) or massive stars, are listed in Table~\ref{Tab9} under ``type''. However, if the radio sources are uncataloged, they have instead been flagged as being possibly compact, diffuse, or arc/shell structured diffuse emission. Two \chic\ sources, ChI J181116--1828\_5 and ChI J184741--0219\_3, appear to be previously unidentified AGN as their coincident radio sources show core-lobe morphologies in the MAGPIS 1.4 GHz survey images.

Each of the \chic\ sources with a possible radio association is then listed in Table~\ref{Tab9} as being either coincident to, adjacent to, or on the limb of the radio source (such as on the limb of a SNR, diffuse emission or H {\sc ii} region). If a \chic\ source is listed as either coincident or on the limb of a SNR then this means it is within the extent of the SNR based on the SNR's size quoted in the \citet{green09} catalog. In the cases of the two candidate AGN, these \chic\ sources appear to be directly coincident with the core of the AGN. The name of each radio source and the corresponding reference are listed in Table~\ref{Tab9}.

The 10 \chic\ sources observed with the ATCA are also included in Table~\ref{Tab9}. The ATCA-detected compact radio counterparts to both ChI J144701--5919\_1 and ChI J163252--4746\_2, aided in their identification as X-ray emitting massive stars or CWBs \citep[see][]{anderson11}. No radio counterparts were detected for the other 8 \chic\ sources observed with the ATCA.

In summary, Table~\ref{Tab9} shows there to be 16 \chic\ sources, from the \cxo\ observations of 8 different AGPS source regions, that are coincident, or on the limb, of 9 SNRs. There are 54 \chic\ sources, from 7 different AGPS source regions, that fall within the extent of 6 H {\sc ii} regions. There are also 4 massive stars, all of which are confirmed or candidate CWBs \citep[see][]{anderson11,motch10}, with radio counterparts. Only two \chic\ sources, both toward the same single AGPS target, are coincident with an IRDC.

Several \chic\ sources also fall within regions of uncataloged extended radio emission. These include 15 \chic\ sources, from 5 different AGPS source regions, falling within the extent of 5 regions of diffuse radio emission. Of these 15 \chic\ sources, there are 6 (from 3 different AGPS source regions) that are coincident with uncataloged diffuse emission with an arc or shell structure. Excluding the two AGN candidates, there is only one other \chic\ source coincident with an unidentified compact radio source.

There are 74 \chic\ sources (14 sources with $>20$ X-ray counts and 60 with $<20$ X-ray counts), out of the 253 detected, with no optical or infrared counterparts, making them possible compact object candidates and therefore potentially detectable in the radio. We therefore searched for any possible pulsar counterparts in the Australia Telescope National Facility Pulsar Catalogue \citep[Version 1.44\footnote{http://www.atnf.csiro.au/research/pulsar/psrcat}][]{manchester05} but no known pulsars exist within $0\farcm6$ of the \texttt{wavdetect} position of any of the 74 sources. 

\section{Discussion}

\subsection{Unidentified \chic\ Sources with Radio Counterparts}

Table~\ref{Tab9} lists 16 \chic\ sources that fall within the extent of 9 SNRs. X-ray point sources within SNRs could be associated compact objects. Identification of an optical or infrared counterpart discounts such a possibility, since the optical/IR counterparts to neutron stars and other compact objects are expected to be very faint \citep[for the details on this approach see][]{kaplan04}. Those SNR coincident \chic\ sources with a random chance of association $>0.01$ in the USNO B1 and GLIMPSE catalogs are listed in Table~\ref{Tab10}. (The chance of a random association between one of these 16 \chic\ sources and a 2MASS catalog source is $<0.01$.)

Of the 16 \chic\ sources inside SNRs, only 5 have no optical or infrared counterparts. These 5 X-ray sources are all very faint, with $<8$ X-ray counts detected for each in the \chic\ \cxo\ observations. These are ChI J145732--5901\_2 in SNR G318.2+0.1 \citep{whiteoak96}, ChI J182435--1311\_2,3,4 in SNR G18.1--0.1 \citep{helfand06,brogan06} and ChI J184447--0305\_1 in SNR G29.3667+0.1000 \citep{helfand06}. \citet{bocchino01} has already reported on three X-ray sources within SNR G318.2+0.1 but not at the position of ChI J145732--5901\_2. Both SNR G18.1--0.1 and SNR G29.3667+0.1000 are newly discovered SNRs so little X-ray analysis has been done on these objects. Further investigation is required to determine if any of these 5 \chic\ sources are compact objects and if they are associated with the surrounding SNRs. 

There are 54 \chic\ sources coincident with 6 different H {\sc ii} regions that were detected in the \chic\ \cxo\ observations of 7 AGPS sources (see Table~\ref{Tab9}). Based on the results from X-ray observations of other H {\sc ii} regions \citep[for example see][]{broos07} it is possible that many of these 54 \chic\ sources could be pre-main sequence (PMS) stars, massive OB and WR stars, and CWBs. \citet{simpson04} also found optically obscured star clusters in 2MASS images within $1'$ of 3 of these AGPS sources; AX J144519--5949, AX J151005--5824, and AX J162208--5005, which supports a H {\sc ii} region and young and massive star interpretation for these \asca\ sources. Of these 54 \chic\ sources, 43 have optical and/or infrared counterparts supporting a possible stellar origin (see Tables~\ref{Tab2}, \ref{Tab5} and \ref{Tab7}). In most cases the chance of random association with a field source is low ($<0.01$). Those H {\sc ii} region coincident \chic\ sources for which the chance of random association with a 2MASS or GLIMPSE source is $>0.01$ are listed in Table~\ref{Tab11}.

The possible nature of these H {\sc ii} region coincident \chic\ sources could be further investigated by comparing their luminosities to those of PMS stars, massive OB and WR stars, and CWBs. A spectral analysis of the 54 \chic\ sources is extremely difficult given that they all have $\le32$ X-ray counts. However, the primary goal is to obtain a wide-band flux that can then be converted into a luminosity. Quantile analysis was therefore performed to obtain absorbed Mewe-Kaastra-Liedahl \citep[Mekal;][]{mewe85,mewe86,kaastra92,liedahl95} spectral interpolations of all the H {\sc ii} region coincident \chic\ sources with $\ge5$ X-ray counts. A Mekal model was chosen as thin thermal plasma emission is expected from hot X-ray emitting stars \citep[for example see][]{wolk05,sana06}. The absorbed Mekal spectral interpolations can be found in Table~\ref{Tab11}.

The luminosity of these \chic\ sources were calculated using kinematic distance estimates to the H {\sc ii} regions of which they are coincident. Kinematic distances calculated by \citet{russeil03} were used to calculate luminosities for the \chic\ sources coincident with G320.3--0.3, G333.6--0.2, and G59.5--0.2. (Note that there is a more distant kinematic distance estimate of 6.3 kpc to G59.5--0.2 calculated by \citet{kuchar94} but we have decided to use the more recent estimate from \citet{russeil03}.) The kinematic distances used in the luminosity calculations for G326.96+0.03 and G62.9+0.1 were obtained from \citet{mcclure01} and \citet{fich84}, respectively. The kinematic distance to the massive young stellar object G316.8112--00.0566 \citep{busfield06}, likely embedded within GAL 316.8--00.1, was used in the luminosity calculations for those \chic\ sources coincident with this H {\sc ii} region. (The distance from \citet{busfield06} agrees reasonably well with the near kinematic distance to GAL 316.8--00.1 calculated by \citet{caswell87} when revised for a modern Galactic center distance of 8.5 kpc.) Table~\ref{Tab11} lists the kinematic distance and corresponding absorbed and unabsorbed luminosities calculated for each H {\sc ii} region coincident \chic\ source.

The range of absorbed luminosities calculated for all the \chic\ sources coincident with the 6 H {\sc ii} regions span the range $30.6 \mathrm{~erg s}^{-1} <\mathrm{Log~}L_{x,abs}<32.4 \mathrm{~erg s}^{-1}$ ($0.3-8$ keV). This absorbed luminosity range is similar to that observed from the H {\sc ii} region M17 \citep[$29.3 \mathrm{~erg s}^{-1} <\mathrm{Log~}L_{x,abs}<32.8 \mathrm{~erg s}^{-1}$ ($0.5-8$ keV);][]{broos07}. With the exception of ChI J151005--5824\_8, the H {\sc ii} region coincident \chic\ sources in Table~\ref{Tab11} have unabsorbed luminosities between $L_{x,unab} \sim 10^{31} \mathrm{~to~} 10^{35}$ erg s$^{-1}$. The unabsorbed luminosities of these \chic\ sources cover the ranges of what has been observed from flaring PMS stars \citep[$L_{x,unab} \sim 10^{30} \mathrm{~to~} 10^{33}$ erg s$^{-1}$;][]{favata05,wolk05}, single and binary massive O type stars \citep[$L_{x,unab} \sim 10^{31} \mathrm{~to~} 10^{33}$ erg s$^{-1}$;][]{oskinova05,sana06}, WR stars \citep[$L_{x,unab} \sim 10^{31} \mathrm{~to~} 10^{34}$ erg s$^{-1}$;][]{oskinova05,mauerhan10}, and CWBs \citep[$L_{x,unab} \sim 10^{32} \mathrm{~to~} 10^{34}$ erg s$^{-1}$;][]{oskinova05,mauerhan10}. In fact, we were able to identify ChI J194310+2318\_5, which is within G59.5--0.2, as the O7V((f)) type star \object{HD 344784} \citep{walborn73} using the SIMBAD Astronomical Database. It is therefore likely that the AGPS sources AX J144519--5949, AX J151005--5824, AX J154905--5420, AX J162208--5005, AX J194310+2318, AX J194332+2323, and AX J195006+2628 are young and massive stars within H {\sc ii} regions. Deeper X-ray, radio and IR observations are required to determine the precise nature of the individual \chic\ sources. 

\subsection{Unidentified \chic\ Sources with Infrared Counterparts}

The X-ray and infrared population statistics performed in this section are just conducted using those \chic\ sources with $>20$ X-ray counts, therefore concentrating on the persistent populations that fall within the AGPS flux range ($F_{x} \sim 10^{-13} \mathrm{~to~} 10^{-11}$ \erg). However, discarding the \chic\ sources with $<20$ X-ray counts limits our analysis as we are excluding the populations of sources that exhibit long term variability or transient behavior. Sources with $<20$ X-ray counts will need to be investigated in future work using archival X-ray observations at different epochs.

\subsubsection{X-ray and Infrared Populations Statistics}

Using the X-ray and infrared properties of the unidentified \chic\ sources, it is possible to classify some of the sources detected in the AGPS into possible populations. We chose to focus on the IR counterparts as this waveband is less affected by interstellar extinction when compared to the optical band. This group of sources therefore makes up a larger subset of the unidentified \chic\ sources than those with optical counterparts (see Section 3.3). For those unidentified \chic\ sources observed with the \cxo\ HRC instrument (for which there is no spectral information), we generated fake spectra, using \texttt{XSpec} and the \texttt{CIAO} spectral fitting tool \texttt{Sherpa}. These spectra are based on the absorbed power-law fits reported in \citet{sugizaki01}, allowing the absorbed X-ray flux and median energy ($E_{50}$) of the unidentified \chic\ source in question to be calculated. These values are used in the statistical plots described below. 

To help identify possible distinct populations in the statistical plots, we have also included both archival sources \citep[those AGPS source identified by][or in the literature and so were not observed with \cxo\ as part of the \chic\ survey]{sugizaki01} and the previously identified \chic\ sources \citep[for example those investigated by][]{anderson11,anderson12}. Fake spectra were generated using \texttt{Sherpa} and \texttt{XSpec} for these archival sources, using spectral fits in the literature, to determine their absorbed X-ray fluxes and $E_{50}$ values for the energy ranges investigated. All archival AGPS sources are summarized and tabulated in Section 4.5 (see Table~\ref{Tab12}) and are individually described in Appendix A. 

The identified sources were divided into the following categories: AGN, CVs, CWBs, HMXBs, magnetars, massive stars and stars. The ``AGN'' category includes ChI J184741--0219\_3, which we identified by positional comparison with MAGPIS radio data (for the radio identification and further details on this source see Sections 3.4 and 4.3.15, respectively). The ``CWB'' category includes AX J163252-4746 and AX J184738-0156, which were identified in \citet[][listed as ChI J163252--4746\_2 and ChI J184738--0156\_1 in Table~\ref{Tab1}, respectively]{anderson11}, and \object[AX J1831.2-1008]{AX J183116--1008} and \object[AX J1832.1-0938]{AX J183206--0938}, which were identified in the XGPS \citep[][listed as ChI J183116-1008\_1 and ChI J183206-0938\_1 in Table~\ref{Tab1}, respectively]{motch10}. The ``HMXB'' category includes the archival AGPS sources that are supergiant HMXBs \citep{mcclintock06}, supergiant fast X-ray transients \citep[SFXTs;][]{sguera06}, and SyXBs \citep{masetti07}. The infrared and X-ray fluxes from magnetars are variable and correlated \citep{durant05} so the fluxes we used in the ``magnetar'' category are from infrared and X-ray observations that occurred close together in time. PSR J1622--4950, which was determined through the \chic\ \cxo\ observation to be the main contributor to AX J162246--4946 has also been included as an identified magnetar \citep[see][and Section 4.3.6 for further details]{anderson12}. The sources included in the ``massive star'' category are massive stars that are Wolf-Rayet (WR), luminous blue variable (LBV) stars, and massive O-type stars, which emit X-rays through instability-driven wind-shocks \citep{lucy80,lucy82} and possibly through colliding-winds in a CWB. These include \object[AX J1445.7-5931]{AX J144547-5931} and AX J144701-5919, which were identified by \citet[][listed as ChI J144547-5931\_1 and ChI J144701-5919\_1 in Table~\ref{Tab1}, respectively]{anderson11}. All other non-degenerate stars are in the ``star'' category, and most likely correspond to active stellar coronae or PMS stars.

\begin{figure*}[htp]
\begin{center}
\includegraphics[width=0.8\textwidth]{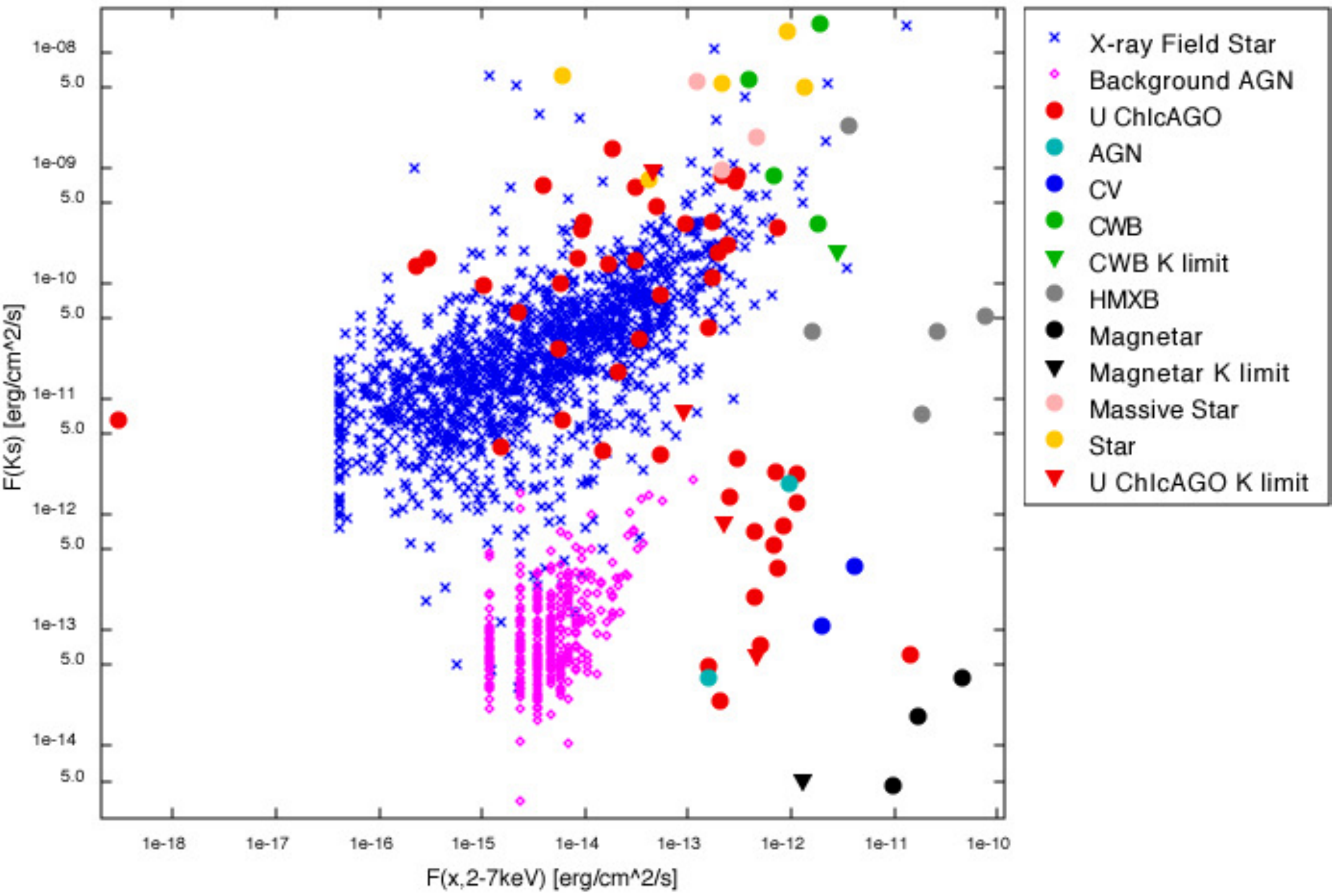}
\caption{Observed $K_{s}$-band flux ($F_{Ks}$) vs absorbed $2-7$ keV X-ray flux ($F_{x,2-7\mathrm{keV}}$) of unidentified ChIcAGO sources (``U ChIcAGO:'' red points) and identified \chic\ and archival sources. Triangular data points represent those sources that only have upper limits on their $K_{s}$-band fluxes. The \cxo\ Orion Ultradeep Project sources (blue crosses) and the XBootes Survey sources (magenta diamonds), which are predominantly PMS stars and background AGN, respectively, are also included. The lone red data point on the far left of the diagram (ChI J154557--5443\_3) should be treated as a X-ray flux upper limit as \cxo\ detected very few hard X-ray counts ($>2$ keV) from this source. This plot is based upon Figure 5 of \citet{gelfand07}.}
\label{Fig7}
\end{center}
\end{figure*}

We first investigated the relationship between the X-ray and infrared flux of the \chic\ sources by comparing these properties to those of known stars and AGN. Figure 5 of \citet{gelfand07} shows the X-ray versus $K_{s}$-band flux of sources from the \cxo\ Orion Ultradeep Project \citep[COUP;][]{getman05} and XBootes Survey \citep{jannuzi04,kenter05}, which are stars (predominantly in the PMS) and AGN, respectively. In Figure~\ref{Fig7} we create a similar plot that includes the unidentified \chic\ sources (U ChIcAGO; red data points) along with the COUP stars (blue crosses) and the XBootes Survey AGN (magenta diamonds) \citep[for further details on the data from these surveys see][and references therein]{gelfand07}. The X-ray flux ($F_{x,2-7keV}$) is over the $2.0-7.0$ keV energy range and the $K_{s}$-band flux ($F_{Ks} = \lambda F_{\lambda,Ks}$) is derived from $F_{\lambda,Ks}$ (\erg $\mu$m$^{-1}$) where the effective wavelength is $\lambda=2.159$ $\mu$m. The archival sources and the identified \chic\ sources with $K$-band counterparts have also been included in Figure~\ref{Fig7} in order to further distinguish between possible X-ray populations. 

Two groups and two outliers are apparent in Figure~\ref{Fig7} among the unidentified \chic\ sources: group 1, those coincident with the COUP stars, and group 2 that sits to the right of the AGN detected in the XBootes survey. The two outliers are ChI J154557-5443\_3, on the very left of this Figure with a hard X-ray flux limit of $F_{x,2-7keV} \lesssim 3 \times 10^{-19}$ \erg,\footnote{The majority of X-ray counts that \cxo\ detected from ChI J154557-5443\_3 were soft ($<2$ keV). This hard flux is therefore an upper-limit.} and the source on the bottom right, ChI J153818-5541\_1, with an X-ray flux of $F_{x,2-7keV} \sim 1 \times 10^{-11}$ \erg, located near the identified magnetars. 

Group 1 is distributed similarly to the COUP stars, following a ``track'' indicating an increase in $K$-band flux with X-ray flux. This comparison demonstrates that a large number of the \chic\ stellar population could be PMS stars. While the overall Galactic X-ray population is dominated by field stars, we would expect that the ChIcAGO survey (and the AGPS) is bias towards PMS stars as such objects are brighter and harder X-ray emitters \citep[for example see][]{wolk05}. The identified \chic\ and archival sources that have been categorized as massive stars (light pink dots) and CWBs (green data points) congregate near the top right of group 1, beyond most of the COUP stars. These massive stars and CWBs are composed of massive late-type WR stars in the nitrogen sequence that are hydrogen rich \citep[WNH:][]{smith08}, or their massive O star progenitors \citep[Of:][]{crowther95}, all of which are expected to be bright in the infrared and potentially harder in X-rays than other types of X-ray emitting stars (yellow data points). \citep[See][for further details on these WR and massive O stars.]{anderson11} It is therefore possible that the unidentified \chic\ sources located near these massive stars and CWBs in Figure~\ref{Fig7} could be similar objects. It should also be noted that the identified AGPS HMXBs sit to the right of group 1, with no unidentified \chic\ sources near their positions to draw comparison with.

Group 2 sits at a similar $K_{s}$-band flux but higher X-ray flux than the AGN from the XBootes survey. The identified \chic\ and archival AGN are coincident with the unidentified \chic\ sources in group 2, indicating that at least part of this group could also be AGN. Such AGN would have to be very X-ray bright in order to be detected through the high foreground column density in the Galactic plane. The identified archival CVs sit adjacent to group 2, at a slightly higher X-ray flux, indicating another possible population identification. 

The unidentified \chic\ source ChI J154557-5443\_3, has a similar $K$-band flux to the COUP stars but has an extremely faint hard X-ray flux. No identified \chic\ or archival sources are located near ChI J154557-5443\_3 in Figure~\ref{Fig7} that suggest a clue to its nature. ChI J154557-5443\_3 is discussed further in Section 4.3.4.

The unidentified \chic\ source that sits at the bottom right of Figure~\ref{Fig7}, ChI J153818-5541\_1, is very faint in the near-infrared but bright in high energy X-rays. It is therefore very similar in both X-ray and $K_{s}$-band flux to the identified archival magnetars, suggesting a similar identification that warrants further investigation. ChI J153818-5541\_1 is further discussed in Section 4.3.3. 

\begin{figure*}[htp]
\begin{center}
\includegraphics[width=0.8\textwidth]{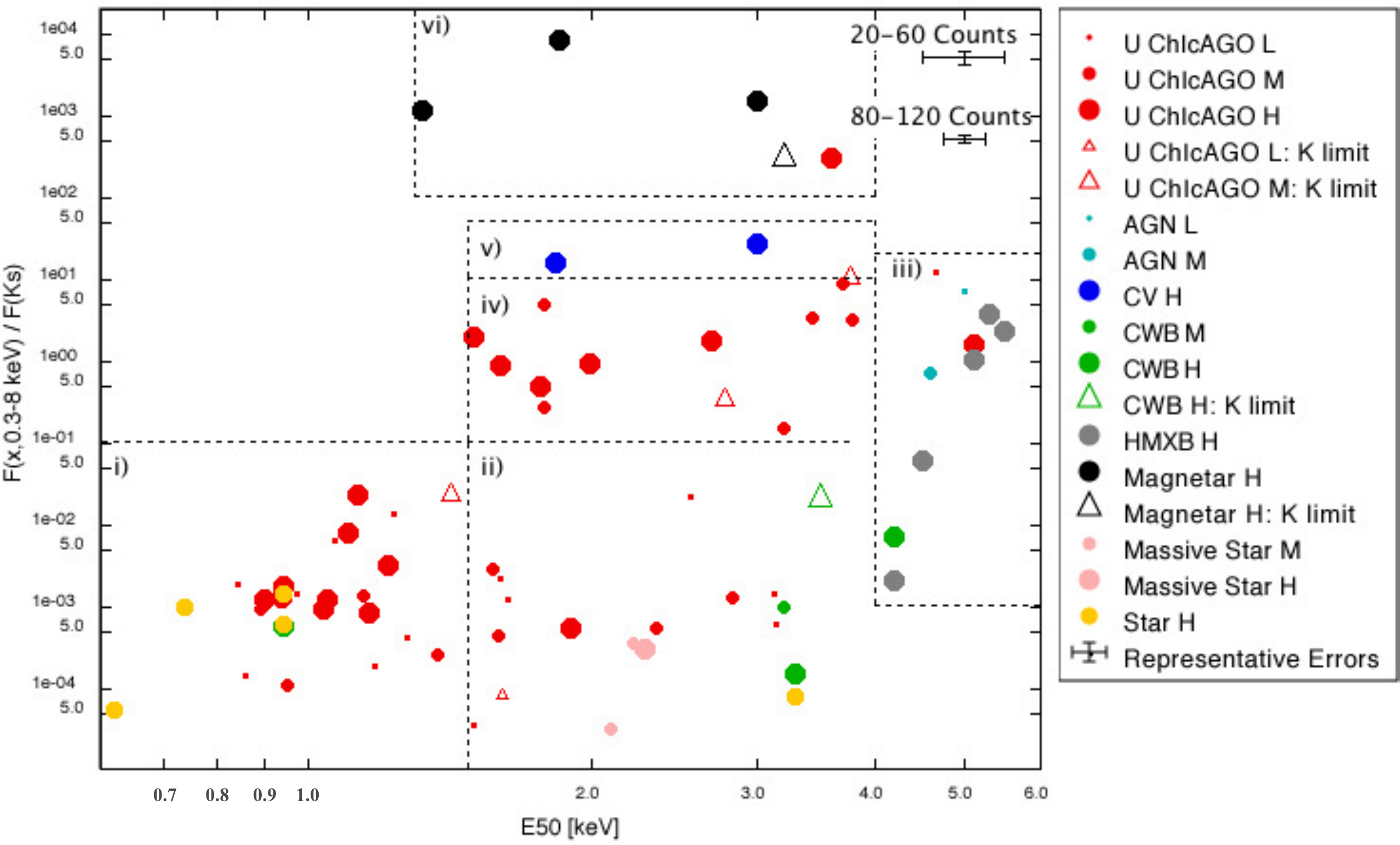}
\caption{Observed X-ray-to-$K_{s}$-band flux ratio ($F_{x,0.3-8\mathrm{keV}}/F_{Ks}$) vs median energy ($E_{50}$ keV) of unidentified \chic\ sources and the identified \chic\ and AGPS sources. The different sizes of data points represent whether a given source has a low count-rate (L as listed in the legend), medium count-rate (M) or a high count-rate (H). The unidentified \chic\ sources (``U ChIcAGO'') are represented by red data points and the identified \chic\ and archival sources are in other colors. The triangles represent those sources that only have upper limits on their $K_{s}$-band fluxes, implying lower limits on their $F_{x}/F_{Ks}$ ratios. The representative error bars demonstrate the sizes of errors expected from a $20-60$ count X-ray source or a $80-120$ count X-ray source. This plot has been divided into 6 regions, indicated by the dashed lines and Roman numerals, in order to further explore the source populations (see Section 4.2).}
\label{Fig8}
\end{center}
\end{figure*}

In order to further identify the possible populations that make up the above described groups, we created a statistical plot that also takes into account the hardness of each of the unidentified \chic\ sources. Figure~\ref{Fig8} plots the X-ray-to-$K_{s}$-band flux ratio ($F_{x,0.3-8.0 \mathrm{keV}} / F_{Ks}$) versus the median energy ($E_{50}$ keV), in the $0.3-8.0$ keV energy band, of each unidentified \chic\ source (``U \chic;'' red data points). We further separated the \chic\ sources into the categories ``low'', ``medium'' and ``high'', based on their ACIS-S and HRC-I X-ray count rates, using symbol sizes to distinguish these categories. The smallest data points (low) have count rates $<15$ counts s$^{-1}$, the medium sized data points (medium) have count rates between 15 and 40 counts s$^{-1}$, and the largest data points (high) have count rates $>40$ counts s$^{-1}$. The identified \chic\ and archival sources, also divided into categories based on the X-ray count-rate expected from a \cxo\ ACIS-S observation, have also been included in Figure~\ref{Fig8}. As the X-ray fluxes and median energy errors are fundamentally based on the number of X-ray counts detected in the \cxo\ observations, we show two representative error bars for the low count \chic\ sources ($20-60$ counts) and the high count \chic\ sources ($80-120$ counts). The error associated with the $K_{s}$-band flux is greater for PANIC magnitudes than for 2MASS magnitudes. We therefore adopt the PANIC $K_{s}$-band flux errors so that the vertical error bars represent the maximum possible error in the $F_{x}/F_{Ks}$ ratio for low- and high-count unidentified \chic\ sources.

In Figure~\ref{Fig8}, group 1 plus the source ChI J154557-5443\_3 from Figure~\ref{Fig7} have flux ratios $F_{x}/ F_{Ks} < 0.1$ but have a wide range of median energies. Group 2 is harder than group 1, with $E_{50} > 1.5$ keV but with a flux ratio $0.1 < F_{x} / F_{Ks} < 20$. The unidentified \chic\ source ChI J153818-5541\_1, from Figure~\ref{Fig7} is quite hard ($E_{50} > 3$ keV), with $F_{x} / F_{Ks} \approx 300$. Using Figure~\ref{Fig7} as a guide, as well as the relative positions between the unidentified \chic\ sources and the identified \chic\ and archival sources in Figure~\ref{Fig8}, we divided the groups and isolated sources from Figure~\ref{Fig7} into six different population regions. These regions are marked by dashed lines and labeled with Roman numerals in Figure~\ref{Fig8}. The region boundaries in this plot are based on the observed X-ray properties and $K_{s}$ band flux. If these values were extinction and absorption corrected, both the $F_{x} / F_{Ks}$ and $E_{50}$ region boundaries would lower.

Region i ($E_{50} < 1.5$ keV and $F_{x} / F_{Ks} < 0.1$) in Figure~\ref{Fig8} contains unidentified \chic\ sources with low, medium and high count rates. The majority of the low count rate unidentified \chic\ sources included in this Figure fall into this region. These low count rate sources could be either very nearby objects that were only detected due to their close proximity to the Solar System or they are simply a more distant version of the medium and high count rate sources in Region i. All the unidentified \chic\ sources in this region are unlikely to be extragalactic given their softer X-ray spectra and bright $K_{s}$-band counterparts compared to the XBootes AGN in Figure~\ref{Fig7}. Within this region are three of the archival X-ray stars (yellow data points), of which there is an RS CVn star, a PMS star and a multiple system. The unidentified \chic\ sources in Region i are therefore likely to be soft X-ray stars with active stellar coronae or PMS stars, similar to the three archival stars, but at a variety of distances in the Galaxy.

The sources in Region ii ($E_{50} > 1.5$ keV and $F_{x} / F_{Ks} < 0.1$) of Figure~\ref{Fig8} have similar $F_{x}/F_{Ks}$ ratios, but slightly harder X-ray emission, to those in Region i. This region also has fewer unidentified \chic\ sources than Region i, implying that it may contain a slightly rarer X-ray source population. The majority of the unidentified \chic\ sources in Region ii have low or medium count-rates, with only one in the high count rate category. The identified \chic\ and archival sources in this region are CWBs (green data points) and massive stars (pink dots), all of which sit at the top of the stellar track in Figure~\ref{Fig7}. We defined $E_{50} > 1.5$ keV as the lower energy cutoff of Region ii by looking at the closest unidentified \chic\ sources to the CWBs and massive stars in Figure~\ref{Fig7}. The CWBs and massive stars are WNH and Of stars, which produce X-rays through instability-driven wind-shocks but can also, in the case of CWBs, produce hard X-rays due to colliding-winds \citep[for example see][]{anderson11}. It is therefore likely that many of the unidentified \chic\ sources in Region ii are WR and Of stars, some of which may also be CWBs.

Region iii ($E_{50} > 4.0$ keV and $1 \times 10^{-3} < F_{x} / F_{Ks} < 20$) in Figure~\ref{Fig8} encompasses the archival HMXBs (grey dots) and the archival and identified \chic\ AGN (cyan dots). The only two unidentified \chic\ sources in this region are ChI J170017-4220\_1 (high) and ChI J172550-3533\_1 (low), and are therefore quite hard X-ray sources, making HMXB or AGN identifications a strong possibility. 

Region iv ($1.5 < E_{50} < 4.0$ keV and $0.1 < F_{x} / F_{Ks} < 10$) in Figure~\ref{Fig8} contains 12 medium and high count rate unidentified \chic\ sources. However, there are no identified \chic\ or archival sources in this region that could indicate any likely source populations. The best clue comes from Figure~\ref{Fig7}, which shows that these unidentified sources are in the same region of this plot as the identified \chic\ and archival AGN. In the Galactic plane, the \ns\ relation of X-ray sources in the $2.0-10.0$ keV energy range \citep[see Figure 15 of][]{hands04}, demonstrates that within the X-ray flux range $1 \times 10^{-13} < F_{x} < 2 \times 10^{-12}$ of Region iv, between 0.06 and 8 extragalactic sources are expected per square degree. These number densities are consistent with more recent \ns\ modeling conducted by \citet{mateos08} who used 1129 \xmm\ observations at $|b| > 20^{\circ}$ to demonstrate that sources in the $2-10$ keV energy range, at high Galactic latitudes, agree with AGN models to better than 10\%. As the \chic\ sources in Region iv have a number density $< 8$ deg$^{-2}$ it is not unreasonable to speculate that the many of the unidentified \chic\ sources in this region could be AGN. However, we compared the $N_{H}$ values of the Region iv \chic\ sources calculated from the power law quantile analysis and spectral fits in Table~\ref{Tab3} to the Galactic column densities in their direction from surveys conducted by \citet{kalberla05} and \citet{dickey90}.\footnote{The Galactic column densities were obtained using the online HEASARC calculator http://heasarc.gsfc.nasa.gov/cgi-bin/Tools/w3nh/w3nh.pl} In all by four cases the \chic\ sources have $N_{H}$ values an order of a magnitude lower than the Galactic $N_{H}$, suggesting a possible Galactic origin. The \chic\ sources ChI J181116--1828\_2, ChI J181213--1842\_7, ChI J190749+0803\_1, and ChI J194152+2251\_2 have $N_{H}$ values of the same order as the Galactic column density \citep[$N_{H} = 1.2 \times 10^{22}, 1.3 \times 10^{22}, 1.5\times10^{22}, \rm{~and~} 1.1\times10^{22} \rm{~cm}^{-2}$, respectively,][]{kalberla05}, which is more indicative of an extragalacitc origin and therefore an AGN identification. Further investigation is required to confirm the nature of the Region iv population. 

Region v ($1.5 < E_{50} < 4.0$ keV and $10 <  F_{x} / F_{Ks} < 50$) of Figure~\ref{Fig8} encompasses the two identified archival CVs and the lower $F_{x} / F_{Ks}$ ratio limit of unidentified \chic\ source ChI J181852-1559\_2. As its flux ratio is only a lower limit (corresponding to an upper-limit on the $K$-band flux), it is possible that ChI J181852-1559\_2 may instead be a member of Region vi (described below).

Region vi ($1.3 < E_{50} < 4.0$ keV and $F_{x} / F_{Ks} > 1 \times 10^{2}$) contains four identified archival magnetars. The only unidentified \chic\ source that sits within this region is ChI J153818-5541\_1. Based on its proximity to these magnetars, it is possible that ChI J153818-5541\_1 could also be a magnetar, however, its X-ray emission is much harder in comparison. Regardless, the position of ChI J153818-5541\_1 in Figure~\ref{Fig8} indicates that this source is definitely worthy of further study.

\subsubsection{Infrared Population Statistics}

In order to further refine the possible populations within Figure~\ref{Fig8}, we investigated the infrared colors of the unidentified \chic\ sources, once again drawing comparisons with the identified \chic\ and archival sources. As mentioned above, there is strong evidence in Figure~\ref{Fig8} that some of the unidentified \chic\ sources in Region ii could be massive stars such as WR and Of stars, and perhaps even CWBs. It is also possible that many of the \chic\ sources, particularly in Region i, are PMS stars given their similar X-ray-to-infrared flux ratio to the COUP stars in Figure~\ref{Fig8}. However, for the purposes of this paper we have chosen to concentrate on the selection criteria for the hard X-ray emitting massive stars. We therefore leave the investigation of the PMS star population in the \chic\ survey for future work \citep[for infrared selection criteria for PMS stars see][]{favata05,maercker05,maercker06}.

\begin{figure*}[htp]
\begin{center}
\subfigure{\includegraphics[width=0.8\textwidth]{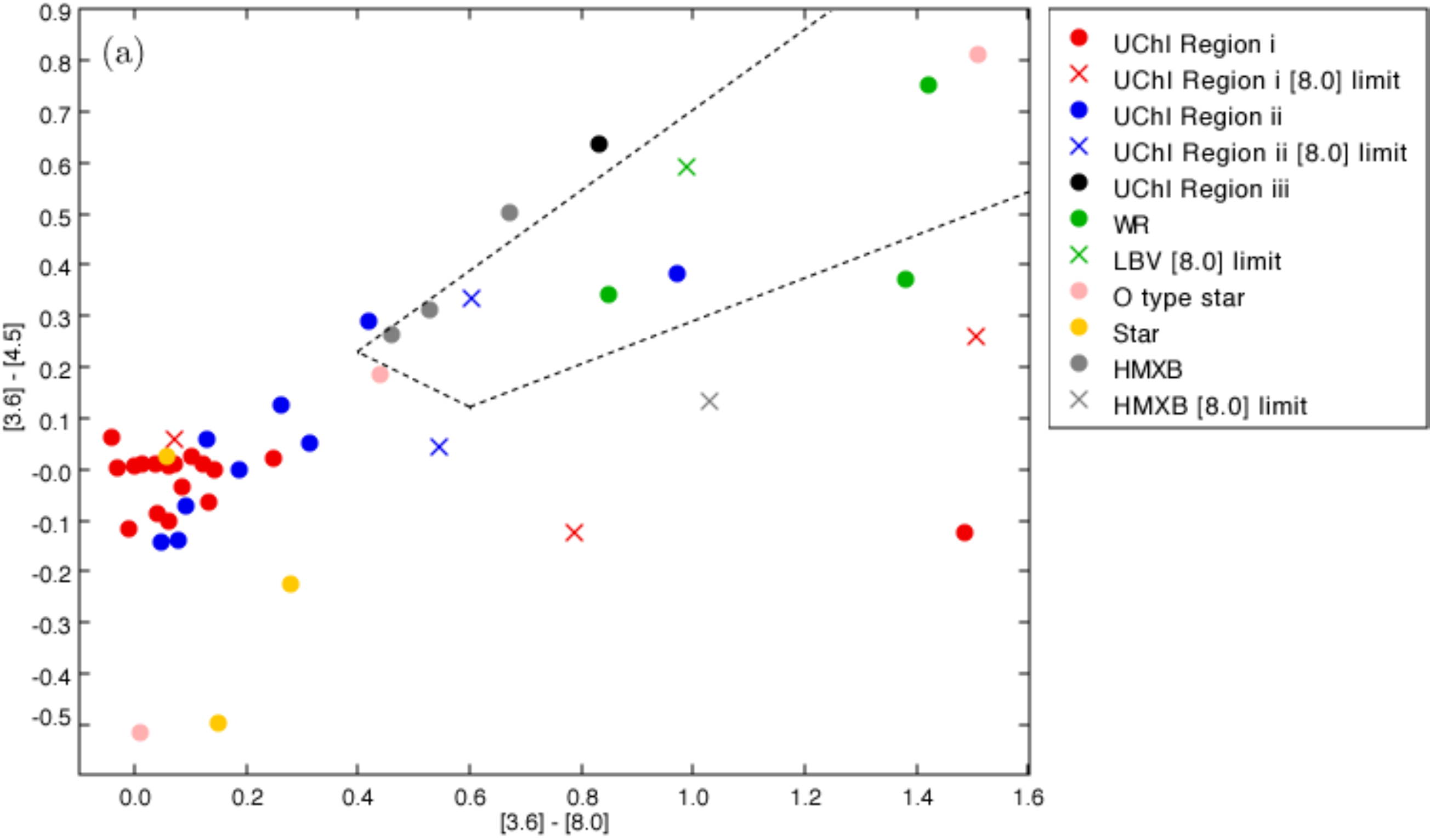}\label{Fig9a}}
\subfigure{\includegraphics[width=0.8\textwidth]{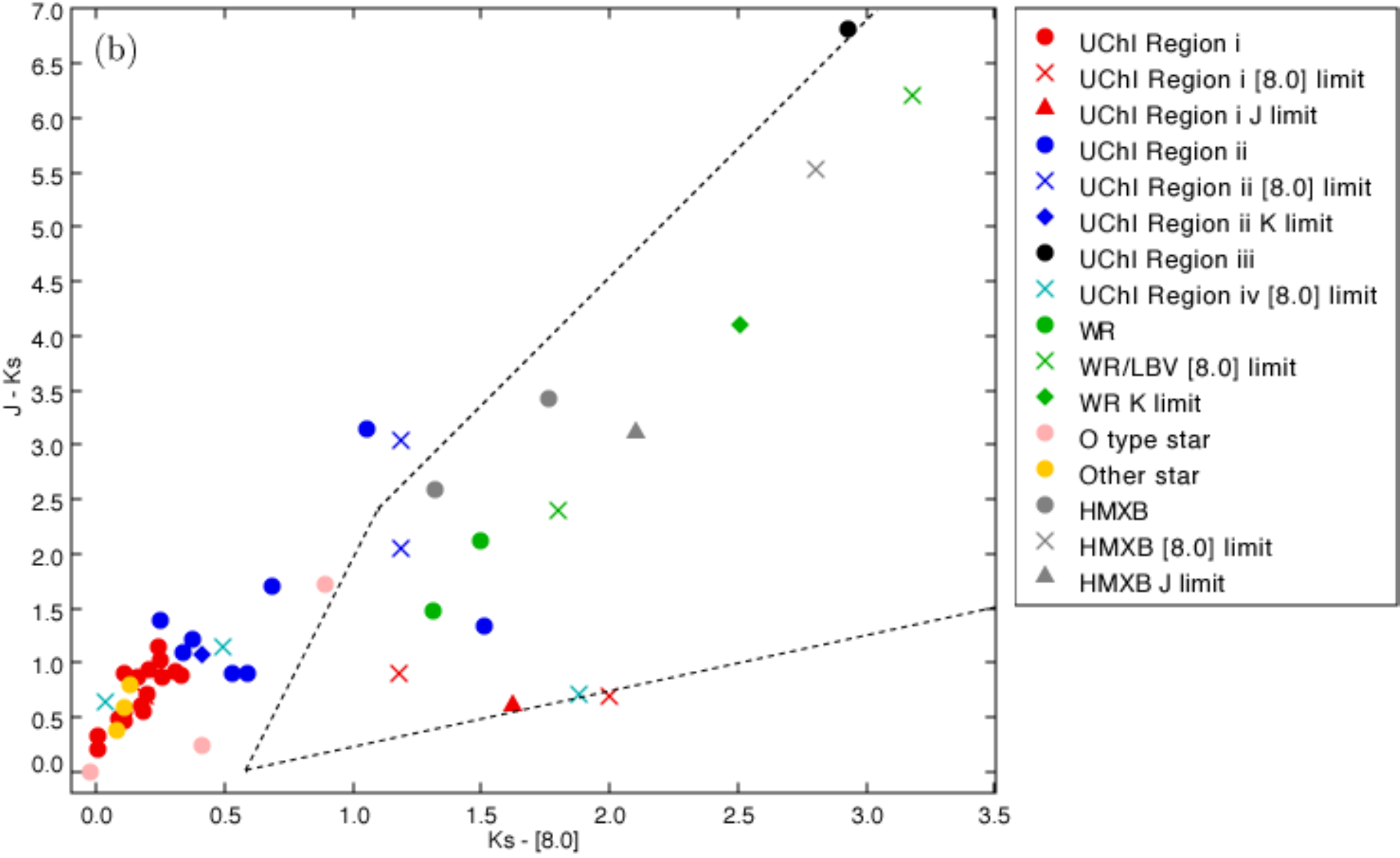}\label{Fig9b}}
\caption{Infrared and near-infrared color-color diagrams illustrating the selection criteria of \citet{hadfield07} and \citet{mauerhan11} for WR stars. (a) 3.6 $\mu$m $-~4.5$ $\mu$m vs 3.6 $\mu$m $-~8.0$ $\mu$m. (b) $J - K_{s}$ vs $K_{s} - 8.0$ $\mu$m. In both plots the dashed lines indicated the WR color space used by \citet{mauerhan11}. The unidentified \chic\ sources (``UChI'') are represented by red, blue and black for Regions i, ii, and iii, respectively. The identified \chic\ and archival AGPS sources that are WRs and LBVs (green), massive O-type stars (pink), all other non-degenerate stars (yellow) and HMXBs (grey) are also represented. The dots represent those sources that have accurate photometric information in the magnitude bands plotted. The crosses indicate those source for which we have used their GLIMPSE 5.8 $\mu$m magnitude in the absence of a GLIMPSE 8.0 $\mu$m detection. The Region iv sources in Figure~\ref{Fig9b} (cyan crosses) were not detected at 5.8 or 8.0 $\mu$m, so their 4.5 $\mu$m magnitudes were used instead. The triangles mark sources for which the $J$ magnitude is a lower-limit. The diamonds mark sources for which the $K_{s}$ magnitude is a lower-limit.}
\label{Fig9}
\end{center}
\end{figure*}

\citet{hadfield07} created a selection criterion for WR stars using GLIMPSE and 2MASS magnitudes, which was further refined by \citet{mauerhan11}. Figures~\ref{Fig9a} and \ref{Fig9b} are recreations of the \citet{hadfield07} [3.6]-[4.5] vs [3.6]-[8.0] and $J-K_{s}$ vs $K_{s} -$ [8.0] color-color plots, showing the unidentified \chic\ sources. (The numbers in brackets correspond to the effective wavelength in $\mu$m of the GLIMPSE magnitude bands.) The dashed lines in Figures~\ref{Fig9a} and \ref{Fig9b} indicate the color space used by \citet{mauerhan11} to select WR candidates. These color spaces indicate where WR stars are expected to fall, in comparison to field stars, due to the infrared excess of WR stars resulting from free-free emission generated in their strong, dense stellar winds \citep[see][and references therein]{mauerhan11}. 

In Figures~\ref{Fig9a} and \ref{Fig9b}, the unidentified \chic\ sources (UChI) have been separated into different colored symbols based on whether they are in Region i (red), Region ii (blue), or Region iii (black). In both these Figures, the majority of unidentified \chic\ sources, particularly from Region i, do not have an infrared excess so cluster together near the origin where \citet{hadfield07} state the general stellar locus is located. These unidentified \chic\ sources are therefore unlikely to have strong stellar winds, so are more likely to have active stellar coronae. There are, however, several Region ii sources, and three Region i sources, that have more unusual colors, falling within, very close or below the indicated WR color spaces. (It should be noted that many of these unusually colored unidentified \chic\ sources only have magnitude lower limits in one or more of the filter bands, making their positions in these color-color diagrams uncertain. See Figures~\ref{Fig9a} and \ref{Fig9b} for details on these magnitude limited sources.)

Rather than keeping the original identified source categories we have instead separated the identified \chic\ and archival massive stars and CWBs into categories based on their dominant stellar components. These categories are WR and LBVs (green data points), massive O type stars (pink data points), and all other non-degenerate stars (yellow data points). The HMXBs are still indicated by grey data points. 

The identified WR stars fall within one or both of the \citet{mauerhan11} WR color spaces in Figures~\ref{Fig9a} and \ref{Fig9b}. Based on their positions within one or both WR color spaces, it is possible that the unidentified Region ii \chic\ sources ChI J181915--1601\_2, ChI J180857--2004\_2 and ChI J183345--0828\_1 may also be WR stars. In fact, using this color space technique \citet{hadfield07} identified the 2MASS counterpart of ChI J181915--1601\_2, \object[2MASS J18192219-1603123]{2MASS J18192219−-1603123}, as a WR star and further classified it as a WN7o star using spectroscopy. There are also three Region i unidentified \chic\ sources, ChI J154557--5443\_1, ChI J154557--5443\_3, and ChI J172550--3533\_2, whose magnitude limits fall within the WR color space of Figure~\ref{Fig9b} and below the WR color space of Figure~\ref{Fig9a}, which may also be worth further investigation.

The identified \chic\ source AX~J144547-5931 \citep[an Of type star, which is also shown as a pink dot in Figures~\ref{Fig7} and \ref{Fig8}, see][]{anderson11} sits to the left of both WR color spaces in Figures~\ref{Fig9a} and \ref{Fig9b}. The unidentified Region ii \chic\ sources ChI J182435--1311\_1, ChI J182651--1206\_4, ChI J183356--0822\_3, and ChI J184652--0240\_1 are also located in this same vicinity as AX~J144547--5931 in one or both Figures (within $0.2<[3.6]-[8.0]<0.5$ in Figure~\ref{Fig9a} and within $0.5<K_{s}-[8.0]<1.2$ in Figure~\ref{Fig9b}). While these four unidentified \chic\ sources have unusual infrared colors, they are not as extreme as those of typical WR stars. Given their proximity in Figure~\ref{Fig9} to the Of star AX~J144547-5931, it is possible that these unidentified \chic\ sources could be similar massive O-type stars. The remaining Region ii \chic\ sources lie within the general stellar locus of Figures~\ref{Fig9a} and \ref{Fig9b}, so could be massive stars, active stellar coronae, or PMS stars.

 The archival HMXBs also have unusual infrared colors as they fall within or above the top edge of the WR color spaces in Figures~\ref{Fig9a} and \ref{Fig9b}. These infrared colors may be due to intrinsic absorption \citep[for example see][]{rodriguez06}. Several unidentified Region ii \chic\ sources also lie close to the HMXBs in these Figures, but as they have much softer median X-ray energies (see Figure~\ref{Fig8}), such an identification is unlikely. ChI J170017--4220\_1 is the only Region iii source with sufficient infrared magnitude information to be displayed in Figures~\ref{Fig9a} and \ref{Fig9b}. This unidentified \chic\ source falls within the vicinity of the HMXBs in Figure~\ref{Fig8} but has much more extreme infrared colors, and lies above the WR color space in both color-color plots. Further investigation is required to confirm the X-ray binary nature of ChI J170017--4220\_1.

None of the 11 unidentified Region iv \chic\ sources have been detected at 8 $\mu$m, with only three, ChI J163751--4656\_1, ChI J165420--4337\_1, and ChI J190818+0745\_1, detected with GLIMPSE at 3.6 and 4.5 $\mu$m, all of which are likely to be Galactic in origin (see Section 4.2.1). In Figure~\ref{Fig9b} we have plotted these three unidentified \chic\ sources using their 4.5 $\mu$m magnitude in place of 8 $\mu$m magnitude. Only ChI J165420-4337\_1 falls within the WR color space. Given the extreme uncertainties associated with the $K_{s} - [8.0]$ color of ChI J165420-4337\_1, no conclusions can be drawn about its possible nature.

\subsection{Further Discussion of Interesting \chic\ and AGPS Sources}

We have flagged several of the unidentified \chic\ sources as interesting and worthy of future follow-up based on their X-ray and infrared properties explored in Section 4.2. These population statistics have also allowed us to make some tentative identifications. The details of these sources are described below. There are several unidentified \chic\ sources that are not detailed in this Section but are listed, along with their tentative identification based on the population statistics, in Table~\ref{Tab12}.

\subsubsection{ChI J144519--5949\_2} 

ChI J144519--5949\_2 is one of the six \chic\ sources detected within the H {\sc ii} region GAL 316.8--00.1 (see Table~\ref{Tab9}). As it does not have a GLIMPSE counterpart, it is not included in Figures~\ref{Fig9a} and \ref{Fig9b}. However, its 2MASS counterpart, 2MASS~14452143-5949251, has unusual colors and falls within the WR color space of the $J-H$ vs $H-K_{s}$ plot in Figure 1 of \citet{mauerhan11}. We therefore tentatively identify ChI J144519--5949\_2 as a candidate WR star. Given the location of ChI J144519--5949\_2 in Region ii of Figure~\ref{Fig8}, particularly near two identified CWBs, it is very likely that this star has very strong winds that are generating X-rays through instability-driven wind-shocks. This object could even possibly be a CWB \citep[for example see][]{anderson11}.

\subsubsection{ChI J150436-5824\_1} 

Several unidentified AGPS sources have been investigated by \citet{degenaar12} using the X-ray telescope (XRT) on the \textit{Swift} satellite \citep{gehrels04}. \citet{degenaar12} suggest that ChI J150436-5824\_1 (which they publish under the AGPS name \object[AX J1504.6-5824]{AX J1504.6--5824}) could possibly be associated with a main sequence star that falls within the 90\% XRT error circle. However, the \chic\ \cxo\ observation of ChI J150436-5824\_1 demonstrates that there is no cataloged counterpart within the $<1''$ position error circle of this X-ray source, arguing against this tentative identification.

\subsubsection{ChI J153818--5541\_1} 

In Figure~\ref{Fig7}, ChI J153818--5541\_1 has properties that are clearly not consistent with the stellar or AGN populations. It sits near the bottom right-hand corner of Figure~\ref{Fig7}, having similar X-ray and infrared fluxes to the archival magnetars (black data points). ChI J153818--5541\_1 also falls within Region vi in Figure~\ref{Fig8} along with these same magnetars, further indicating a possible magnetar nature. However, during the writing of this paper, recent studies with \textit{Swift} have indicated that ChI J153818--5541\_1 (published under its AGPS name \object[AX J1538.3-5541]{AX J1538.3--5541}) may in fact be a low mass X-ray binary \citep[LMXB;][]{degenaar12}. If this is the case, then X-ray and infrared statistical analysis is not a foolproof diagnostic. That being said, this same statistical analysis has allowed us to show that interesting and unusual Galactic X-ray sources fall within Region vi of Figure~\ref{Fig8}. (It should be noted that a magnetar identification for this source has not been completely ruled out.)

\subsubsection{ChI J154557--5443\_3} 

ChI J154557--5443\_3 is a soft X-ray source as demonstrated by its isolated position in Figure~\ref{Fig7}. Only 2 X-ray counts, from a total of 22 detected with \cxo, had an energy $>2$ keV, but otherwise the source's median energy ($E_{50} = 0.8$ keV) is consistent with the other sources in Region i of Figure~\ref{Fig8}. The X-ray-to-optical flux ratio for this source is log[$F_{(x,0.3-8\mathrm{keV})}/F_{R}]=-2.5$ (where $F_{R}=\lambda F_{R,\lambda}$ using the $R$ magnitude band), which is consistent with what is expected from stars and normal galaxies \citep{mainieri02}. An unusual property of this source is its position in WR color space in Figure~\ref{Fig9b}, but this may result from using its $5.8$~$\mu$m GLIMPSE magnitude in place of a lacking $8.0~\mu$m magnitude. An optical or infrared spectrum would likely reveal the identity of this source. ChI J154557--5443\_3 is therefore unlikely to be related to \object[AX J1545.9-5443]{AX J154557--5443} as \asca\ only detected hard counts ($>2$ keV) from this AGPS source.

\subsubsection{ChI J162011--5002\_1} 

\citet{degenaar12} tentatively identified ChI J162011--5002\_1 (published under its AGPS name AX J1620.1--5002) as a candidate accreting magnetized white dwarf based on its hard X-ray spectrum described by a flat power law, and lack of any cataloged optical/infrared counterpart. The power-law index obtained using quantile spectral interpolation (see Table~\ref{Tab3}) agrees with the result from \citet{degenaar12} and is consistent with the spectra of magnetically accreting white dwarfs \citep[$\Gamma < 1$;][]{muno03}. However, our analysis resulted in a higher column density than the value calculated by \citet[][$N_{H} \lesssim 3 \times 10^{21}$ cm$^{-2}$]{degenaar12}. Deep imaging in the $H$-band with PANIC resulted in the detection of a faint counterpart ($H=17.01 \pm 0.14$). Further follow-up is required to confirm this classification. 

\subsubsection{AX J162246--4946} 

The newly discovered radio and X-ray emitting magnetar, PSR J1622--4950 \citep[also known as CXOU J162244.8--495054,][]{evans10,levin10}, was not cataloged by \chic\ MAP. This is because it lies $4'$ from the position of AX J162246--4946, and is therefore outside the $3'$ radius for which \chic\ MAP searches for X-ray point sources. However, PSR J1622--4950 was detected in this \chic\ \cxo\ observation and further investigation in \citet{anderson12} demonstrates that this magnetar may have contributed up to $75\%$ of the X-ray emission originally detected by \asca\ from AX J162246--4946. We therefore identify AX J162246--4946 as PSR J1622--4950, and this source has been included as an identified magnetar in Figures~\ref{Fig7} and \ref{Fig8}. (For further discussion on X-ray point sources detected beyond the $3'$ search radius surrounding the AGPS position see Appendix B.)

\subsubsection{AX J165951--4209} 

While \citet{sugizaki01} reported an absorbed flux of $F_{x}=4.04 \times 10^{-12}$ \erg\ from AX J165951--4209 in the $0.7-10.0$ keV band, this AGPS source was not detected in the \chic\ \cxo\ observation on 2008 June 21, with an upper limit on any X-ray flux of $F_{x} \sim 5 \times 10^{-14}$ \erg\ ($0.3-8$ keV). AX J165951--4209 was also not detected with the \textit{Swift} XRT on 2008 January 23 \citep{degenaar12} at a flux level of $F_{x} \sim 3 \times 10^{-13}$ \erg\ ($0.3-10$ keV). While variability of several orders of magnitude (in this case a factor of $\sim100$) is typical behavior of both Be X-ray binaries \citep[BeXs;][]{reig11} and black hole transients \citep{mcclintock06}, the hard power law index calculated from the \asca\ spectrum of this source \citep[$\Gamma \sim 0.47$;][]{sugizaki01} is quite hard. This power law index is harder than what is usually observed from black hole transients \citep{mcclintock06} but is consistent with spectra from BeXs \citep[for example see][]{haberl08}. We therefore suggest that AX J165951--4209 could be a transient BeX.

\subsubsection{ChI J170017--4220\_1} 

ChI J170017--4220\_1 (also known as \object[AX J1700.2-4220]{AX J1700.2--4220}) has long been assumed to be a HMXB \citep[see][]{liu06,bird07,krivonos07,bird10}. This is, however, unconfirmed and is partly based upon the assumption that the Be star \object{HD 153295} (2MASS J17002524-4219003) is the counterpart to ChI J170017--4220\_1 \citep{negueruela07}. The \chic\ \cxo\ observation shows that HD 153295 is actually the counterpart to ChI J170017--4220\_2 (see Table~\ref{Tab7}) and that ChI J170017--4220\_1 has no 2MASS counterpart. The main clue to the nature of ChI J170017--4220\_1 comes from \citet{markwardt10}, who detected a 54s X-ray pulse period and 44 day orbital period using archival \xmm\ and \textit{Rossi X-ray Timing Explorer} data. These period values are suggestive of a Be HMXB \citep{markwardt10}. 

We obtained PANIC NIR observations of ChI J170017--4220\_1 and identified its GLIMPSE counterpart, allowing this \chic\ source to be represented in Figures~\ref{Fig7}, \ref{Fig8}, and \ref{Fig9}. ChI J170017--4220\_1 is part of Group 2 in Figure~\ref{Fig7}, and is therefore consistent with an AGN, but this source is also situated in the same region as the archival HMXBs in Figure~\ref{Fig8}. Its placement on Figures~\ref{Fig9a} and \ref{Fig9b} demonstrates that it has very unusual infrared colors, even more extreme than the archival HMXBs, which could indicate strong winds or a large amount of circumstellar absorption. Further investigation is required to determine the true nature of ChI J170017--4220\_1. 

\subsubsection{ChI J172050--3710\_1} 

\citet{degenaar12} suggested that ChI J172050--3710\_1 (also known as \object[AX J1720.8-3710]{AX J1720.8--3710}) may be associated with the NIR source 2MASS J17205180-3710371, which has colors similar to a main sequence star. They also suggested that the low absorption inferred from a power-law X-ray fit indicates that ChI J172050--3710\_1 may be a foreground star. We agree that this 2MASS source is the likely counterpart to ChI J172050--3710\_1 based on the position obtained with the \chic\ \cxo\ observation (see Table~\ref{Tab2}). We also agree that the counterpart colors are unremarkable (see Figures~\ref{Fig9a} and \ref{Fig9b}), given that they are consistent with the general stellar locus depicted in Figure 1 of \citet{hadfield07}. Our quantile thermal bremsstrahlung spectral interpolation of ChI J172050--3710\_1 (see Table~\ref{Tab4}) also predicts a low value of $N_{H}$. This, combined with the unremarkable colors of 2MASS J17205180-3710371, and the fact that this \chic\ source is situated in Region i of Figure~\ref{Fig8} with the other soft X-ray emitting stars, supports the claim by \citet{degenaar12} that ChI J172050--3710\_1 is likely a foreground main sequence star. We therefore argue that the likely source of X-ray emission from ChI J172050--3710\_1 is generated in an active stellar corona.

\subsubsection{ChI J180857--2004\_2} 

ChI J180857--2004\_2 falls within the inner edge of the infrared dust bubble CN 148 \citep{churchwell07}. Such bubbles are formed by the stellar winds of young hot stars impacting the interstellar medium \citep{churchwell06}. The morphology of the dust cloud immediately surrounding ChI J180857--2004\_2 demonstrates a shell-type structure. This could indicate that its stellar winds are impacting the environment and generating a small secondary bubble within CN 148. If this is the case, then ChI J180857--2004\_2 could be a young massive star with strong stellar winds. This is already indicated by its position within the WR color spaces in Figures~\ref{Fig9a} and \ref{Fig9b}, and by the detection of hard X-ray emission with \cxo, indicated by its position within Region ii of Figure~\ref{Fig8}. ChI J180857--2004\_2 could therefore be a WR star, for which the X-ray emission is generated through instability-driven wind-shocks or in a CWB \citep[see][]{anderson11}.

\subsubsection{ChI J181116--1828\_5}

ChI J181116--1828\_5 is a faint point source detected in the \cxo\ observation of \object[AX J1811.2-1828]{AX J181116--1828} so it is unlikely to be the main contributor of X-ray emission to this AGPS source. This source is, however, quite hard, with all but one count having an energy $>2$ keV, and is coincident with a MAGPIS radio source with a core-lobe morphology (see Section 3.4). Given the morphology of its likely radio counterpart, along with the detection of predominately hard X-rays ($E_{50}=3.4$ keV), we tentatively identify ChI J181116-1828\_5 as an AGN.

\subsubsection{ChI J181852--1559\_2} 

During the writing of this paper, ChI J181852--1559\_2 (also known as \object[AX J1818.8-1559]{AX J1818.8--1559}) was proposed as a magnetar candidate through the analysis of several X-ray observations \citep{mereghetti12}. This proposed identification is encouraging as the lower limit of ChI J181852--1559\_2 in Figure~\ref{Fig8} is consistent with the archival magnetars in Region vi, demonstrating the usefulness of this statistical plot for identifying interesting sources. 

\subsubsection{ChI J181915--1601\_2} 

As mentioned in Section 4.2.2, \citet{hadfield07} identified \object[2MASS J18192219-1603123]{2MASS J18192219--1603123} (the counterpart to ChI J181915--1601\_2, see Table~\ref{Tab2}) as a WR star of type WN7o using the same selection criteria that we have adopted in Figures~\ref{Fig9a} and \ref{Fig9b}. The mechanism behind the production of X-ray emission is still unknown, but is likely caused by massive stellar winds as demonstrated by the position of ChI J181915--1601\_2 within Region ii of Figure~\ref{Fig8}. Further investigation is required to determine if this WR star is mainly producing X-ray emission through instability-driven wind-shocks, or if it is part of a CWB.

\subsubsection{ChI J183345--0828\_1} 

\citet{kargaltsev12} identified \object[2MASS J18334038-0828304]{2MASS J18334038--0828304} as the counterpart to ChI J183345--0828\_1 (also known as CXOU J183340.3--082830), but the nature of this X-ray source is still unknown. ChI J183345--0828\_1 falls within an extended X-ray source in SNR G23.5+0.1, which \citet{kargaltsev12} have tentatively identified as the PWN powered by \object[PSR B1830-08]{PSR B1830--08}. It is possible that both this PWN and ChI J183345--0828\_1 may have contributed to the X-ray emission originally detected by \asca\ from \object[AX J1833.7-0828]{AX J183345--0828}. While ChI J183345--0828\_1 is unlikely to be associated with the PWN, given the relative brightness of its IR counterpart, it does fall within the WR color space in Figure~\ref{Fig9b}. ChI J183345--0828\_1 could therefore be a massive star with strong stellar winds. 

\subsubsection{ChI J184741--0219\_3}

As mentioned in Section 3.4, ChI J184741--0219\_3 is coincident with a MAGPIS radio source that has a core-lobe morphology. Deep PANIC observations also allowed the detection of its infrared counterpart ($K=18.46$), which places it in Region iii of Figure~\ref{Fig8} (represented by the smallest cyan data point) along with the other archival AGN \object[AX J1830.6-1002]{AX J183039--1002} \citep{bassani09}, which is the middle sized cyan data point. (We obtained the NIR magnitudes $J>20.6 \pm 0.3$, $H=17.3 \pm 0.2$ and $K_{s}=14.3 \pm 0.2$ on 2007 July 29 for AX J183039--1002 using PANIC.) As ChI J184741--0219\_3 had by far the highest count-rate of all sources detected in the $3'$ \cxo\ region around the aim-point for this target (see Table~\ref{Tab1}), it is likely the main contributor to the X-ray emission detected from \object[AX J1847.6-0219]{AX J184741--0219} in the AGPS. 

ChI J184741--0219\_3 has a hard X-ray spectrum (see Table~\ref{Tab3}) where the resulting absorbed X-ray flux from the power-law spectral interpolation is $F_{x}=2.6 \pm 0.5 \times 10^{-13}$ \erg\ in the $0.3-8$ keV energy band. Using the same spectral interpolation parameters and the \asca\ count-rates \citep{sugizaki01}, \cxo\ PIMMS estimates that ChI J184741--0219\_3 had an absorbed X-ray flux of $F_{x} \approx 1 \times 10^{-12}$ \erg\ during the 1999 AGPS observation. \citet{degenaar12} also observed the field of ChI J184741--0219\_3 with the \textit{Swift} XRT in 2007 March but only obtained an absorbed X-ray flux upper-limit of $F_{x} < 1 \times 10^{-13}$ \erg\ ($0.3-10$ keV). These flux measurements suggest likely long-term variability from this source. We therefore tentatively identify ChI J184741--0219\_3 as an X-ray variable AGN.

\subsubsection{AX J185905+0333} 

\textit{Suzaku} observations have demonstrated that AX J185905+0333 is likely an X-ray luminous cluster of galaxies behind the Galactic Plane \citep{yamauchi11}. The extended nature of this source is the likely reason it was not detected in the short \chic\ \cxo\ observation.

\subsubsection{AX J191046+0917} 

AX J191046+0917 (also known as AX J1910.7+0917), while not detected in the \chic\ \cxo\ observation, has been detected intermittently in a small number of \asca, \xmm, and \textit{Integral} observations, allowing \citet{pavan11} to identify it as a likely HMXB candidate. Further X-ray observations are required to confirm such an identification and further refine its class.

\subsubsection{ChI J194939+2631\_1} 

During the writing of this paper, ChI J194939+2631\_1 (also known as AX J194939+2631) was identified as a CV by \citet{zolotukhin11} using the \chic\ \cxo\ observation and the Isaac Newton Telescope Photometric H$\alpha$ Survey of the northern Galactic plane \citep[IPHAS;][]{drew05}. Further follow-up is required to confirm their tentative intermediate polar (IP) classification.

\subsection{\chic\ Sources Identified with SIMBAD}

We were also able to identify four of the bright ($>20$ X-ray counts) \chic\ sources using the SIMBAD Astronomical Database. These four \chic\ sources are described below. We also discuss the secondary \chic\ source ChI J165420--4337\_2.

\subsubsection{ChI J144042--6001\_1} 

The position obtained with the \chic\ \cxo\ observation confirms the \citet{sugizaki01} identification of ChI J144042-6001\_1 (also known as \object[AX J1440.7-6001]{AX J144042--6001}) as the PMS star \object{HD 128696}. This source appears as an identified star (yellow data point) in group 1 of Figure~\ref{Fig7}, in Region i of Figure~\ref{Fig8}, and is consistent with the general stellar locus in Figures~\ref{Fig9a} and \ref{Fig9b}. 

\subsubsection{ChI J165420--4337\_2}

\citet{sugizaki01} identified \object[AX J1654.3-4337]{AX J165420--4337} as the K5 type star \object{HD 326426}. After observing the \asca\ position of AX J165420--4337 with \cxo\ HRC, we detected two \chic\ sources ChI J165420--4337\_1 (123 X-ray counts detected) and ChI J165420--4337\_2 (16 X-ray counts detected). As ChI J165420--4337\_1 is by far the brightest of the \chic\ sources detected in this \cxo\ observation, it is therefore likely to be the main source of X-ray emission originally detected in the AGPS. The fainter (secondary) \chic\ source in the field, ChI J165420--4337\_2, is in fact the X-ray counterpart to HD 326426 based on position comparisons using SIMBAD. Therefore the identification of AX J165420--4337 as HD 326426 by \citet{sugizaki01} is incorrect.

\subsubsection{ChI J172550--3533\_2} 

Using the position obtained with the \chic\ \cxo\ observation, we identified ChI J172550--3533\_2 as the dwarf Nova V478 Sco \citep{vogt82}. ChI J172550--3533\_2 has a soft X-ray spectrum, as indicated by its position within Region i of Figure~\ref{Fig8}, but has quite unusual infrared colors based on its position below the WR color-spaces in Figures~\ref{Fig9a} and \ref{Fig9b}.

\subsubsection{ChI J172642--3540\_1} 

Using SIMBAD, we identified ChI J172642--3540\_1 as the X-ray counterpart to the K2V type star CD-35 11565 \citep{torres06}. ChI J172642--3540\_1 is soft, like the other stars in Region i of Figure~\ref{Fig8}. Its position in Figures~\ref{Fig9a} and \ref{Fig9b} is also consistent with the general stellar population.

\subsubsection{ChI J194310+2318\_5} 

ChI J194310+2318\_5 is one of the 18 \chic\ sources detected within the H {\sc ii} region G59.5--0.2 (see Section 4.1 and Table~\ref{Tab9}). Using SIMBAD and the \cxo\ position for this source, we determined that ChI J194310+2318\_5 is the X-ray counterpart to the O7V((f)) type star \object{HD 344784} \citep{walborn73}. This source is quite soft (within Region i of Figure~\ref{Fig8}) and has unremarkable infrared colors (see Figures~\ref{Fig9a} and \ref{Fig9b}). The placement of ChI J194310+2318\_5 in the aforementioned Figures, combined with its stellar classification, make this \chic\ source compatible with an active stellar corona identification.

\subsection{Confirmed and Tentative Identifications of the AGPS Sources}

Table~\ref{Tab12} reproduces the original list of the 163 AGPS sources from \citet{sugizaki01}, but now including the corresponding \chic\ sources with $>20$ X-ray counts in column two. The confirmed or tentative identification of these \chic\ sources are listed in the third column, where the abbreviations are given in the Table caption. Those AGPS sources and corresponding \chic\ sources with confirmed identifications were either obtained from the literature, and are therefore called ``archival AGPS'' sources, or were obtained through the \chic\ survey's \cxo\ observations \citep[i.e. this paper and][]{anderson11,anderson12}. The tentative identifications were made through the multi-wavelength follow-up and population statistics conducted on the \chic\ sources in Section 4.2. The forth column gives the most common name for the AGPS and/or \chic\ source while column five lists the references from which the X-ray and infrared properties of a given source was obtained for use in the statistical plots (Figures~\ref{Fig7}, \ref{Fig8}~and \ref{Fig9}).

The tentative identifications of the \chic\ sources are based on the statistical plots in Section 4.2. For example, if a \chic\ source falls within Region i of Figure~\ref{Fig8}, then its type in Table~\ref{Tab12} has been listed as being either an active stellar corona (ASC) or PMS star (ASC/PMS). If the \chic\ source falls within Region ii of Figure~\ref{Fig8}, and also within one or both of the WR color spaces in Figures~\ref{Fig9a} and \ref{Fig9b}, then its type has been classified as a WR star. Those \chic\ sources that fall within Region ii of Figure~\ref{Fig8} and near AX~J144547-5931 in Figures~\ref{Fig9a} and \ref{Fig9b} (see text in Section 4.2.2) have been classified as massive O-type stars (MS-O). All Region ii stars that fall within the general stellar loci of Figures~\ref{Fig9a} and \ref{Fig9b} could be either massive stars (MS) or ASC so are listed as MS/ASC. (A PMS star interpretation is also possible for these Region ii sources.)

Those Region ii \chic\ sources listed as Wolf-Rayet (WR), massive stars (MS), or massive O-type stars (MS-O), still need to be properly investigated for evidence of X-ray emission emanating from colliding-winds in a CWB. Such an identification requires further X-ray spectroscopic follow-up to identify plasma temperatures between 1 and 10 keV \citep{usov92}. There are also 7 AGPS sources that are listed as H {\sc ii} regions. Those AGPS sources identified as ``H {\sc ii}'' are actually made up of many X-ray point sources that were detected in our \chic\ \cxo\ observations (see Section 4.1) and are therefore young and massive stars in the H {\sc ii} region listed. If there is an identified X-ray star that is a significant contributor to the X-ray emission, the stellar type is listed before the H {\sc ii} abbreviation in column 3, along with its \chic\ source name in column 2. The suffixes of the other \chic\ sources coincident with the H {\sc ii} region are listed in parentheses. All other AGPS and \chic\ sources for which there is no identification information are classified as unknown (U). If no sources were detected in a \chic\ \cxo\ observation, then the AGPS source falls in the no detection (ND) category. A detailed description of some of the AGPS sources identified through the \chic\ survey and of the archival AGPS sources (i.e. those sources identified in the literature and therefore not observed with \cxo\ as part of the \chic\ survey) can be found in Section 4.3 and Appendix A, respectively.

The final column in Table~\ref{Tab12} contains a flag for columns one and two that indicates whether an AGPS and/or \chic\ source has a confirmed identification (I) obtained through work in the \chic\ survey or a tentative identification (T) using the population statistics in Section 4.2. Those AGPS sources for which no \chic\ sources were detected within $3'$ of the \asca\ position, or only faint ($<20$ X-ray counts) \chic\ sources were detected (F), are also indicated. (This flag excludes those AGPS sources that have been identified as H {\sc ii} regions.) The unidentified population of \chic\ sources that fall in Region iv of Figure~\ref{Fig8} (R) are also included. There is also a flag for column three that indicates whether the type identification for a \chic\ source is unconfirmed (N) due to it being based on its tentative statistical identification in Section 4.2.

\section{Conclusions}

The main aim of the \chic\ survey is to identify the Galactic plane X-ray source populations that make up the $F_{x} \sim 10^{-13} \mathrm{~to~} 10^{-11}$ \erg\ flux range. To achieve this we have used new observations from the \cxo\ X-ray telescope, along with extensive multi-wavelength follow-up, to identify sources from the \asca\ Galactic Plane Survey \citep{sugizaki01}. We have reported observations of the \asca\ positions of 93 unidentified AGPS sources with \cxo, from which a total of 253 X-ray point sources, termed ``\chic\ sources'', have been detected.

Through visual inspection of Galactic plane radio surveys, we have found 5 \chic\ sources within supernova remnants that have no cataloged optical or infrared counterparts. These sources could potentially be compact objects associated with their surrounding SNRs. Further radio analysis has also demonstrated that the \chic\ sources detected in the \cxo\ observations of the AGPS sources AX J144519--5949, AX J151005--5824, AX J154905--5420, \object[AX J1622.1-5005]{AX J162208--5005}, AX J194310+2318, AX J194332+2323, and AX J195006+2628 are all coincident with H {\sc ii} regions. Table~\ref{Tab11} demonstrates that the range of luminosities, which are calculated using kinematic distances to the H {\sc ii} regions, are consistent with the luminosities we expect from flaring PMS stars, massive stars, and CWBs. We therefore identify the $54$ separate \chic\ sources seen in these \cxo\ observations as young and massive stars within H {\sc ii} regions.

Of the 93 \cxo-observed AGPS fields, 62 have one or more sources with $>20$ X-ray counts, resulting in the detailed study of 74 \chic\ sources in this paper. The multi-wavelength follow-up of these \chic\ sources demonstrates the need for \cxo's subarcsecond localization capabilities to correctly identify likely infrared and optical counterparts. The main focus of this paper has been on those unidentified \chic\ sources with $>20$ X-ray counts and with near-infrared or infrared counterparts. This has allowed us to perform population statistics to identify some of the likely objects that make up the $F_{x} \sim 10^{-13} \mathrm{~to~} 10^{-11}$ \erg\ Galactic plane X-ray source populations. 

We have developed a new statistical diagnostic for identifying likely populations of X-ray emitting sources using $K$-band fluxes and upper-limits (see Figure~\ref{Fig8}). The unidentified \chic\ sources in Region i of Figure~\ref{Fig8} have soft X-ray emission and low X-ray-to-infrared flux ratios, making them consistent with many of the archival and identified \chic\ stars. Their X-ray-to-infrared flux ratios are also similar to the COUP stars (see Figure~\ref{Fig7}), which are predominantly PMS stars. The majority of the Region i sources also fall within the general stellar locus that is expected in Figures~\ref{Fig9a} and \ref{Fig9b} \citep{hadfield07}. They are therefore likely to be active stellar coronae, which is consistent with the main soft X-ray populations expected in the Galactic plane \citep{hong05}, or PMS stars.

Many of the \chic\ sources in Region ii of Figure~\ref{Fig8} have infrared colors similar to known Wolf-Rayet stars, as demonstrated in Figure~\ref{Fig9}, which indicate the presence of excess infrared emission resulting from strong, dense stellar winds. These sources are therefore likely to be massive stars generating X-rays through instability-driven wind shocks or even colliding winds in CWBs \citep[for example see][]{anderson11}.

Only two unidentified \chic\ sources are located within Region iii of Figure~\ref{Fig8}, along with the archival high-mass and symbiotic X-ray binaries (and AGN). As such X-ray binaries (XRBs) are rare, only a few unidentified \chic\ sources are expected to fall within this group. This result therefore demonstrates that Figure~\ref{Fig8} may be a very useful diagnostic for identifying XRBs.

Region vi contains four identified magnetars and a candidate LMXB. Even though there are likely two different source populations in Region vi, Figure~\ref{Fig8} demonstrates that hard X-ray sources ($E_{50}>1.3$ keV), with an X-ray-to-infrared flux ratio $F_{x}/F_{Ks} > 10^{2}$, are very rare and interesting Galactic X-ray sources.

The population of \chic\ sources in Region iv of Figure~\ref{Fig8} remains unidentified. Based on their position relative to the identified AGN in Figure~\ref{Fig7} and they high $N_{H}$ values compared to the Galactic column density, we suggest that the 4 \chic\ sources ChI J181116--1828\_2, ChI J181213--1842\_7, ChI J190749+0803\_1, and ChI J194152+2251\_2 could be background AGN. The remaining 8 unidentified Region iv \chic\ sources have far lower $N_{H}$ values than the Galactic column densities indicating that they could be located in our own Galaxy. Optical and infrared spectroscopic follow-up is required to identify the true nature of this population. 

With further source identifications, a full \ns\ model of the hard ($2-10$ keV) Galactic plane X-ray populations between $F_{x} \sim 10^{-13} \mathrm{~and~} 10^{-11}$ \erg\ will be able to be constructed. This \ns\ will be more complete than those constructed from previous X-ray surveys in the same flux range, as it will be representative of 40 deg$^{2}$ of the Galactic plane. It will also show individual contributions from different Galactic X-ray source populations including non-accretion powered sources such as CWBs, SNRs, PWNe, and magnetars, which have not been a focus of previous work. Using the \ns\ distribution and distance estimates, it will then be possible to construct luminosity functions and three-dimensional spatial distributions of each class of X-ray source in the Galactic plane.

\appendix

\section{Source Descriptions of the Archival AGPS Sources}

As mentioned in Section 1, approximately one third of the AGPS sources were identified by \citet{sugizaki01} or were classified by other research groups prior to the \chic\ survey. It is these identified AGPS sources, referred to as ``archival sources'' in Section 4.2.1, which have been used to narrow down the possible unidentified \chic\ source populations. The archival AGPS sources are listed in Table~\ref{Tab12} and are briefly described in below.

\begin{description}

\item[AX J143416--6024] RS CVn type variable star, HD 127535, with spectral type K1IIIe \citep{sugizaki01}.  

\item[AX J155052--5418] X-ray and radio emitting magnetar 1E 1547.0-5408, associated with the possible radio SNR G327.24-0.13 \citep{gelfand07}.

\item[AX J155644-5325] KOIIIe type star, TYC 8697--1438-1 \citep{torres06}.

\item[AX J161929--4945] SFXT, a sub-class of HMXBs that display fast X-ray outbursts \citep{sguera06,tomsick06}.

\item[AX J162155--4939] K3III type star, HD 147070 \citep{sugizaki01}; this identification needs to be confirmed by follow-up X-ray observations

\item[AX J163159--4752] Accretion driven 1300s X-ray pulsar in a supergiant HMXB \citep{rodriguez06,walter06}. This system is one of the highly absorbed HMXBs identified by \textit{Integral} \citep{negueruela07}.

\item[AX J163351--4807] The magnetic Of?p star HD 148937 \citep{naze12}.

\item[AX J163555--4719] The X-ray emission associated with SNR G337.2+0.1 and its PWN \citep{combi06}. This system is also associated with the \textit{Fermi}-LAT source 1FGL J1635.7--4715 \citep{abdo10}.

\item[AX J163904--4642] Originally identified as a 912s pulsating, heavily-absorbed, HMXB \citep{bodaghee06,thompson06}, this source has now been reclassified as a SyXB \citep{nespoli10}.

\item[AX J164042--4632] X-ray PWN associated with the radio SNR G338.3--0.0 and the very high energy $\gamma$-ray source HESS J1640--465 \citep{funk07,lemiere09}. This system is also associated with the \textit{Fermi}-LAT source 1FGL J1640.8--4634 \citep{abdo10,slane10}.

\item[AX J165437--4333] The F7V type star HD 152335 \citep{sugizaki01}.

\item[AX J165904--4242] Herbig Be star V921 Sco, where the X-ray emission may arise from magnetic activity \citep{hamaguchi05}. \citep[][incorrectly assigned this star as the counterpart to AX J165901--4208.]{sugizaki01}

\item[AX J170006--4157] Magnetized CV, likely of the IP class, with 715s X-ray pulsations \citep{torii99,kaur10}.

\item[AX J170047--4139] 38s pulsating HMXB with an Ofpe/WNL type mass donor \citep{chakrabarty02,mason09}.

\item[AX J170349--4142] SNR G344.7--0.1 and its possible central compact object \citep[CCO;][]{combi10}. There is a possible $\gamma$-ray counterpart, HESS J1702--420 \citep{giacani11}.

\item[AX J171804--3726] SNR G349.7+0.2 and its possible CCO \citep{slane02a,lazendic05}. This remnant is also associated with the \textit{Fermi}-LAT source 1FGL J1717.9--3729 \citep{castro10}.

\item[AX J172105--3726] SNR G350.1--0.3 and its CCO \citep{gaensler08,lovchinsky11}. 

\item[AX J172743--3506] SNR G352.7--0.1 \citep{giacani09}.

\item[AX J173441--3234] Colliding-wind binary (CWB) HD 159176 (07V+07V) in the young open cluster NGC 6383. The short period of this binary implies that the winds likely collide well before reaching their terminal velocities, limiting the hardness of the resulting thermal X-ray emission \citep{debecker04}.

\item[AX J173518--3237] SNR G355.6--0.0 \citep{yamauchi08}.

\item[AX J180225--2300] X-ray emission associated with the OB type and pre-main-sequence stars in the Trifid Nebula. The main X-ray contributor is the HD 164492 multiple system of OB stars \citep{rho04}. 

\item[AX J180838--2024] Magnetar SGR 1806--20 \citep{kouveliotou98}. 

\item[AX J180902--1948] SNR G10.5--0.0. Other than this \asca\ detection \citep{sugizaki01}, no other X-ray papers exist on this source. The radio SNR was discovered by \citet{brogan06}.

\item[AX J180948--1918] PSR J1809--1917 and its PWN, which are likely associated with HESS J1809--193 \citep{kargaltsev07,aharonian07}. 

\item[AX J180951--1943] The X-ray and radio emitting magnetar XTE J1810--197 \citep{ibrahim04,halpern05}.

\item[AX J181211--1835] SNR G12.0--0.1 \citep{yamauchi08}.

\item[AX J182104--1420] SNR G16.7+0.1 and its central PWN \citep{helfand03}.

\item[AX J183039--1002] A Compton-thick active galactic nucleus \citep{bassani09}. A $K_{s}$-band magnitude of $14.3 \pm 0.2$ was obtained for this source with PANIC on 2007 July 29 (see Section 4.3.15). 

\item[AX J183221--0840] Magnetized CV, likely of the IP class, with 1549.1s X-ray period pulsations \citep{sugizaki00,kaur10}.

\item[AX J183528--0737] The 112s pulse period X-ray binary, Scutum X-1. This system is likely to be a SyXB \citep{kaplan07}. 

\item[AX J183800--0655] The 70.5 ms pulsar, PSR J1838--0655, and its PWN. This system is possibly associated with HESS J1837--069 \citep{gotthelf08,kargaltsev12}.

\item[AX J183931--0544] The luminous blue variable (LBV) candidate G26.47+0.02. This source is possibly in a CWB \citep{paron12}. It is assumed that the  2MASS J18393224--0544204 is the correct NIR counterpart for the purpose of the statistical analysis in Section 4.4. 

\item[AX J184121--0455] The magnetar 1E 1841--045 and its associated SNR G27.4+0.0 \citep[Kes 73;][]{gotthelf97,morii03}.

\item[AX J184355--0351] X-ray emission associated with the non-thermal SNR G28.6--0.1/AX J1843.8--0352 and the thermal source CXO J184357--035441 \citep[which may or may not be part of SNR G28.6--0.1,][]{bamba01,ueno03}.

\item[AX J184629--0258] X-ray emission from the SNR G29.7--0.3 (Kes 75), its central pulsar PSR J1846--0258 and associated PWN \citep{helfand03b}.

\item[AX J184848--0129] X-ray sources in the Galactic globular cluster GLIMPSE--C01 \citep{pooley07} and the nearby diffuse source CXOU J184846.3--013040 \citep[either a PWN or the globular cluster's bow shock;][]{mirabal10}. These sources may also be associated with the \textit{Fermi}-LAT source 0FGL J1848.6--0138 \citep{luque09}.

\item[AX J184930--0055] X-ray emission associated with the thermal composite SNR G31.9+0.0 \citep[3C 391;][]{chen01,chen04}. This SNR is likely associated with 1FGL J1849.0--0055 \citep{castro10}.

\item[AX J185015--0025] The X-ray synchrotron-dominated SNR G32.4+0.1 \citep{yamaguchi04}.

\item[AX J185240+0038] X-ray emission associated with the SNR G33.6+0.1 (Kes 79) and the 105 ms pulsar PSR J1852+0040 \citep[there is no detectable PWN;][]{gotthelf05}. This pulsar has been described as an ``anti-magnetar'' \citep{halpern10}.

\item[AX J185551+0129] X-ray emission from SNR G34.7--0.4 (also known as W44 and 3C392) and the PWN associated with its central pulsar, PSR B1853+0.1 \citep{petre02}. This system may be associated with the \textit{Fermi}-LAT source 0FGL J1855.9+0126/1FGL 1856.1+0122 \citep{abdo09,abdo10}.

\item[AX J190734+0709] SNR G41.1--0.3 \citep[3C 397;][]{safiharb05}.

\item[AX J191105+0906] SNR G43.3--0.2 \citep[W49B;][]{hwang00}.

\item[AX J194649+2512] X-ray emission likely associated with the H$\alpha$ emission line star VES 52 \citep[][this identification still needs to be properly confirmed by follow-up X-ray observations]{kohoutek97,sugizaki01}.

\end{description}

\section{\textit{CHANDRA} detected X-ray point sources beyond the $3'$ search region}

Table~\ref{Tab13} lists those 14 \cxo\ detected X-ray point sources with $>20$ X-ray counts that lie within $3'-5'$ of the \asca\ position of 11 AGPS sources and therefore outside the \chic\ MAP default search radius. This Table includes the name of the AGPS sources for which the above applies, the position of the X-ray source, the offset from the original \asca\ position, the net number of counts, and the most likely 2MASS counterpart.

A thorough analysis of these 14 X-ray point sources is beyond the scope of this paper, however, we have done a preliminary investigation of their possible contribution to the fluxes originally measured in the AGPS \citep{sugizaki01}. Using a similar technique to that described in Section 2.1 we entered the power law spectral fit of the AGPS source measured by \citet{sugizaki01} into \cxo\ PIMMS in order to estimate the number of source counts expected to be detected in the corresponding \chic\ \cxo\ observation. (Once again the photon index and absorption were set to $\Gamma=2$ and $N_{H}=10^{22}\mathrm{cm}^{-2}$ if no power law fit was provided.) These count predictions are listed in Table~\ref{Tab13}. The total number of counts detected from all the X-ray sources within $3'$ of the AGPS position are also included alongside these values. In the case of AX J154557--5420, AX J154905--5420, AX J172642--3504, AX J181915--1601, AX J184741--0129, AX J194332+2323, the total number of X-ray counts detected within $3'$ of the AGPS position contribute $\leq65\%$ of the predicted number of counts. It is therefore possible that the 8 X-ray point sources detected between $3'-5'$ from these AGPS sources could have contributed to the X-ray flux originally detected with \asca. Such a result would not be unexpected for AX J154905--5420 and AX J194332+2323 as their corresponding \cxo\ resolved point sources are stars in H~{\sc ii} regions, which were not individually resolved with \asca. This analysis demonstrates that few X-ray sources beyond $3'$ of the \asca\ position contributed to the overall flux of a given AGPS source. The $3'$ search radius used in \chic\ MAP is therefore reasonable for detecting the majority of AGPS associated point sources detected with \cxo. \citep[Note that this Table does not include the magnetar PSR J1622--4950 that was detected in the \cxo\ observation of AX J162246--4946, $4'$ for its \asca\ position, as it was well investigated in][]{anderson12}.

\begin{acknowledgements}

G.E.A acknowledges the support of an Australian Postgraduate Award. B.M.G. acknowledges the support of an Australian Laureate Fellowship through ARC grant FL100100114. P.O.S. acknowledges partial support from NASA Contract NAS8-03060. D.T.H.S. acknowledges a STFC Advanced Fellowship. J.J.D was supported by NASA contract NAS8-39073 to the Chandra X-ray Center (CXC). Support for this work was also provided by NASA through \cxo\ Award Number GO9-0155X issued by the CXC, which is operated by the Smithsonian Astrophysical Observatory for and on behalf of NASA. The access to major research facilities program is supported by the Commonwealth of Australia under the \textit{International Science Linkages program}. This research makes use of data obtained with the \cxo\ \textit{X-ray Observatory}, and software provided by the CXC in the application packages \texttt{CIAO}. The ATCA is part of the Australia Telescope, funded by the Commonwealth of Australia for operation as a National Facility managed by CSIRO. The MOST is operated with the support of the Australian Research Council and the Science Foundation for Physics within the University of Sydney. Observing time on the 6.5m Clay and Baade Magellan Telescopes, located at Las Campanas Observatory, was allocated through the Harvard-Smithsonian Center for Astrophysics and the Massachusetts Institute of Technology. 2MASS is a joint project of the University of Massachusetts and the IPAC/Caltech, funded by the NASA and NFS. GLIMPSE survey data are part of the Spitzer Legacy Program. The \textit{Spitzer Space Telescope} is operated by JPL/Caltech under a contract with NASA. This research has made use of NASA's Astrophysics Data System.

\end{acknowledgements}

{\it Facilities:} \facility{ASCA}, \facility{ATCA}, \facility{CXO (ACIS,HRC)}, \facility{CTIO:2MASS}, \facility{FLWO:2MASS}, \facility{Magellan:Baade (IMACS, MagIC, PANIC)}, \facility{Molonglo}, \facility{Spitzer}, \facility{VLA}, \facility{XMM (EPIC)}

\clearpage
\LongTables
\begin{landscape}
\begin{deluxetable}{llccccccccccccccc}
\tablewidth{0pt}
\tabletypesize{\scriptsize}
\setlength{\tabcolsep}{0.03in}
\tablecaption{\chic\ MAP X-ray Results, including the position, position errors, net counts and quantile analysis values for each \chic\ source \label{Tab1}}
\tablehead{
\colhead{Chandra Obs\tablenotemark{a}} & \colhead{ChIcAGO\tablenotemark{b}} & \multicolumn{2}{c}{\texttt{wavdetect} Position} & \colhead{Offset\tablenotemark{c}} & \multicolumn{2}{c}{Positional Errors\tablenotemark{d}} & &  \multicolumn{3}{c}{Net Counts\tablenotemark{e}} & & \multicolumn{2}{c}{Energy Quantiles\tablenotemark{f}} & \colhead{CSC\tablenotemark{g}} & \colhead{XMMi-DR3\tablenotemark{h}}
\\
\cline{3-4} \cline{6-7} \cline{9-11} \cline{13-14}
\\
\colhead{(ID, Date,} & \colhead{Source ID} & \colhead{RA} & \colhead{Dec} & \colhead{from} & \colhead{wav} & \colhead{Total} & & \colhead{0.3-8} & \colhead{0.5-2} & \colhead{2-8 keV} & & \colhead{$E_{50}$} & \colhead{Quartile} & \colhead{Name} & \colhead{Name}
\\
\colhead{Inst, Exp)} & \colhead{ChI} & \colhead{(J2000)} & \colhead{(J2000)} & \colhead{\asca\ ($'$)} & \colhead{($''$)} & \colhead{($''$)} & & \colhead{} & \colhead{} & \colhead{} & & \colhead{(keV)} & \colhead{Ratio} & \colhead{CXO J} & \colhead{2XMMi}
}
\startdata
\multicolumn{13}{l}{AX J143148-6021 8239 2007-09-05 ACIS-S 2.62ks}\\
 & J143148-6021\_1 & 14:31:50.18 & $-$60:22:08.3 & 0.97 & 0.6 & 0.9 & & $4.0_{-1.9}^{+3.2}$ & $4.0_{-1.9}^{+3.2}$ &  $0.0_{-0.0}^{+1.9}$ & &   &   &  & \\
 & J143148-6021\_2 & 14:31:48.56 & $-$60:21:44.5 & 0.56 & 0.5 & 0.8 & & $10_{-3}^{+4}$ & $7.0_{-2.6}^{+3.8}$ & $2.0_{-1.3}^{+2.7}$ & & $1.1 \pm 0.4$ & $1.4 \pm 0.9$ &  & \\
 & J143148-6021\_3 & 14:31:48.06 & $-$60:19:59.5 & 1.19 & 0.3 & 0.8 & & $146_{-12}^{+13}$ & $51_{-7}^{+8}$ & $95_{-10}^{+11}$ & & $2.7 \pm 0.2$ & $1.1 \pm 0.1$ &  143148.0-601959 & \\
 & J143148-6021\_4 & 14:31:41.52 & $-$60:19:55.2 & 1.56 & 0.6 & 0.9 & & $5.9_{-2.4}^{+3.6}$ & $2.9_{-1.6}^{+2.9}$ & $0.0_{-0.0}^{+1.9}$ & & $0.6 \pm 0.3$ & $0.5 \pm 0.4$ &  & \\
 & J143148-6021\_5 & 14:32:02.80 & $-$60:18:57.4 & 2.81 & 0.7 & 1.0 & & $20_{-4}^{+6}$ & $0.0_{-0.0}^{+1.9}$ & $20_{-4}^{+6}$ & & $4.4 \pm 0.5$ & $1.8 \pm 0.2$ &  143202.8-601857 & \\
\hline
\multicolumn{13}{l}{AX J144042-6001 9599  2008-05-04  HRC-I  1.18ks}\\
 & J144042-6001\_1 & 14:40:38.40 & $-$60:01:36.8 & 0.56 & 0.3 & 0.8 & & $326_{-18}^{+19}$ & &  &   &   &   &  144038.4-600136 & \\
\enddata
\end{deluxetable}
\clearpage
\end{landscape}

\clearpage
\LongTables
\begin{landscape}
\begin{deluxetable}{lccccccccccc}
\tablewidth{0pt}
\tabletypesize{\scriptsize}
\setlength{\tabcolsep}{0.02in}
\tablecaption{\chic\ sources with $>20$ X-ray counts and their infrared properties \label{Tab2}}
\tablehead{
\colhead{\chic\ Source} & \colhead{NIR\tablenotemark{a}} & \colhead{2MASS name} & \multicolumn{3}{c}{Near-Infrared Magnitudes} & \colhead{PANIC date} & \colhead{GLIMPSE name} & \multicolumn{4}{c}{GLIMPSE Magnitudes\tablenotemark{b}}
\\ 
\cline{4-6} \cline{9-12} 
\\
\colhead{ChI} & \colhead{Data} & \colhead{2MASS J} & \colhead{$J$} & \colhead{$H$} & \colhead{$K$} & \colhead{yyyy-mm-dd} & \colhead{SSTGLMC G} & \colhead{$M_{3.6}$} & \colhead{$M_{4.5}$} & \colhead{$M_{5.8}$} & \colhead{$M_{8.0}$} 
}
\startdata
  J143148-6021\_3 & P &                  &  $17.1   \pm 0.1  $ & $16.3    \pm 0.1    $ & $15.6    \pm 0.1    $ & 2005-05-12    & 			    &         	      	      &                        &        	       &                      
\\
  J144042-6001\_1 & M & 14403847-6001368 &  $8.28   \pm 0.02  $ & $7.88   \pm 0.02  $ & $7.70   \pm 0.02  $ &         	 & 316.1785+00.0060 &   $ 7.65  \pm 0.06 $ & $  7.63  \pm  0.05 $ & $  7.64  \pm  0.03 $ & $  7.59 \pm  0.03 $ 
\\
  J144519-5949\_2 & M & 14452143-5949251 &  $14.12  \pm 0.04  $ & $10.95   \pm 0.03  $ & $9.30   \pm 0.03  $ &          	 &      	    &   		      &                        &   	               &	          		      
\\ 
  J144547-5931\_1 & M & 14454369-5932050 &  $9.23   \pm 0.04  $ & $8.10   \pm 0.03  $ & $7.52   \pm 0.02  $ &          	 & 316.9656+00.1864 &   $ 7.07  \pm 0.05 $ & $  6.89  \pm  0.04 $ & $  6.708 \pm  0.03 $ & $  6.63 \pm  0.02 $ 
\\ 
  J144701-5919\_1 & M & 14465358-5919382 &  $8.93   \pm 0.02  $ & $7.60   \pm 0.04   $ & $6.82   \pm 0.02  $ &          	 & 317.1882+00.3106 &   $ 6.70  \pm 0.09^{\dagger} $ & $  6.33  \pm  0.09^{\dagger} $ & $  5.574 \pm  0.03 $ & $  5.32 \pm  0.03 $ 
\\ 
  J150436-5824\_1 &   &                  &   	    	       	 &  	    	         &   	    	       	 &          	 &      	    &    	    	      &                        &         	       &                                    
\\ 
  J153818-5541\_1 & P** &                &  $>21.9   \pm 0.3    $ & $>19.5   \pm 0.15   $ & $18.0    \pm 0.5    $ & 2009-04-13 	 &      	    &   		      &                        &   	               &	          		      
\\ 
  J154122-5522\_1 & M & 15411580-5521286 &  $9.60   \pm 0.02  $ & $8.93    \pm 0.06  $ & $8.68    \pm 0.02  $ &        	 & 325.4962-00.1082 &   $ 8.61  \pm 0.05 $ & $  8.68  \pm  0.06 $ & $  8.51 \pm  0.04 $ & $  8.47 \pm  0.03  $ 
\\
\enddata																         
\tablecomments{The \texttt{wavdetect} position of ChI J165646-4239\_1 and the position of the quoted 2MASS source agree within $4\sigma$. We have listed this 2MASS source as a possible counterpart as it is considered to be the same as the quoted GLIMPSE source, which does agree within $3\sigma$ of the \texttt{wavdetect} position. The quoted NIR magnitudes of ChI J181116-1828\_2 and ChI J185643+0220\_2, obtained with PANIC, are limits due to the counterparts being blended. The $H$-band PANIC observation of ChI J194152+2251\_2 took place on 2007-07-31, whereas the other bands were observed on 2007-07-29.}

\tablenotetext{a}{Data from which the counterpart near-infrared magnitudes were obtained. M = 2MASS Point Source Catalog \citep{skrutskie06}, P = PANIC Magellan observations. If three letters are listed this indicates that the magnitudes were obtained from different datasets. This order corresponds to the $J$, $H$ and $K_{s}$ magnitude order. Those datasets marked with * are likely blended or, particularly in the case of ChI J194152+2251\_2, perhaps extended. Those datasets marked with ** means that the counterpart has a low sigma detection of $>1.5\sigma$ for the respective background plus RMS estimate.}
\tablenotetext{b}{$M_{3.6},~M_{4.5},~M_{5.8},~\mathrm{and}~M_{8.0}$ refer to the counterpart infrared magnitudes at 3.6, 4.5, 5.8, and 8.0~${\mu}$m, respectively. Those magnitudes marked with $^{\dagger}$ are from the GLIMPSE archives (more complete, less reliable) rather than the catalogs (which are more reliable). In these cases the GLIMPSE archive source's names begin with SSTGLMA G. }

\end{deluxetable}
\clearpage
\end{landscape}

\begin{deluxetable*}{lccccccccccc}
\tablewidth{0pt}
\tabletypesize{\scriptsize}
\tablecaption{Quantile analysis power law interpolation and Cash statistics power law spectral fit to ACIS \cxo\ detections of the \chic\ sources \label{Tab3}}
\tablehead{
\colhead{\chic\ Source} & \multicolumn{5}{c}{Quantile absorbed power law interpolation\tablenotemark{a}} & \colhead{} & \multicolumn{5}{c}{Cash statistics absorbed power law fit\tablenotemark{b}}
\\
\cline{2-6} \cline{8-12}
\\
\colhead{ChI} & \colhead{$\Gamma$} & \colhead{$N_{H}$} & \colhead{$F_{x,abs}$} & \colhead{$F_{x,unabs}$} & \colhead{$\chi^{2}_{red}$} & \colhead{} & \colhead{$\Gamma$} & \colhead{$N_{H}$} & \colhead{$F_{x,abs}$} & \colhead{$F_{x,unabs}$} & \colhead{$C$}
}
\startdata
  J143148-6021\_3 &  $0.91_{-0.35}^{+0.44}$  &  $0.61_{-0.34}^{+0.54}$   &   $9.12 \pm 0.75$  &   $10.93 \pm 0.90$  &  0.3   &  &  $0.70_{-0.25}^{+0.26}$  &  $0.42_{-0.17}^{+0.20}$  &  $9.85_{-1.15}^{+1.35}$  &  $11.17_{-1.31}^{+1.29}$         &  0.5   \\
  J144547-5931\_1 &                          &                           &                 &                   &        &  &  $4.98_{-0.94}^{+1.06}$  &  $4.53_{-1.07}^{+1.25}$  &  $2.38_{-0.59}^{+0.45}$  &  $1802.71_{-1388.60}^{+4414.31}$ &  1.0   \\
  J144701-5919\_1 &                          &                           &                 &                   &        &  &  $3.39_{-0.61}^{+0.67}$  &  $2.70_{-0.64}^{+0.73}$  &  $5.55_{-1.90}^{+1.15}$  &  $177.46_{-139.42}^{+157.69}$    &  0.6   \\
  J150436-5824\_1 &  $-0.37_{-0.00}^{+0.17}$  &  $0.01_{-0.00}^{+0.06}$   &   $8.45 \pm 0.71$  &   $8.47 \pm 0.71$  &  0.9   &  &  $-0.05_{-0.27}^{+0.28}$ &  $0.49_{-0.27}^{+0.32}$  &  $8.46_{-0.84}^{+1.07}$  &  $9.03_{-0.98}^{+1.05}$          &  1.2   \\
  J154122-5522\_1 &  $2.53_{-0.49}^{+0.62}$  &  $0.14_{-0.11}^{+0.14}$   &   $2.50 \pm 0.27$  &   $4.46 \pm 0.49$  &  0.6   &  &  $2.35_{-0.39}^{+0.43}$  &  $0.15_{-0.09}^{+0.10}$  &  $2.86_{-0.46}^{+0.38}$  &  $4.82_{-1.22}^{+0.99}$          &  1.0   \\
  J154557-5443\_1 &  $2.98_{-0.70}^{+1.30}$  &  $0.29_{-0.17}^{+0.19}$   &   $0.17 \pm 0.04$  &   $0.56 \pm 0.12$  &        &  &                          &                          &                          &                                  &         \\
  J155035-5408\_3 &  $2.73_{-0.46}^{+0.55}$  &  $0.67_{-0.21}^{+0.24}$   &   $2.90 \pm 0.20$  &   $11.47 \pm 0.80$  &  0.9   &  &  $2.72_{-0.27}^{+0.28}$  &  $0.61 \pm 0.11$  &  $2.88_{-0.33}^{+0.36}$  &  $11.33_{-3.55}^{+3.17}$         &  1.8   \\
\enddata

\tablecomments{The quantile analysis spectral interpolation and Cash statistics spectral modeling are performed on those \chic\ sources with $>20$ and $>50$ X-ray counts, respectively.}

\tablenotetext{a}{The quantile analysis interpolated absorbed power law model parameters including the spectral index, $\Gamma$, and absorption column density $N_{H}$ ($10^{22}$ cm$^{-2}$), as well as the absorbed and unabsorbed X-ray fluxes, $F_{x,abs}$ and $F_{x,unabs}$ ($10^{-13}$ erg cm$^{-2}$ s$^{-1}$), respectively, in the $0.3-8.0$ keV energy range. The quoted $\chi^{2}_{red}$ value is what is expected if these interpolated spectral parameters are used to fit the X-ray spectrum of the source.}
\tablenotetext{b}{The absorbed power law fit model parameters using \citet{cash79} statistics including the spectral index, $\Gamma$, and absorption column density $N_{H}$ ($10^{22}$ cm$^{-2}$), as well as the absorbed and unabsorbed X-ray fluxes, $F_{x,abs}$ and $F_{x,unabs}$ ($10^{-13}$ erg cm$^{-2}$ s$^{-1}$), respectively, in the $0.3-8.0$ keV energy range. The quoted $C$ value is the reduced Cash statistic, which is the observed statistic divided by the number of degrees of freedom. A good fit is indicated by a $C$ value of order 1.}

\end{deluxetable*}

\clearpage
\begin{deluxetable*}{lccccccccccc}
\tablewidth{0pt}
\tabletypesize{\scriptsize}
\tablecaption{Quantile analysis Bremsstrahlung spectral interpolation and Cash Statistic Bremsstrahlung spectral fit to ACIS \cxo\ detections of the \chic\ sources \label{Tab4}}
\tablehead{
\colhead{\chic\ Source} & \multicolumn{5}{c}{Absorbed Bremsstrahlung interpolation\tablenotemark{a}} & \colhead{} & \multicolumn{5}{c}{Cash statistics absorbed Bremsstrahlung fit\tablenotemark{b}}
\\
\cline{2-6} \cline{8-12}
\\
\colhead{ChI} & \colhead{$kT$} & \colhead{$N_{H}$} & \colhead{$F_{x,abs}$} & \colhead{$F_{x,unabs}$} & \colhead{$\chi^{2}_{red}$} & \colhead{} & \colhead{$kT$} & \colhead{$N_{H}$} & \colhead{$F_{x,abs}$} & \colhead{$F_{x,unabs}$} & \colhead{$C$}
}
\startdata
  J144519-5949\_2 &  $1.38_{-0.65}^{+2.15}$  &  $4.80_{-2.2}^{+3.60}$     &   $2.17 \pm 0.44$  &    $21.59 \pm 4.37$  &         &  &                            &                           &                           &                            &             \\ 
  J144547-5931\_1 &  $0.84_{-0.21}^{+0.30}$  &  $4.10_{-0.9}^{+1.20}$     &   $2.56 \pm 0.31$  &    $57.74 \pm 6.95$  &   0.9   &  &  $1.01_{-0.23}^{+0.37}$    &   $3.27_{-0.74}^{+0.86}$  &   $1.93_{-1.24}^{+0.68}$  &   $37.28_{-22.51}^{+9.98}$ &   0.9      \\ 
  J144701-5919\_1 &  $1.00_{-0.39}^{+0.70}$  &  $3.50_{-1.2}^{+1.90}$     &   $5.49 \pm 0.60$  &    $74.01 \pm 8.12$  &   0.6   &  &  $2.04_{-0.54}^{+0.95}$    &   $1.93_{-0.43}^{+0.49}$  &   $5.78_{-2.03}^{+1.16}$  &   $23.52_{-7.51}^{+4.04}$  &   0.9      \\ 
  J154122-5522\_1 &  $1.88_{-0.70}^{+0.96}$  &  $0.04_{-0.03}^{+0.09}$    &   $2.35 \pm 0.26$  &    $2.72 \pm 0.3$   &   0.6   &  &  $2.62_{-0.80}^{+1.68}$    &   $0.05_{-0.05}^{+0.06}$   &   $2.46_{-0.47}^{+0.45}$  &   $3.08_{-0.70}^{+0.47}$   &   1.0      \\ 
  J154557-5443\_1 &  $1.38_{-0.73}^{+1.32}$  &  $0.14_{-0.1}^{+0.15}$     &   $0.16 \pm 0.03$  &    $0.25 \pm 0.05$  &         &  &                            &                           &                           &                            &             \\ 
  J154557-5443\_2 &  $1.77_{-1.33}^{+96.43}$ &  $24.0_{-15.8}^{+46.0}$    &   $0.63 \pm 0.16$  &    $17.65 \pm 4.49$  &         &  &                            &                           &                           &                            &             \\ 
  J154557-5443\_3 &  $0.10_{-0.00}^{+0.30}$    &  $0.97_{-0.88}^{+1.23}$    &   $0.11 \pm 0.03$  &    $224.95 \pm 51.46$  &         &  &                            &                           &                           &                            &             \\
\enddata     

\tablecomments{The quantile analysis spectral interpolation and Cash statistics spectral modeling are performed on those \chic\ sources with $>20$ and $>50$ X-ray counts, respectively. The \texttt{CIAO} task \texttt{sample\_flux} failed for ChI J190814+0832\_2 and so the quoted flux is taken from the best fit model parameters.}

\tablenotetext{a}{The quantile analysis interpolated absorbed thermal bremsstrahlung model parameters including the temperature, $kT$ (keV), and absorption column density $N_{H}$ ($10^{22}$ cm$^{-2}$), as well as the absorbed and unabsorbed X-ray fluxes, $F_{x,abs}$ and $F_{x,unabs}$ ($10^{-13}$ erg cm$^{-2}$ s$^{-1}$), respectively, in the $0.3-8.0$ keV energy range. The quoted $\chi^{2}_{red}$ value is what is expected if these interpolated spectral parameters are used to fit the X-ray spectrum of the source.}
\tablenotetext{b}{The absorbed thermal bremsstrahlung fit model parameters using \citet{cash79} statistics including the temperature, $kT$ (keV), and absorption column density $N_{H}$ ($10^{22}$ cm$^{-2}$), as well as the absorbed and unabsorbed X-ray fluxes, $F_{x,abs}$ and $F_{x,unabs}$ ($10^{-13}$ erg cm$^{-2}$ s$^{-1}$), respectively, in the $0.3-8.0$ keV energy range. The quoted $C$ value is the reduced Cash statistic, which is the observed statistic divided by the number of degrees of freedom. A good fit is indicated by a $C$ value of order 1.}

\end{deluxetable*}

\clearpage
\begin{deluxetable*}{lccccccc}
\tablewidth{0pt}
\tabletypesize{\scriptsize}
\tablecaption{Properties of optical counterparts, where they exist, for all \chic\ sources with $>20$ X-ray counts\label{Tab5}}
\tablehead{
\colhead{\chic\ Source} & \colhead{USNO Name} & \colhead{Optical\tablenotemark{a}} & \multicolumn{4}{c}{Optical Magnitudes\tablenotemark{b}} & \colhead{MagIC or IMACS Date\tablenotemark{c}}
\\
\cline{4-7}
\\
\colhead{ChI} & \colhead{USNO-B1.0} & \colhead{Data Source} & \colhead{$B$} & \colhead{$V$} & \colhead{$R$} & \colhead{$I$} & \colhead{yyyy-mm-dd}
}
\startdata
J143148-6021\_3 & 0296-0521275 & U,A,A,A & 19.71 $\pm$ 0.3 & 20.49 $\pm$ 0.03 & 19.31 $\pm$ 0.03 & 18.27 $\pm$ 0.03 & 2008-05-13
\\
J144042-6001\_1 & 0299-0503336 & U & 10.26 $\pm$ 0.3 &       & 9.36 $\pm$ 0.3 & 8.99 $\pm$ 0.3 &   
\\
J144547-5931\_1 & 0304-0487344 & U & 20.54 $\pm$ 0.3 &       & 15.79 $\pm$ 0.3 & 12.30 $\pm$ 0.3 &   
\\
J144701-5919\_1 & 0306-0492632 & U &       &       & 16.61 $\pm$ 0.3 & 12.95 $\pm$ 0.3 &   
\\
J154122-5522\_1 & 0346-0518164 & U & 13.17 $\pm$ 0.3 &     & 11.87 $\pm$ 0.3 & 10.87 $\pm$ 0.3 &   
\\
J154557-5443\_1 & 0352-0535858 & U & 16.16 $\pm$ 0.3 &     & 15.10 $\pm$ 0.3 & 14.67 $\pm$ 0.3 &   
\\
J154557-5443\_3 & 0352-0535969 & U & 17.91 $\pm$ 0.3 &     & 16.51 $\pm$ 0.3 & 15.42 $\pm$ 0.3 &   
\\
J155035-5408\_1 & 0358-0537979 & U & 11.40 $\pm$ 0.3 &     & 10.71 $\pm$ 0.3 & 10.43 $\pm$ 0.3 &   
\\
\enddata

\tablecomments{ChI J170017--4220\_1 was observed with both MagIC and IMACS. The dates of both observations are listed and prefixed by the corresponding dataset symbol described below. The MagIC observation of ChI J181116-1828\_2 shows that it may be a blend.}

\tablenotetext{a}{Catalog or observations from which the counterpart optical magnitudes were obtained. U = USNO B1 Catalog \citep{monet03}, A = MagIC Magellan observations, I = IMACS Magellan observations, NA = no data for the corresponding magnitude. If four symbols are listed this indicates that the magnitudes were obtained from different datasets. This order corresponds to the $B$, $V$, $R$ and $I$ magnitude order.}
\tablenotetext{b}{The USNO-B1.0 $B$ and $R$ band magnitudes listed are the second epoch $B$ and $R$ magnitudes \citep{monet03}}
\tablenotetext{c}{Date of the MagIC and IMACS observations: yyyy-mm-dd.}

\end{deluxetable*}

\clearpage
\begin{deluxetable*}{lcc}
\tablewidth{0pt}
\tabletypesize{\scriptsize}
\tablecaption{The chance of a random association between the listed \chic\ sources (all of which have $>20$ X-ray counts) and a 2MASS and GLIMPSE source\label{Tab6}}
\tablehead{
\colhead{\chic\ source} & \colhead{2MASS $N(<K_{s})$\tablenotemark{a}} & \colhead{GLIMPSE $N(<M_{3.6})$\tablenotemark{b}}
\\
\colhead{ChI} & \colhead{$(\times10^{-2})$} & \colhead{$(\times10^{-2})$}
}
\startdata
J155831--5334\_1 & 1.76 &
\\
J163751--4656\_1 & 1.67 & 3.16
\\
J165420--4337\_1 &      & 2.29
\\
J172550--3533\_2 & 1.08 & 1.71
\\
J173548--3207\_1 & 1.22 & 3.02
\\
J181116--1828\_2 &      & 3.15
\\
J183345--0828\_1 & 1.25 & 
\\
J190818+0745\_1 &      & 1.17
\enddata

\tablecomments{Only those \chic\ sources where the chance of random alignment with a survey source is $>0.01$ are listed.}

\tablenotetext{a}{Chance of finding a 2MASS source with a $K_{s}$ magnitude brighter than the counterpart listed in Table~\ref{Tab2} within the \chic\ source's 95\% position error circle.}

\tablenotetext{b}{Chance of finding a GLIMPSE source with a $3.6 \mu$m magnitude brighter than the counterpart listed in Table~\ref{Tab2} within the \chic\ source's 95\% position error circle.}

\end{deluxetable*}

\begin{deluxetable*}{lcccccc}
\tablewidth{0pt}
\tabletypesize{\scriptsize}
\tablecaption{Likely optical and infrared counterparts of secondary ($<20$ X-ray counts) \chic\ Sources \label{Tab7}}
\tablehead{
\colhead{\chic\ Source} & \colhead{USNO Name} & \colhead{offset\tablenotemark{a}} & \colhead{2MASS name} & \colhead{offset\tablenotemark{a}} & \colhead{GLIMPSE name\tablenotemark{b}} & \colhead{offset\tablenotemark{a}}
\\
\colhead{ChI} & \colhead{USNO-B1.0} & \colhead{(arcsec)} & \colhead{2MASS J} & \colhead{(arcsec)} & \colhead{SSTGLMC G} & \colhead{(arcsec)}
}
\startdata
J143148-6021\_1 & 0296-0521300 & 0.4 & 14315021-6022087 & 0.56 & G315.0382+00.1212 & 0.41
\\
J143148-6021\_2 & 0296-0521279 & 0.35 & 14314859-6021444 & 0.25 & G315.0377+00.1287 & 0.36
\\
J143148-6021\_4 & 0296-0521208 & 0.32 & 14314161-6019552 & 0.7 & G315.0358+00.1622 & 0.56
\\
J144519-5949\_1 &   &   & 14451506-5949286 & 0.51 & G316.7879-00.0504 & 0.61
\\
J144519-5949\_4 & 0301-0471052 & 0.36 & 14451937-5948007 & 0.47 & G316.8065-00.0321 & 0.64
\\
J144519-5949\_5 &   &   & 14445787-5950387 & 0.3 & G316.7471-00.0527 & 0.66
\\
J144519-5949\_6 &   &   & 14450584-5949317 & 0.64 & G316.7701-00.0429 & 0.78
\\
J145732-5901\_1 &   &   & 14573495-5859538 & 0.18 & G318.5603-00.0084 & 0.22
\\
\enddata

\tablenotetext{a}{The offset, in arcseconds, between the catalogs and \chic\ source position for the likely USNO B1, 2MASS and GLIMPSE counterparts, respectively.}
\tablenotetext{b}{Those GLIMPSE counterpart names marked with a * are from the archives (more complete, less reliable) rather than the catalogs (more reliable). In these cases the archival GLIMPSE source's names begins with ``SSTGLMA G.''}

\end{deluxetable*}

\begin{deluxetable*}{lccc}
\tablewidth{0pt}
\tabletypesize{\scriptsize}
\tablecaption{Summary of the fraction of \chic\ sources with a coincident cataloged source from 2MASS, GLIMPSE and USNO B1. These fractions are for the total number of \chic\ sources and for the \chic\ sources with $>20$ X-ray counts.\label{Tab8}}
\tablehead{
\colhead{\chic\ source} & \colhead{2MASS fraction} & \colhead{GLIMPSE fraction} & \colhead{USNO B1 fraction}
\\
\colhead{samples} & \colhead{(\%)} & \colhead{(\%)} & \colhead{(\%)}
}
\startdata
Total (253 sources) & 57 & 61 & 47
\\
$>20$ X-ray counts (74 sources) & 59 & 61 & 55
\enddata

\end{deluxetable*}

\begin{deluxetable*}{lccccc}
\tablewidth{0pt}
\tabletypesize{\scriptsize}
\tablecaption{Radio counterparts to \chic\ Sources\label{Tab9}}
\tablehead{
\colhead{ChIcAGO Source (ChI)\tablenotemark{a}} & \colhead{Survey\tablenotemark{b}} & \colhead{Radio source} & \colhead{Type\tablenotemark{c}} & \colhead{Location\tablenotemark{d}} & \colhead{Reference\tablenotemark{e}}
}
\startdata
J144042--6001\_1 & MGPS & SNR G316.3--0.0 & SNR & L & 1
\\
J144519--5949\_(1-6) & MGPS & GAL 316.8-00.1 & H {\sc ii} & C & 2
\\
J144701--5919\_1 & ATCA & AX J144701--5919 & MS & C & 3
\\
J145732--5901\_(1-2) & MGPS & SNR G318.2+0.1 & SNR & L & 4,5
\\
J151005--5824\_(1-11) & MGPS & G320.3--0.3 & H {\sc ii} & C & 6
\\
J153818--5541\_1 & ATCA & & ND & & 7
\\
J154905--5420\_(1-11) & MGPS & G326.96+0.03 & H {\sc ii} & A & 8
\\
J155331--5347\_1 & ATCA &  & ND &  & 7
\\
\enddata
\tablenotetext{a}{The names of the \chic\ sources with radio counterparts. If all the \chic\ sources in a \cxo\ observation of an AGPS source are coincident with the radio counterpart, then the range of suffixes are listed in parentheses. If only a subset of the \chic\ sources detected in a given \cxo\ observation are coincident with the radio counterpart, then only those suffixes are listed.}
\tablenotetext{b}{The radio Galactic plane surveys or observations visually inspected- MGPS: Molonglo Galactic Plane Survey, 90: 90cm Multi-configuration Very Large Array survey of the Galactic Plane, MAGPIS: The Multi-Array Galactic Plane Imaging Survey, VGPS: VLA Galactic Plane Survey. ATCA: Australia Telescope Compact Array observations obtained for radio follow-up of \chic\ sources (see Sections 2.4.2 and 3.4). }
\tablenotetext{c}{Type of radio source- SNR: Supernova Remnant; H {\sc ii}: H {\sc ii} region; MS: Massive star; CWB: colliding-wind binary; IRDC: Infrared Dark Cloud; AGN: Likely Active Galactic Nuclei based on a core-lobe morphology; D: Diffuse; Arc: Diffuse emission with an arc or shell morphology; Com: Compact; ND: No detection in the case of the ATCA observations.}
\tablenotetext{d}{Position of the \chic\ source relative to the radio source- C: Coincident, A: Adjacent; L: Limb (usually the limb or edge of a SNR, arc/shell structured diffuse emission or H {\sc ii} region); K: Core of a possible AGN given the apparent core-lobe morphology.}
\tablenotetext{e}{Many additional references to the SNRs are compiled in \citet{green09}.}

\tablerefs{1: \citet{shaver70b}; 2: \citet{haverkorn06}; 3: \citet{anderson11}; 4: \citet{whiteoak96}; 5: \citet{bocchino01}; 6: \citet{kuchar97}; 7: This paper; 8: \citet{mcclure01};  9: \citet{peretto09}; 10: \citet{helfand06}; 11: \citet{brogan06}; 12: \citet{caswell75}; 13: \citet{motch10}; 14: \citet{helfand89}; 15: \citet{condon98}; 16: \citet{ramoslarios10}}

\end{deluxetable*}

\begin{deluxetable*}{lcc}
\tablewidth{0pt}
\tabletypesize{\scriptsize}
\tablecaption{The chance of a random association between the listed \chic\ sources coincident with SNRs and a USNO B1 and GLIMPSE source\label{Tab10}}
\tablehead{
\colhead{\chic\ source} & \colhead{USNO B1 $N(<R)$\tablenotemark{a}} & \colhead{GLIMPSE $N(<M_{3.6})$\tablenotemark{b}}
\\
\colhead{ChI} & \colhead{$(\times10^{-2})$} & \colhead{$(\times10^{-2})$}
}
\startdata
J181213-1842\_1   & & 1.15
\\
J181213-1842\_2   & 3.05 & 7.15
\\
J181213-1842\_5   & 4.04 &
\\
J182435-1311\_1   & & 3.17
\\
J183356-0822\_6   & 1.68 &
\enddata

\tablecomments{Only those \chic\ sources where the chance of random alignment with a survey source is $>0.01$ are listed.}

\tablenotetext{a}{Chance of finding a USNO source with a $R$ magnitude brighter than the counterpart listed in Table~\ref{Tab7} within the \chic\ source's 95\% position error circle.}

\tablenotetext{b}{Chance of finding a GLIMPSE source with a $3.6 \mu$m magnitude brighter than the counterpart listed in Tables~\ref{Tab2} or \ref{Tab7} within the \chic\ source's 95\% position error circle.}

\end{deluxetable*}

\clearpage
\LongTables
\begin{landscape}
\begin{deluxetable*}{lccccccccccccc}
\tablewidth{0pt}
\tabletypesize{\scriptsize}
\setlength{\tabcolsep}{0.02in}
\tablecaption{Quantile spectral interpolations and luminosity estimates for all \chic\ sources within HII regions that have $>5$ X-ray counts\label{Tab11}}
\tablehead{
\colhead{\chic\ Source} & \multicolumn{4}{c}{Absorbed Mekal interpolation\tablenotemark{a}} & \colhead{} & \multicolumn{2}{c}{Distance\tablenotemark{b}} & \colhead{} & \multicolumn{2}{c}{Luminosities\tablenotemark{c}} & \colhead{} & \multicolumn{2}{c}{False associations ($\times10^{-2}$)\tablenotemark{d}} 
\\
\cline{2-5} \cline{7-8} \cline{10-11} \cline{13-14}
\\
\colhead{ChI} & \colhead{$kT$} & \colhead{$N_{H}$} & \colhead{$F_{x,abs}$} & \colhead{$F_{x,unabs}$} & \colhead{} & \colhead{(kpc)} & \colhead{Ref} & \colhead{} & \colhead{Log$L_{x,abs}$} & \colhead{Log$L_{x,unabs}$} & \colhead{} & \colhead{2MASS $N(<K_{s})$} & \colhead{GLIMPSE $N(<M_{3.6})$}
}
\startdata
J144519-5949\_2 & $1.52_{-0.73}^{+1.38}$ & $5.00_{-2.10}^{+3.60}$ & $23.70 \pm 4.80$ & $218.44 \pm 44.20$ & & 2.8 & 1 & & 32.35 & 33.31 & &   & \\  
J144519-5949\_3 & $0.78_{-0.57}^{+2.69}$ & $22.00_{-15.00}^{+51.00}$ & $13.05 \pm 4.73$ & $7207.60 \pm 2610.31$ & & 2.8 & 1 & & 32.09 & 34.83 & &  & \\  
J144519-5949\_5 & $>1.94$ & $0.14_{-0.14}^{+0.68}$ & $3.37 \pm 2.29$ & $3.98 \pm 2.71$ & & 2.8 & 1 & & 31.50 & 31.57 & &  & \\  
J151005-5824\_1 & $0.90_{-0.77}^{+1.80}$ & $7.60_{-4.60}^{+27.40}$ & $0.74 \pm 0.45$ & $47.76 \pm 28.64$ & & 4.7 & 2 & & 31.29 & 33.10 & &  & \\  
J151005-5824\_2 & $82.45$ & $2.80_{-2.07}^{+0.00}$ & $1.72 \pm 0.85$ & $2.92 \pm 1.45$ & & 4.7 & 2 & & 31.66 & 31.89 & &  & \\  
J151005-5824\_3 & $0.21_{-0.11}^{+1.42}$ & $1.60_{-1.35}^{+1.30}$ & $0.18 \pm 0.12$ & $65.24 \pm 44.33$ & & 4.7 & 2 & & 30.68 & 33.24 & &  & \\  
J151005-5824\_4 & $82.45$ & $2.20_{-1.53}^{+-0.10}$ & $1.21 \pm 0.73$ & $1.96 \pm 1.18$ & & 4.7 & 2 & & 31.51 & 31.71 & &  & \\  
J151005-5824\_5 & $82.45$ & $1.80_{-1.50}^{+0.00}$ & $1.14 \pm 0.68$ & $1.76 \pm 1.06$ & & 4.7 & 2 & & 31.48 & 31.67 & &  & \\
\enddata
\tablenotetext{a}{The quantile analysis interpolated absorbed Mekal model parameters including the temperature, $kT$, and absorption column density $N_{H}$ ($10^{22}$ cm$^{-2}$), as well as the absorbed and unabsorbed X-ray fluxes, $F_{x,abs}$ and $F_{x,unabs}$ ($10^{-14}$ erg cm$^{-2}$ s$^{-1}$), respectively, in the $0.3-8.0$ keV energy range. (If $kT=82.45$ then this indicates that the spectral interpolation of this \chic\ source has hit the hard limit for this parameter.)}
\tablenotetext{b}{The distances used to estimate the absorbed and unabsorbed luminosities. The corresponding reference is also listed.}
\tablenotetext{c}{The Log of the absorbed and unabsorbed luminosities, $L_{x,abs}$ and $L_{x,unabs}$ (erg s$^{-s}$), respectively, in the $0.3-8.0$ keV energy range.}
\tablenotetext{d}{Chance of finding a 2MASS (or GLIMPSE) source with a $K_{s}$ (or $3.6 \mu$m) magnitude brighter than the counterpart listed in Tables~\ref{Tab2} or \ref{Tab7}, within the \chic\ source's 95\% position error circle. This value is only quoted for the \chic\ sources where the chance of random alignment with a survey source is $>0.01$.}

\tablerefs{1: \citet{busfield06}; 2: \citet{russeil03}; 3: \citet{mcclure01}; 4: \citet{fich84}}

\end{deluxetable*}
\clearpage
\end{landscape}

\begin{deluxetable*}{llcccc}
\tablewidth{0pt}
\tabletypesize{\scriptsize}
\setlength{\tabcolsep}{0.03in}
\tablecaption{Confirmed and tentative identifications of the AGPS sources, including both literature and \chic\ survey identifications \label{Tab12}}
\tablehead{
\colhead{AGPS ID} & \colhead{\chic\ ID\tablenotemark{a}} & \colhead{Type\tablenotemark{b}} & \colhead{ID\tablenotemark{c}} & \colhead{Reference\tablenotemark{d}} & \colhead{Flag\tablenotemark{e}}
}
\startdata
AX J143148--6021 &  ChI J143148--6021\_3  & U & & 1 & n,R,n
\\
  AX J143416--6024  & &  RS CVn                  &    HD 127535                                    &   2,3*   & n,n,n
\\
  AX J144042--6001  & ChI J144042-6001\_1  & PMS              	     &    HD 128696                         &    2*  & n,n,n    
\\
   AX J144519--5949   &  ChI J144519-5949\_2(1,3-6)  &  WR and H {\sc ii}  &    GAL 316.8--00.1     & 1 & T,T,N
\\
   AX J144547--5931   &  ChI J144547-5931\_1   & OIf$^{+}$  	     &                    &    4,4      & I,I,n
\\
   AX J144701--5919   &  ChI J144701-5919\_1  & WR       	     &                               &  4,4 & I,I,n
\\
AX J145605--5913 & & U & &          & n,n,n
\\
 AX J145732--5901  & & U & & & F,n,n          
\\
\enddata
\end{deluxetable*}

\clearpage
\begin{deluxetable*}{ccccccccc}
\tablewidth{0pt}
\tabletypesize{\scriptsize}
\tablecaption{\cxo\ point sources with $>20$ X-ray counts detected between $3'-5'$ from the AGPS position \label{Tab13}}
\tablehead{
\colhead{AGPS Source\tablenotemark{a}} & \multicolumn{2}{c}{\texttt{wavdetect} Position} & \colhead{Position\tablenotemark{b}} & \colhead{Offset\tablenotemark{c}} & \colhead{Net Counts\tablenotemark{d}} & \colhead{Predicted Counts\tablenotemark{e}} & \colhead{Total Counts\tablenotemark{f}} & \colhead{2MASS Name}\\
\cline{2-3}
\\
\colhead{AX} & \colhead{RA} & \colhead{Dec} & \colhead{Error} & \colhead{from \asca} & \colhead{0.3-8} & \colhead{} & \colhead{} & \colhead{2MASS J}\\
\colhead{} & \colhead{(J2000)} & \colhead{(J2000)} & \colhead{($''$)} & \colhead{($'$)} & \colhead{(keV)} & \colhead{} & \colhead{$<3'$} & \colhead{}
}
\startdata
  J150436--5824 & 15:04:13.54 & -58:25:07.4 & 0.84 & 3.0 &  47.7 $\pm$ 8.0 & 139 & 154 &                     \\
  J154557--5443 & 15:46:09.08 & -54:39:12.9 & 0.89 & 4.2 &  48.0 $\pm$ 8.1 & 147 & 96  &    15460913--5439128 \\
               & 15:45:54.60 & -54:38:49.6 & 0.94 & 4.3 &  38.9 $\pm$ 7.4  &     &     &                     \\
  J154905--5420 & 15:49:12.40 & -54:16:30.2 & 0.94 & 4.1 &  31.8 $\pm$ 6.8 & 178 & 84  &    15491237--5416301 \\
               & 15:49:14.67 & -54:24:54.6 & 1.03 & 4.6 &  20.2 $\pm$ 5.7  &     &     &    15491463--5424549 \\
  J155035--5408 & 15:50:19.25 & -54:11:27.3 & 0.89 & 3.6 &  35.5 $\pm$ 7.1 & 152 & 242 &    15501925--5411271 \\
               & 15:50:05.97 & -54:07:23.1 & 1.07 & 4.5 &  22.8 $\pm$ 6.0  &     &     &                     \\
  J170444--4109 & 17:04:26.88 & -41:07:34.9 & 0.95 & 3.9 &  26.7 $\pm$ 6.3 & 95  & 170 &    17042687--4107351 \\
\enddata

\tablenotetext{a}{The AGPS source for which one or two X-ray point sources with $>20$ counts were detected in the corresponding \chic\ \cxo\ observation between $3'-5'$.}
\tablenotetext{b}{The total 95\% position error circle (in arcseconds) of the \cxo\ point source taking into account the \texttt{wavdetect} position error and the absolute astrometric accuracy of \cxo.}
\tablenotetext{c}{The offset, in arcminutes, between the AGPS position published in \citet{sugizaki01} and the \texttt{wavdetect} position of the \cxo\ point source.}
\tablenotetext{d}{The net number of counts in the $0.3-8.0$ energy range. The source counts were calculated using a source region with a radius equal to 95\% of the PSF at 1.5 keV. The total counts have been background subtracted and the error corresponds to the upper $1\sigma$ confidence limit calculated using \citet{gehrels86} statistics.}
\tablenotetext{e}{The number of X-ray counts that \cxo\ PIMMS predicts should be detected in the \chic\ \cxo\ observation based on the power law spectral fits measured by \citet{sugizaki01}.}
\tablenotetext{f}{The total number of X-ray counts (rounded to the nearest count) from all the sources detected in the \chic\ \cxo\ observations within $3'$ of the AGPS position.}

\end{deluxetable*}

\end{document}